\documentclass{raa2}   % two columns
\usepackage{graphicx,times}
\usepackage{natbib}
\usepackage{amssymb,amsmath}

\usepackage{CJKutf8}
\usepackage[varg]{txfonts}

\usepackage{tabularx}
\usepackage{array}

\usepackage{amsmath}
\usepackage{units}
\usepackage{ulem}

\usepackage[pagebackref=true]{hyperref}

\usepackage[utf8]{inputenc}
\usepackage{float}

\newcolumntype{L}[1]{>{\raggedright\let\newline\\\arraybackslash\hspace{0pt}}m{#1}}
\newcolumntype{C}[1]{>{\centering\let\newline\\\arraybackslash\hspace{0pt}}m{#1}}
\newcolumntype{R}[1]{>{\raggedleft\let\newline\\\arraybackslash\hspace{0pt}}m{#1}}

\newcommand{\Heii}{\ion{He}{ii\ }}
\newcommand{\Siii}{\ion{Si}{ii\ }}
\newcommand{\Nai}{\ion{Na}{i}\,D\ }

\hypersetup{colorlinks = true, linkcolor=red, citecolor=blue, urlcolor=blue}

\bibpunct{(}{)}{;}{a}{}{,}

\begin{document}

   \title{Type Ia Supernova Explosions in Binary Systems: A Review}

 \volnopage{ {\bf 2020} Vol.\ {\bf 20} No. {\bf XX}, 000--000}
   \setcounter{page}{1}

\author{Zheng-Wei Liu
   \inst{1,2,3,4}, Friedrich K. R\"{o}pke\inst{5,6} and Zhanwen Han\inst{1,2,3,4}}

   \institute{ Yunnan Observatories, Chinese Academy of Sciences (CAS), 
               Kunming 650216, China;\\ {\it zwliu@ynao.ac.cn}\\
%% Please give the E-mail address of the author, to whom future correspondence and
%% offprint requests will be sent.
        \and
             Key Laboratory for the Structure and Evolution of Celestial Objects, CAS, Kunming 650216, China\\
	\and
    International Centre of Supernovae, Yunnan Key Laboratory, Kunming 650216, China\\
\and 
University of Chinese Academy of Science, Beijing 100012, China\\
     \and
     Zentrum f{\"u}r Astronomie der Universit{\"a}t Heidelberg, Institut f{\"u}r Theoretische Astrophysik, Philosophenweg 12, 69120 Heidelberg, Germany\\
     \and
     Heidelberger Institut f{\"u}r Theoretische Studien, Schloss-Wolfsbrunnenweg 35, 69118 Heidelberg, Germany\\
\vs \no
  % {\small Received 2020 June 30; accepted 2020 XXX}
}
 
\abstract{Type Ia supernovae (SNe~Ia) play a key role in the fields of astrophysics and cosmology. It is widely accepted that SNe~Ia arise from thermonuclear explosions of white dwarfs (WDs) in binary systems. However, there is no consensus on the fundamental aspects of the nature of SN~Ia progenitors and their actual explosion mechanism. This fundamentally flaws our understanding of these important astrophysical objects. In this review, we outline the diversity of SNe~Ia and the proposed progenitor models and explosion mechanisms. We discuss the recent theoretical and observational progress in addressing the SN~Ia progenitor and explosion mechanism in terms of the observables at various stages of the explosion, including rates and delay times, pre-explosion companion stars, ejecta--companion interaction, early excess emission, early radio/X-ray emission from circumstellar material (CSM) interaction, surviving companion stars, late-time spectra and photometry, polarization signals, and supernova remnant properties, etc. Despite the efforts from both the theoretical and observational side, the questions of how the WDs reach an explosive state and what progenitor systems are more likely to produce SNe~Ia remain open. No single published model is able to consistently explain all observational features and the full diversity of SNe~Ia. This may indicate that either a new progenitor paradigm or the improvement of current models is needed if all SNe~Ia arise from the same origin.  An alternative scenario is that different progenitor channels and explosion mechanisms contribute to SNe~Ia. In the next decade, the ongoing campaigns with the \textit{James Webb Space Telescope (JWST)}, \textit{Gaia} and the \textit{Zwicky Transient Facility (ZTF)}, and upcoming extensive projects with the Vera C.\ Rubin Observatory \textit{Legacy Survey of Space and Time (LSST)} and the \textit{Square Kilometre Array (SKA)} will allow us to conduct not only studies of individual SNe~Ia in unprecedented detail but also systematic investigations for different subclasses of SNe~Ia. This will advance theory and observations of SNe~Ia sufficiently far to gain a deeper understanding of their origin and explosion mechanism.
}

   \authorrunning{{\it Z. Liu et al.}: SN~Ia explosion in binary systems}            %author_head in even pages
   \titlerunning{{\it Z. Liu et al.}: SN~Ia explosion in binary systems}  % title_head in odd pages
   \maketitle

%________________________________________________ sections below
%
\section{Introduction}  %% first-level sections will be auto-capitalized
\label{sect:introduction}

Supernovae (SNe) are highly energetic explosions of some stars, that are so bright that they can outshine an entire galaxy. Their typical bolometric luminosities reach the order of $10^{43}\,\mathrm{erg\,s^{-1}}$, which is about ten billion times the solar luminosity. SNe play an important role in the fields of astrophysics and cosmology because they have been used as cosmic distance indicators, and they are heavy-element factories (especially for intermediate mass and iron-group elements), kinetic-energy sources, and cosmic-ray accelerators in galaxy evolution. SNe are also key players in the formation of new-generation stars by triggering the collapse of molecular clouds. SNe are generally classified into two main categories according to spectroscopical features: Type~I and Type~II SNe \citep{Minkowski1941,Filippenko1997, Parrent2014}. Type~I SNe have no hydrogen (H) lines in their spectra whereas Type~II SNe  contain obvious H lines. Type Ia SNe (SNe~Ia) are a subclass of Type~I which exhibit strong singly ionized silicon (Si) absorption (\Siii at $6150$, $5800$ and $4000\,\AA$) feature in their spectra.

SNe~Ia are widely thought to be thermonuclear explosions of white dwarfs (WDs) in binary systems \citep{Hoyle1960}. They have been found to occur in all galaxy types. Their typical peak luminosity in the $B$-band is about $M_{B}=-19.5\,\mathrm{mag}$, and the typical kinetic energy is ${\sim}\,10^{51}\,\mathrm{erg}$ \citep{Branch1993,Hillebrandt2000}. The light curves of SNe~Ia are powered
by the Comptonization of $\gamma$-rays produced by the radioactive decay of $^{56}\mathrm{Ni}$ $\rightarrow$ $^{56}\mathrm{Co}$ $\rightarrow$ $^{56}\mathrm{Fe}$, with respective half-life times of 6.1 and 77 days \citep{Truran1967,Colgate1969,Arnett1982,Hillebrandt2013}. SNe~Ia have been successfully used as cosmic distance indicators to constrain cosmological parameters, which has led to the discovery of the accelerating expansion of the Universe \citep{Riess1998,Perlmutter1999,Schmidt1998} -- a breakthrough awarded with the 2011 Nobel Prize in physics.  Despite their importance and far-reaching implications, the specific progenitor systems as well
as the explosion mechanism of SNe~Ia remains enigmatic \citep[e.g.][]{Hillebrandt2000,Hillebrandt2013,Wang2012,Maoz2014,Ruiz-Lapuente2014, Maeda2016,Livio2018, Roepke2018,Soker2019}. This affects the reliability of necessary assumptions such as those of universality of their calibration as distance indicators. Recently, it was found that the local measurements of the Hubble constant ($H_{0}$) based on SNe~Ia is inconsistent with the value inferred from the cosmic microwave background radiation observed by the $Planck$ satellite assuming a $\Lambda\mathrm{CDM}$ cosmological world model \citep{Planck2020}. To determine whether this so-called ``$H_{0}$ tension'' hints to new physics, it is critical to improve our understanding of SNe~Ia and, more specifically, their progenitors and explosion mechanisms.

\begin{figure*}[t]
   \centering
   \includegraphics[width=0.95\textwidth, angle=0]{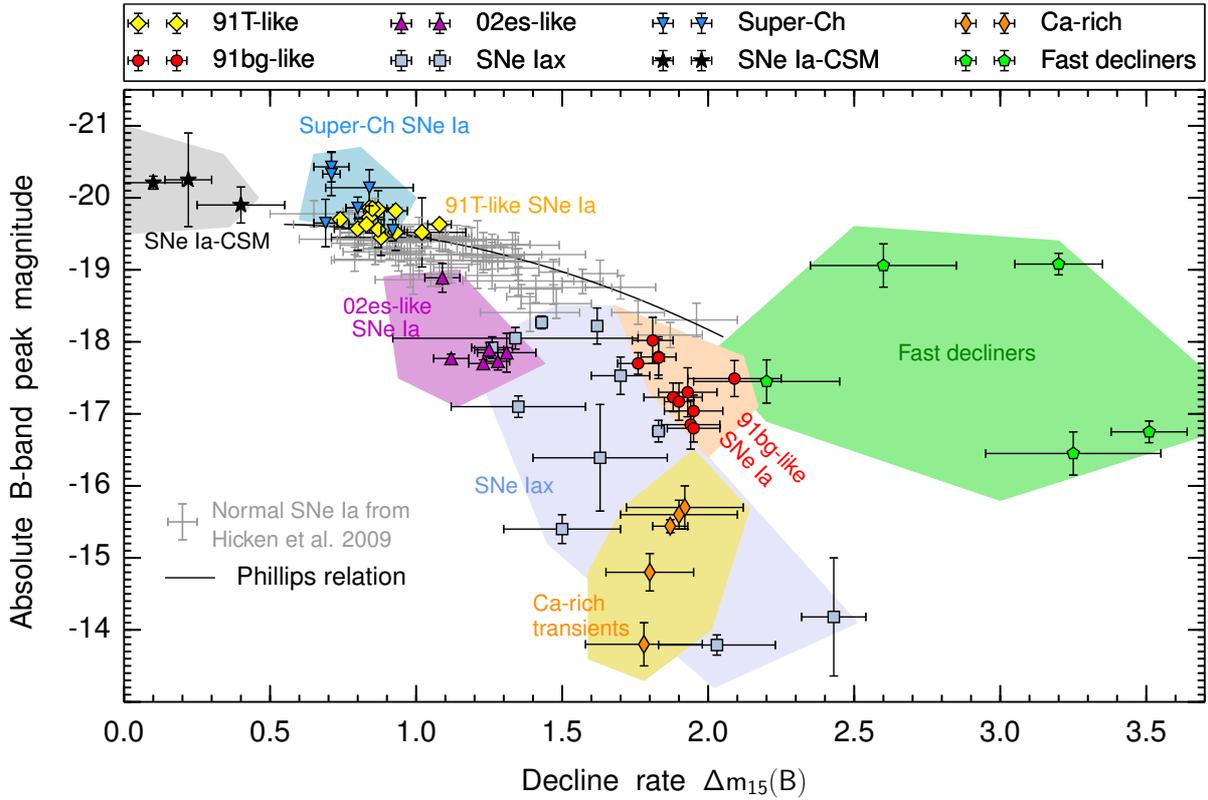}
   \caption{Distributions of normal SNe~Ia and different subclasses in the peak luminosity vs.\ light curve width ($\mathrm{\Delta m_{15}(B)}$; \citealt{Phillips1993}) diagram. Figure is reproduced based on Figure~1 of \citet{Taubenberger2017}.}
   \label{Fig1}
   \end{figure*}

\section{The diversity of SN\lowercase{e}~I\lowercase{a}}
\label{sec:diversity}

A large fraction of observed SNe~Ia (${\sim}\,70\%$) is found to show remarkable homogeneity and quantifiable heterogeneity, and they exhibit a clear empirical relationship between light curve width and peak luminosity, i.e., the so-called ``Phillips relation'' (sometimes known as the width–luminosity relation [WLR], \citealt{Pskovskii1977,Phillips1993, Phillips1999}). These SNe~Ia are usually referred to as ``normal SNe~Ia'', and they have long been used as standardizable candles for measuring cosmological distances \citep{Leibundgut2001,Leibundgut2008,Goobar2011}. However, an increasing number of SNe~Ia has been observed that does not follow the Phillips relation (see Fig.~\ref{Fig1}), and they are diverse in their observational characteristics such as light curve shape, peak luminosity and spectral features \citep[][]{Benetti2005, Blondin2012, Jha2017}. For these reasons, SNe 
Ia have been classified into different sub-classes diverging from normal events, which include 1991T-likes \citep[][]{Filippenko1992t, Phillips1992}, 1991bg-likes \citep[][]{Filippenko1992bg, Leibundgut1993}, SNe~Iax (i.e., SN~2002cx-likes; \citealt{Li2003, Foley2013}), 2002es-likes \citep[][]{Ganeshalingam2012}, Ca-rich objects (i.e., SN~2005E-like; \citealt[][]{Perets2010, Kasliwal2012}), super-Chandrasekhar objects (i.e., SN~2003fg-likes; \citealt[][]{Howell2006, Hicken2007}), SNe~Ia-CSM (\citealt[][]{Hamuy2003}) and fast decliners \citep{Taubenberger2017}. The diversity of SNe~Ia has recently been reviewed by \citet{Taubenberger2017}, we only skim the surface here \citep[see also][]{Jha2019,Ruiter2020}. 

\textbf{1991T-like objects} form a luminous, slow-declining subclass of SNe~Ia, named after the well-observed SN~1991T \citep{Filippenko1992t,Jeffery1992,Phillips1992}. Their optical spectra at pre-maximum phases show extremely weak \ion{Ca}{II}~H~\&~K and \ion{Si}{II}~$\lambda6355$ and strong \ion{Fe}{III} absorption features. 91T-like SNe are expected to be on average $0.2$--$0.5\,\mathrm{mag}$ more luminous than normal SNe~Ia with similar decline rate \citep{Blondin2012,Phillips2022}. 1991T-like SNe are found preferentially in late-type galaxies, suggesting that they are likely associated with young stellar populations \citep{Li2011a}. It has been suggested that 1991T-like SNe could contribute  $2\%$--$9\%$ to all SNe~Ia in the local Universe \citep{Li2011a, Leaman2011,Blondin2012}.

\textbf{1991bg-like objects} are a cool, subluminous, and fast-declining subclass of SNe~Ia with low ejecta velocity \citep{Filippenko1992bg,Ganeshalingam2012}. Typically, they are fainter than normal SNe~Ia in optical band up to $2.5\,\mathrm{mag}$ \citep{Sullivan2011}. Their spectra at maximum light show strong \ion{Ti}{II} absorption, indicating a relatively cool photosphere. 1991bg-like SNe are found preferentially in early-type (i.e., passive) galaxies \citep{Howell2001,Li2011a}. Only few 1991bg-like SNe have been found in spiral galaxies. This suggests old stellar populations for the progenitors of 1991bg-like SNe. There is no agreement about the rates of 1991bg-like SNe in the literature, estimates range from $6\%$--$15\%$ of all SNe~Ia \citep{Ganeshalingam2010,Li2011a,Silverman2012sample}.

\textbf{SNe~Iax} are proposed as a hot, sublumious, subclass of SNe~Ia \citep[e.g.][]{Li2003, Foley2013}. SNe~Iax are fainter than normal SNe~Ia and highly skewed to late-type galaxies \citep[e.g.][]{Foley2013}. Their explosion ejecta are characterized by low expansion velocities and show strong mixing features. Their maximum-light spectra show similar features to those of 1991T-like SNe, which are characterized by weak \ion{Si}{II}~$\lambda6355$ features and dominated by \ion{Fe}{III} lines. In addition, strong He lines are identified in spectra of two events, i.e., SN~2004cs and SN~2007J. The late-time spectra of SNe~Iax are dominated by narrow permitted Fe~II \citep{Jha06,Jha2017}. It has been suggested that they contribute about 1/3 of total SNe~Ia \citep[][]{Li01,Foley2013}.

\textbf{2002es-like objects} are another cool, rapidly fading, subluminous subclass of SNe~Ia which have a peak luminosity and ejecta velocity (${\sim}\,6000\,\mathrm{km\,s^{-1}}$) similar to SN~2002cx \citep{Ganeshalingam2012,White2015}. Their spectra at near maximum light phases share some characteristics in common with the subluminous 1991bg-like SNe, which are clearly characterized by strong \ion{Ti}{II}, \ion{Si}{II}, and \ion{O}{II} absorption features \citep{Taubenberger2017}. However, 2002es-like SNe do not have the fast-declining light curves characteristic of 1991bg-like events. \citet{White2015} suggested that 2002es-like events tend to explode preferentially (but not exclusively) in massive, early-type galaxies \citep{Li2011a}. \citet{Ganeshalingam2012} suggested that SN~2002es-like objects should account for ${\sim}\,2.5\%$ of all SNe~Ia.

\textbf{Calcium-rich objects}  (Ca-rich) constitute a peculiar subclass of SNe~Ia with SN~2005E as a prototype \citep{Perets2010,Kasliwal2012,De2020}. Ca-rich SNe are primarily characterized by peak magnitudes of $-14$ to $-16.5\,\mathrm{mag}$, rapid photometric evolution with typical rise times of $12$--$15\,\mathrm{days}$, and strong Ca features in nebular phase spectra \citep{Kasliwal2012}. They exhibit low ejecta and $^{56}\mathrm{Ni}$ masses of $\lesssim0.5\,M_{\sun}$ and $\lesssim0.1\,M_{\sun}$, respectively \citep{Lunnan2017}. The majority of Ca-rich SNe has been observed in early type galaxies \citep{Kasliwal2012,Lyman2013} and the inferred rates of such SNe are likely in the range of $5\%$--$20\%$ of the normal SN~Ia rates\footnote{The fact that the rate estimates of sub-classes of SNe Ia do not add up to 100\% indicates that there are large uncertainties on the estimated contributions of different sub-classes.} \citep{Perets2010,Kasliwal2012, De2020}.

\textbf{Super-Chandrasekhar objects} are sometimes known as SN~2003fg-like SNe \citep{Howell2006,Hicken2007,Scalzo2010,Silverman2011,Hsiao2020,Srivastav2023}. They are referred to as ``super-Chandrasekhar SNe'' because a differentially rotating WD with a super-Chandrasekhar mass of ${\sim}\,2.0\,M_{\sun}$ was used to interpret the observations of SN~2003fg \citep{Howell2006}. The main features of this subtype are summarized by \citet{Ashall2021}: They are generally characterised by high luminosities ($B$-band peak absolute magnitudes of $\mathrm{-19<M_{B}<-21\,mag}$), broad light curves ($\Delta \mathrm{m_{15}(B)}<1.3\,\mathrm{mag}$, defined as the decline in the $B$-band magnitude light curve from peak to 15 days later), and relatively low ejecta velocities. This is puzzling for a theoretical explanation: the first two properties point to a powerful explosion which seems to be at odds with the low ejecta velocities. They have only one $i$-band maximum which peaks after the epoch of the $B$-band maximum, but with weak (or without) $i$-band secondary maximum. Their maximum-light spectra do not show a \ion{Ti}{II} feature; in addition, their nebular-phase spectra are characterized by a low ionization state. Super-Chandrasekhar SNe seem to be preferentially found in low-mass galaxies, indicating that they prefer a low-metallicity environment \citep{Taubenberger2011}. They seem to make up a small fraction of SNe~Ia, but their exact rates are still unknown \citep{Silverman2011,Taubenberger2017, Ashall2021}.

\textbf{SNe~Ia-CSM}  are a subclass named after the discovery of SN~2002ic \citep{Hamuy2003,Deng2004}, although there is still a debate on whether these objects are SNe~Ia or in fact core-collapse SNe \citep{Benetti2006,Silverman2013,Inserra2014}. A list of several common features of SNe~Ia-CSM has been compiled by \citet{Silverman2013}. They have a range of $R$-band peak absolute magnitudes of $\mathrm{-19<M_{R}<-21.3\,mag}$, and they exhibit narrow hydrogen emission features in their spectra \citep{Dilday2012,Silverman2013}. The presence of narrow H lines is thought to arise from circumstellar-material (CSM), which is strongly indicative of mass loss (or outflows) of the progenitor system prior to the SN explosion. An initial systematic study of this subclass has been presented by \citet{Silverman2013}, and it has been recently updated by \citet{Sharma2023}. SNe~Ia-CSM are preferentially found in late-type spirals and irregular galaxies, indicating the origin from a relatively young stellar population \citep{Silverman2013}. The rate of SNe~Ia-CSM is estimated to be no more than a few per cent of the SN~Ia rates \citep{Han2006,Dilday2012, Silverman2013, Gal-Yam2017, Dubay2022,Sharma2023}.

  \textbf{Fast decliners} are rare and the extremely rapidly declining SNe. So far, this class includes SN~1885A, SN~1939B, SN~2002bj, SN~2005ek, SN~2010X \citep{Poznanski2010,Kasliwal2010,Perets2011apj,Drout2013,Taubenberger2017}. Whether these peculiar objects arise from thermonuclear explosions of  WDs or core-collapse explosions of massive stars remains open \citep{Bildsten2007,Shen2010,Kasliwal2010,Gal-Yam2017}. There is no conclusion on whether or not all of these objects actually belong to the same class of events \citep{Taubenberger2017, Jha2019}.

\begin{figure*}[ht]
   \centering
   \includegraphics[width=0.95\textwidth, angle=0]{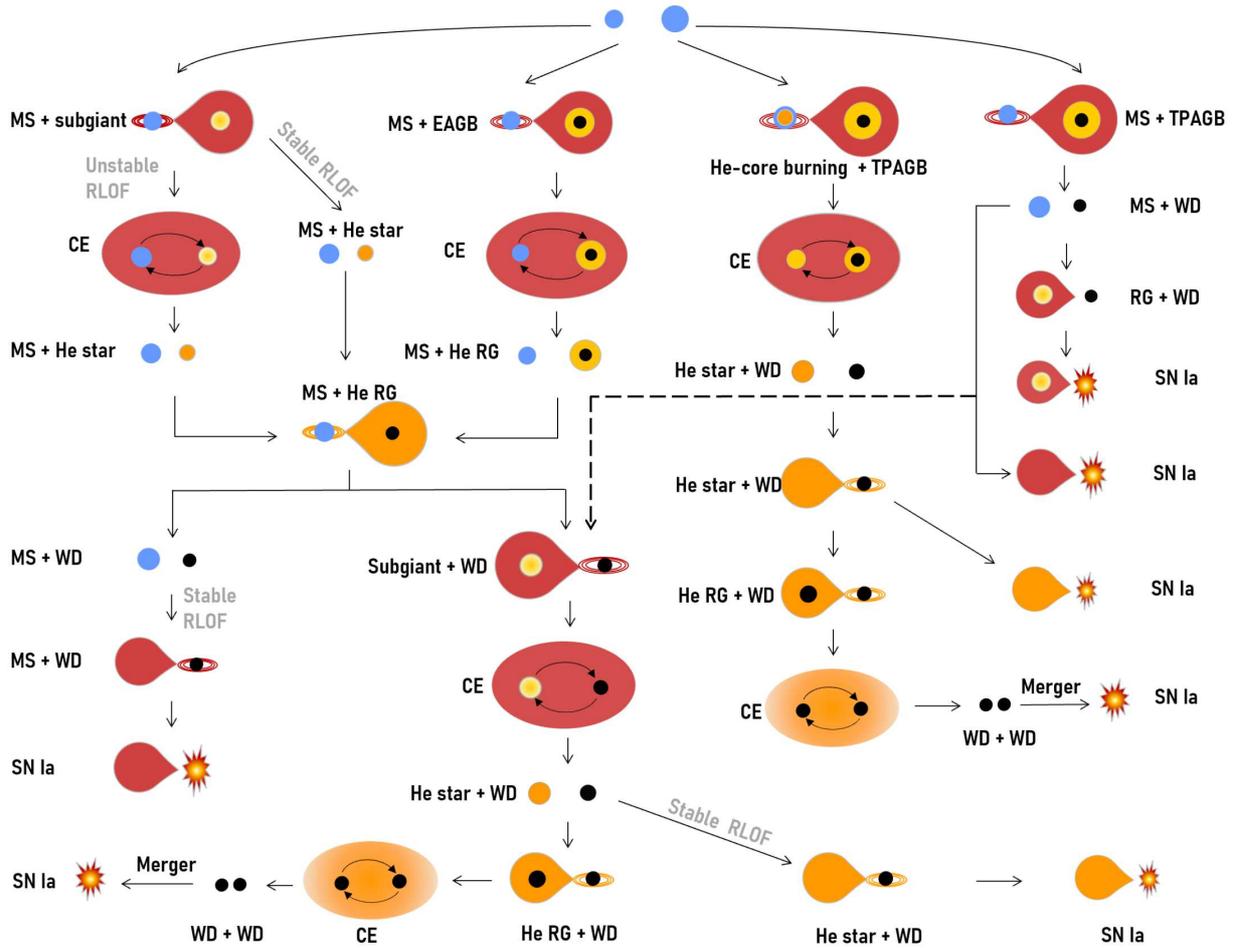}
   \caption{Schematic illustration of binary evolutionary paths for SNe Ia in the SD and DD scenario (see also \citealt{Wang2018}). Note that evolutionary channels here are not complete and that new channels may still be proposed in the future.}
   \label{Fig:channels}
   \end{figure*}

  \section{Progenitors and explosion mechanisms}
  \label{sec:models}

It is widely accepted that SNe~Ia arise from thermonuclear explosions of white dwarfs (WDs) in binary systems \citep{Hoyle1960}. However, there is no consensus on the fundamental aspects of the nature of SN~Ia progenitors and their explosion mechanism from both, the theoretical and observational side \citep[see, e.g.,][for  reviews]{Wang2012, Hillebrandt2013, Maoz2014, Maeda2016,Livio2018, Soker2019}. In this section, potential progenitor models and explosion mechanisms of SNe~Ia are briefly summarized.

\subsection{Progenitor scenarios}
\label{sec:progenitor}

\subsubsection{Single-degenerate scenario}
\label{sec:sd}

In the single-degenerate (SD) scenario, a WD accretes hydrogen-rich or helium-rich  material from a non-degenerate companion star through Roche-lobe overflow (RLOF) or stellar wind until its mass approaches the Chandrasekhar-mass (${\approx}\, 1.4\,M_{\sun}$), at which point a thermonuclear explosion ensues \citep{Whelan1973, Nomoto1982a, Nomoto1982b, Nomoto1984}.  The companion star could be either a main-sequence (MS) star, a subgiant (SG), a red giant (RG), an asymptotic giant-star (AGB), or a He star \citep{Hachisu1999,Han2004,Ruiter2009,Meng2009,Wang2009}. It has been suggested that a Chandrasekhar-mass WD can undergo a deflagration, or a detonation \citep[see, e.g.,][for an overview of thermonuclear combustion in white dwarfs]{Roepke2017hsn}, or a delayed detonation to lead to a SN~Ia explosion \citep[e.g.][]{Arnett1969, Nomoto1984,Woosley1986, Khokhlov1989, Hoeflich1995, Plewa2004, Roepke2005, Roepke2007a, Roepke2007b, Fink2014, Marquardt2015,Seitenzahl2016}. In the SD scenario, SNe~Ia are thought to arise from Chandrasekhar-mass WDs (double detonation explosions of sub-Chandrasekhar mass WDs could happen when accreting from a He-star companion; see Section~\ref{sec:DDet}), the homogeneity of the majority of SNe~Ia therefore can be well explained by this scenario. A schematic illustration of main binary evolutionary paths for producing SNe Ia in the SD scenario is given in Fig.~\ref{Fig:channels}.

One of the key questions in the SD scenario is how the WD retains the accreted companion material and grows in mass to approach the Chandrasekhar limit (i.e. the mass-retention efficiency of onto the WD). The SD scenario requires that the WD accretes material at a relatively narrow range of accretion rates of a few $\times$ ($10^{-8}$--$10^{-7}\,M_{\sun}\,\mathrm{yr^{-1}}$) to allow steady burning of accreted material \citep{Paczynski1976,Fujimoto1982a,Fujimoto1982b,Livio1989,Nomoto2007,Shen2007,Wolf2013,Piersanti2014, Wang2018}, which causes difficulties for explaining the observed nearby SN~Ia rate (see Section~\ref{sec:rates}). Moreover, some recent observations seem to pose a challenge to the SD scenario (see Section~\ref{sec:observables}) such as the missing of surviving companion stars in supernova remnants (SNRs) \citep{Kerzendorf2012,Schaefer2012,Ruiz-Lapuente2018}, the absence of swept-up H/He in their late spectra \citep{Leonard2007,Tucker2020} and low X-ray flux from nearby elliptical
galaxies \citep{Gilfanov2010,Woods2017,Kuuttila2019}.  In addition, although the SD scenario makes the explosion rather homogenious, it turns out to be difficult to cover the observed ranges in brightness and decline rates in this scenario. However, to conclude whether the SD scenario is promising for producing the majority of SNe~Ia requires comparing a full range of predicted observational consequences from this scenario with the observations of SNe~Ia (see discussions in Section~\ref{sec:observables}).

A number of candidate progenitors have been suggested for the SD scenario, including cataclysmic variable stars like classic novae, recurrent novae and dwarf novae \citep{Webbink1987,Hachisu2001,Warner2003}), supersoft X-ray sources \citep{VandenHeuvel1992}, symbiotic systems \citep{Webbink1987,Yungelson1998} and WD~+~hot-subdwarf binaries \citep[see Section~\ref{sec:DDet};][]{Iben1994,Nelemans1998,Geier2013}.

In the SD scenario, a WD accretes and retains companion matter that carries angular momentum. As a consequence the WD spins with a short period which leads to an increase of the critical explosion mass. If the critical mass is higher than the actual mass of the WD, the SN explosion could only occur after the WD increases the spin period with a specific spin-down timescale. This scenario is known as the ``spin-up/spin-down model''  \citep{Di-stefano2011,Justham2011,Hachisu2012a}. In this model, if the spin-down timescale is longer than about $10^{6}\,\rm{yrs}$, the CSM around the progenitor system could become diffuse and reach a density similar to that of the ISM. This could explain the lack of radio and X-ray emission from SNe~Ia in agreement with the current radio and X-ray observations (\citealt{Margutti2012,Chomiuk2016,Lundqvist2020}; see Section~\ref{sec:radio}). Also, the H-rich or He-rich companion star (i.e., MS, subgiant, RG and He star) may shrink rapidly before the SN~Ia explosion occurs by exhausting most of its H-rich or He-rich envelope during a long spin-down ($\gtrsim10^{8}\,\rm{yrs}$) phase to become a WD or a hot subdwarf star \citep{Hachisu2012b,Meng2019,Meng2021}. This would explain the non-detection of a pre-explosion companion star (\citealt{Maoz2008,Bloom2012,Kelly2014}; see Section~\ref{sec:pre-explosion}) in SNe~Ia and the absence of swept-up H/He in their late spectra (\citealt{Leonard2007,Tucker2020}; see Section~\ref{sec:late-spectra}). However, no (or weak) interaction signature of shocked gas is predicted in this scenario, which makes it difficult to explain the early excess luminosity (see Section~\ref{sec:early}) seen in some SNe~Ia such as iPTF14atg \citep{Cao2015} and SN~2012cg \citep{Marion2016}. The exact spin-down timescale of the WD in this model is uncertain \citep{Hachisu2012a,Meng2013,Maoz2014,Wang2014b} but a key to the success of the model.

\subsubsection{Double-degenerate scenario}
\label{sec:dd}

In the original double-degenerate (DD) scenario, two carbon-oxygen (CO) WDs in a binary system are brought into contact by the emission of gravitational wave radiation and merge via tidal interaction into one single object, triggering a SN~Ia explosion if the combined mass exceeds the Chandrasekhar-mass limit \citep{Webbink1984,Iben1984b}. There are a number of evolutionary paths that can lead to SN~Ia explosions in the DD scenario \citep[][see Fig~\ref{Fig:channels}]{Han1998,Toonen2012,Yungelson2017,Liu2018dd}. The key question of the original DD scenario is whether the merger of two WDs could successfully lead to an SN~Ia explosion. Different calculations have predicted that the merger of two WDs would likely cause the formation of neutron stars through accretion-induces collapse (AIC) rather than SN~Ia explosions \citep{Saio1998,Nomoto1985,Kawai1987,Timmes1994,Shen2012,Schwab2016,Schwab2021}. The accretion from the secondary WD onto the primary WD during the merger process may lead to burning in the outer layers of the WD rather than central burning, which would turn the original carbon-oxygen WD into an oxygen-neon-magnesium (ONeMg) WD. A Chandrasekhar-mass ONe WD is thought to be prone to collapse into a neutron star via AIC. However, there are possibilities to avoid AIC after the merger of two CO WDs. For instance, \citet{Yoon2007} concluded that the merger of two CO WDs could avoid off-center C-burning and explode as an SN~Ia in the thermal evolution phase if the rotation of the WDs is taken into account.

In the past decades, a number of numerical simulations have investigated the merger of two WDs \citep{Benz1990,Rasio1995, Segretain1997,Guerrero2004,Loren2009,Fryer2010,Pakmor2010,Pakmor2012b,Dan2011,Raskin2012,Raskin2014,Moll2014,Sato2015}. More importantly, some recent theoretical studies have shown that the merger of two WDs can eventually trigger an SN~Ia explosion in ways that are different from the original DD scenario. For instance, a carbon detonation can be directly triggered by the
interaction of the debris of the secondary WD with the primary WD during the violent merger phase of two CO~WDs to eventually trigger an SN~Ia explosion (i.e. the ``violent merger model''; see Section~\ref{sec:carbon-violent}; \citealt{Pakmor2010,Pakmor2012b,Sato2015}). If the secondary WD in a DD binary system is a pure He WD, an initial He detonation could be triggered by accumulating a He shell on top of the primary CO~WD through stable mass transfer, eventually triggering the C-core detonation near the center to successfully cause an SN~Ia. This corresponds to the sub-Chandrasekhar-mass double-detonation scenario (see Section~\ref{sec:DDet}; \citealt{Fink2007, Fink2010, Moll2013, Gronow2020, Gronow2021, Boos2021}). In addition, unstable mass transfer could also lead to the presence of He in the surface layers of the primary CO~WD if the secondary WD is either a He WD or a hybrid HeCO~WD, which could successfully give rise to an SN~Ia during the coalescence itself through the double detonation mechanism (i.e. the He-ignited violent merger model decribed in Section~\ref{sec:d6}; \citealt{Guillochon2010,Dan2011,Pakmor2013,Pakmor2021,Pakmor2022,Roy2022}).

\begin{table*}\renewcommand{\arraystretch}{2.0}
\fontsize{7}{11}\selectfont
\begin{center}
\caption{Progenitor scenarios of SNe Ia.} \label{table:1}
\centering
\begin{tabular}{|L{1.7cm}||L{1.5cm}|L{1.5cm}|L{1.2cm}|L{1.4cm}|L{1.5cm}|L{1.5cm}|L{1.5cm}|L{1.2cm}|} 
\hline
                      % To combine 4 columns into a single one
Scenario & Bright pre-SN object expected?  & H/He present in late-time spectra & CSM expected  & Reproduction of SN Ia rate?   & Reproduction of SN Ia DTDs & Reproduction of SN Ia brightness distribution & Associated thermonuclear explosion scenario  & Surviving companion?  \\

\hline \hline
   single-degenerate (SD general) &  yes  & yes & yes & no & no & no &  $\mathrm{M_{Ch}}$ & yes  \\
\hline
   SD with MS donor & yes  & yes & yes & no & no & no &  $\mathrm{M_{Ch}}$ & MS star \\
\hline
   SD with giant donor & yes  & yes & yes & no & no & no &  $\mathrm{M_{Ch}}$ & compact stellar core  \\
\hline
   SD with He-star donor & yes  & yes & yes & no & no & no &  $\mathrm{M_{Ch}}$ & He star   \\
\hline
   SD with M-dwarf donor & no  & yes & yes & no & no & no &  $\mathrm{M_{Ch}}$ & M-dwarf star   \\
\hline
   SD spin-up/spin-down & unclear  & unclear & unclear & no & unclear & unclear &  $\mathrm{M_{Ch}}$, super-$\mathrm{M_{Ch}}$ & compact object   \\
\hline
   SD with hybrid CONe WD & yes  & unclear & yes & no & no & no &  $\mathrm{M_{Ch}}$, sub-$\mathrm{M_{Ch}}$ & unclear   \\
\hline
   double-degenerate (DD general) & no  & no & yes/no, depending on explosion mechanism & yes & yes & yes &  sub-$\mathrm{M_{Ch}}$, $\mathrm{M_{Ch}}$, super-$\mathrm{M_{Ch}}$ & yes/no, depending on explosion mechanism   \\
\hline
   core degenerate & yes  & yes & unclear & unclear & unclear & unclear &  $\mathrm{M_{Ch}}$ & no \\
\hline
   triple system & unclear  & unclear & unclear & unclear & unclear & unclear &  unclear & unclear\\
   \hline
   single star & yes  & yes & no & unclear & unclear & unclear &  $\mathrm{M_{Ch}}$, e-capture induced & no \\
\hline
\end{tabular}
\end{center}
\end{table*}

There are some evidences in favour of the DD scenario (see Sections~\ref{sec:rates} and~\ref{sec:observables} for a detailed discussion). Binary population synthesis (BPS) calculations have shown that the predicted SN~Ia rates and delay times from the DD scenario could well reproduce those inferred from the observations \citep[][]{Nelemans2001,Ruiter2009,Ruiter2013,Mennekens2010,Toonen2012,Yungelson2017}. In addition, the non-detection of pre-explosion companion stars in normal SNe~Ia \citep{Li2011c, Bloom2012, Kelly2014}, the lack of radio and X-ray emission around peak brightness \citep{Chomiuk2012,  Horesh2012, Margutti2014}, the absence of a surviving companion star in SN~Ia remnants \citep{Kerzendorf2009, Schaefer2012, Ruiz-Lapuente2018}, and the fact that no signatures of the swept-up H/He have been detected in the nebular spectra of SNe~Ia \citep{Leonard2007, Lundqvist2013, Maguire2016} and the lack of X-ray flux (i.e. supersoft X-ray sources) expected for accreting WDs seem to favour the DD scenario. Also, it has been suggested that some superluminous SNe~Ia that have ejecta masses of $\gtrsim2.0\,M_{\sun}$ may arise from the merger of two WDs \citep{Howell2006,Hicken2007,Silverman2011}. However, the DD scenario predicts a relatively wide range of explosion masses and thus makes it difficult to explain the observed homogeneity of the majority of SNe~Ia.

Double WDs (DWDs) are the primary targets of some upcoming space gravitational-wave missions and observatories such as the \textit{Laser Interferometer Space Antenna} (LISA; \citealt{Amaro-Seoane2017}, \textit{Tianqin} \citep{Luo2016,Huang2020} and \textit{Taiji} \citep{Ruan2018}. Searches for DWDs have been carried out by different surveys like the dedicated ESO \textit{Supernovae type Ia Progenitor surveY} (SPY; \citealt{Napiwotzki2001}), the \textit{Sloan Digital Sky Survey} (SDSS; \citealt{York2000}), the \textit{SWARMS survey} \citep{Badenes2009}, the \textit{Extremely Low Mass (ELM) survey} \citep{Brown2010}, the \textit{Kepler-K2 survey} \citep{Howell2014}, and the large all-sky survey \textit{Gaia} \citep[][]{Gaia2016,Gaia2018}. However, to date, only about 150 DWD systems have been detected with detailed orbital parameters \citep{Badenes2009, Hallakoun2016, Breedt2017, Brown2020,  Burdge2020, Napiwotzki2020, Korol2022}. A comprehensive list of close DWD systems (periods below 35 days) containing two low-mass WDs are given by \citet{Schreiber2022}. Only a few DWDs have been reported to be possible SN~Ia progenitors that would merge in a Hubble time, including two systems with sub-Chandrasekhar total masses obtained by SPY (WD2020-425 and HE2209-1444; \citealt{Napiwotzki2020}), two super-Chandrasekhar progenitor candidates composed of a WD and a hot subdwarf (KPD~1930+2752 and HD~265435; \citealt{Maxted2000,Pelisoli2021}), CD-$30^{\circ}11223$ \citep{Vennes2012,Geier2013}, Henize~2-428 system (\citealt{Santander2015}; but see also \citealt{Reindl2020}), 458~Vulpeculae \citep{Rodriguez-Gil2010}, SBS~1150+599A \citep{Tovmassian2010} and GD~687 \citep{Geier2010}. Besides, \citet{Kawka2017} suggested that NLTT~12758 is a super-Chandrasekhar DWD system, but it would merge in a timescale longer than the Hubble time.

\subsubsection{Other proposed progenitor scenarios}
\label{sec:other-progenitors}

Some subtypes of the SD model and other possible progenitor scenarios have been proposed for SNe~Ia, including: \textit{\textbf{(1) The CE wind model}}, in which the SD models are assumed to drive CE winds rather than optical thick winds when the mass transfer rate exceeds the critical accretion rate \citep{Meng2017}; \textit{\textbf{(2) The hybrid C-O-Ne WD model}}, in which a hybrid carbon-oxygen-neon (C-O-Ne) WD with a mass of $\gtrsim1.3\,M_{\sun}$ accretes material from its companion star to approach the Chandrasekhar-mass limit and explodes as faint SNe~Ia \citep{Garcia1997,Chen2014,Denissenkov2015,Wang2014a,Meng2014,Liu2015b,Kromer2015,Marquardt2015}; \textit{\textbf{(3) The M-dwarf donor model}}, in which the WD accretes material from an M-dwarf star so that it approaches the Chandrasekhar-mass limit and triggers an SN~Ia explosion \citep{Wheeler2012}; \textbf{\textit{(4) The core-degenerate model}}, in which an SN Ia is produced from the merger of a CO~WD with the core of an AGB companion star during a common envelope (CE) evolution \citep{Livio2003,Kashi2011,Ilkov2012,Soker2013,Soker2014,Soker2018}. \textit{\textbf{(5) The triple channel}}, in which thermonuclear explosions in triple-star systems are triggered through both the SD and DD channels \citep{Katz2012,Hamers2013,Toonen2018,Swaruba2022}; \textit{\textbf{6) The single-star model}}, in which AGB stars or He stars with a highly degenerate CO core near the Chandrasekhar mass ignite carbon at the center to subsequently cause an SN~Ia explosion if they have lost their H-rich or He-rich envelopes \citep{Iben1983,Tout2005,Antoniadis2020}. Note that this list may not be complete and that new channels may still be proposed. Ultimately, the question of SN~Ia progenitor systems has to be settled by observations. 

For a coarse and sketchy overview of the different progenitor scenarios of SNe Ia, we compile the different characteristics in Table~\ref{table:1}. We would like to caution here, that usually the arguments to be made in favor or against specific scenarios are more complex than what can be listed in a table. Therefore we emphasize that they are only intended for a quick overview. The main benefit of our table is to highlight open research questions that are marked with ``unclear''.

\begin{figure*}[ht]
   \centering
   \includegraphics[width=0.95\textwidth, angle=0]{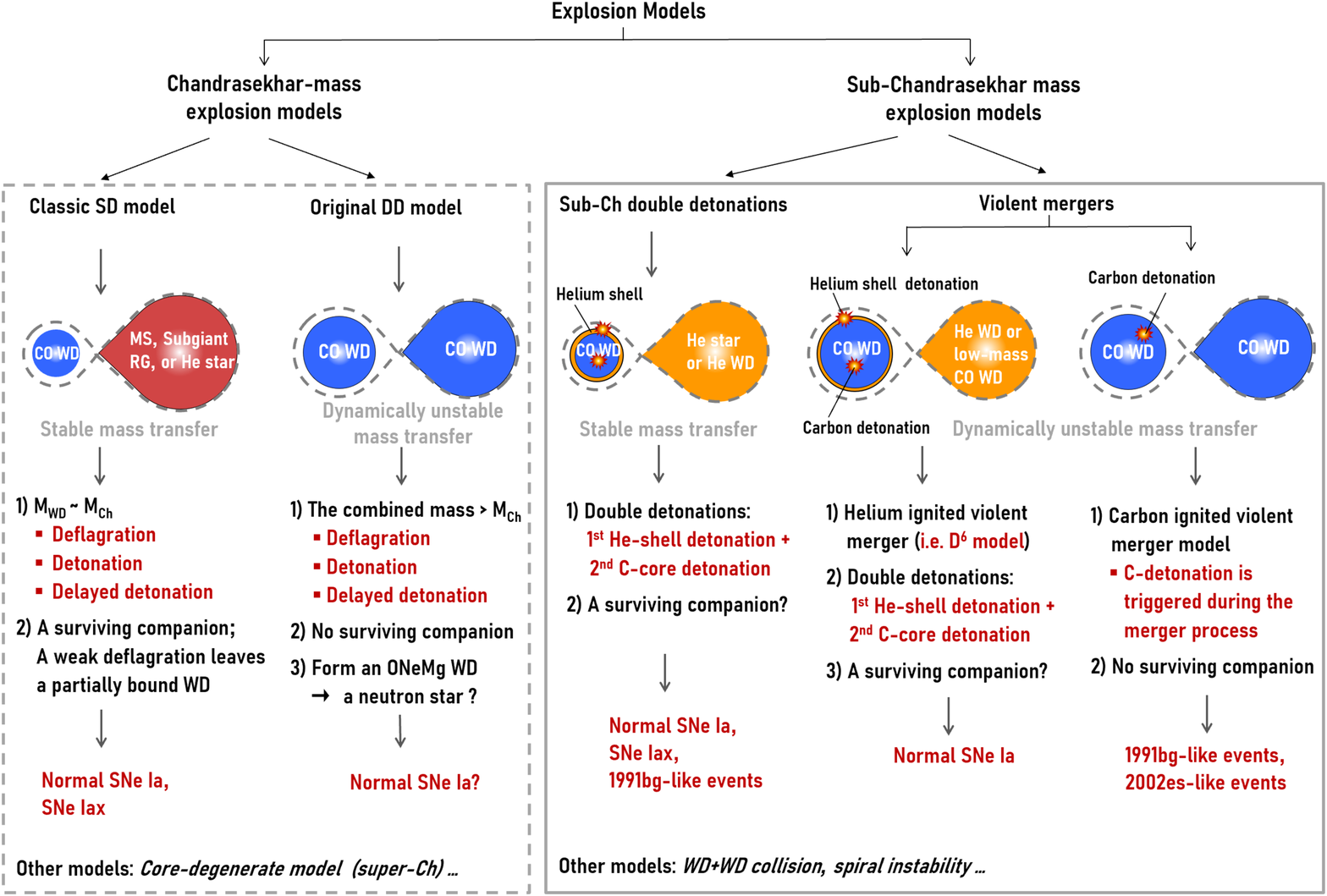}
   \caption{Different explosion models of SNe Ia in the context of either Chandrasekhar-mass or sub-Chandrasekhar-mass explosion \citep[see also][and reference therein]{Hillebrandt2013}. Note that models presented here are not complete.}
   \label{Fig:explosion}
   \end{figure*}

\subsection{Explosion models}
\label{sec:explosion}

The explosion mechanism depends mainly on the question of whether the WD explodes near the Chandrasekhar mass \citep[e.g.][]{Nomoto1984,Livio2003,Roepke2007a,Roepke2007b,Mazzali2007,Kasen2009,Ilkov2012,Rabinak2012,Jordan2012,Seitenzahl2013,Fink2014,Lach2022def,Lach2022gcd} or at a mass below this limit \citep[the ``sub-Chandrasekhar mass'' explosion scenario; e.g.][]{Woosley1986, Benz1989,Fink2007, Shen2007,Rosswog2009b,Raskin2009,Sim2010,Pakmor2010,Kushnir2013,Townsley2019, Gronow2020, Gronow2021}. To provide  clues on the yet poorly understood origin and explosion mechanism of SNe~Ia, one needs to compare the observational features predicted by different explosion mechanism in the context of the progenitor models discussed in Sect.~\ref{sec:progenitor} with the observations. 

A number of explosion models have been proposed to cover various progenitor scenarios of SNe~Ia \citep[][for a recent review]{Hillebrandt2013}, including near Chandrasekhar-mass deflagrations \citep{Nomoto1984, Jordan2012def, Kromer2013, Fink2014,Lach2022def}, near Chandrasekhar-mass delayed detonations \citep[][]{Arnett1969,Khokhlov1991,Gamezo2005, Roepke2007a, Rabinak2012,Seitenzahl2013}, gravitationally-confined detonations \citep{Jordan2008, Seitenzahl2016,Lach2022gcd},  sub-Chandrasekhar-mass double detonations \citep{Nomoto1982a,Nomoto1982b,Woosley1986,Iben1987,Livne1990a, Woosley1994, Fink2007,Moll2013, Gronow2020, Gronow2021, Boos2021}, and violent mergers \citep{Pakmor2010, Pakmor2011}. A schematic overview of various SN Ia explosion models proposed in the framework of either Chandrasekhar-mass or sub-Chandrasekhar-mass explosion is shown in Fig.~\ref{Fig:explosion}.

\begin{figure*}[ht]
   \centering
   \includegraphics[width=0.9\textwidth, angle=0]{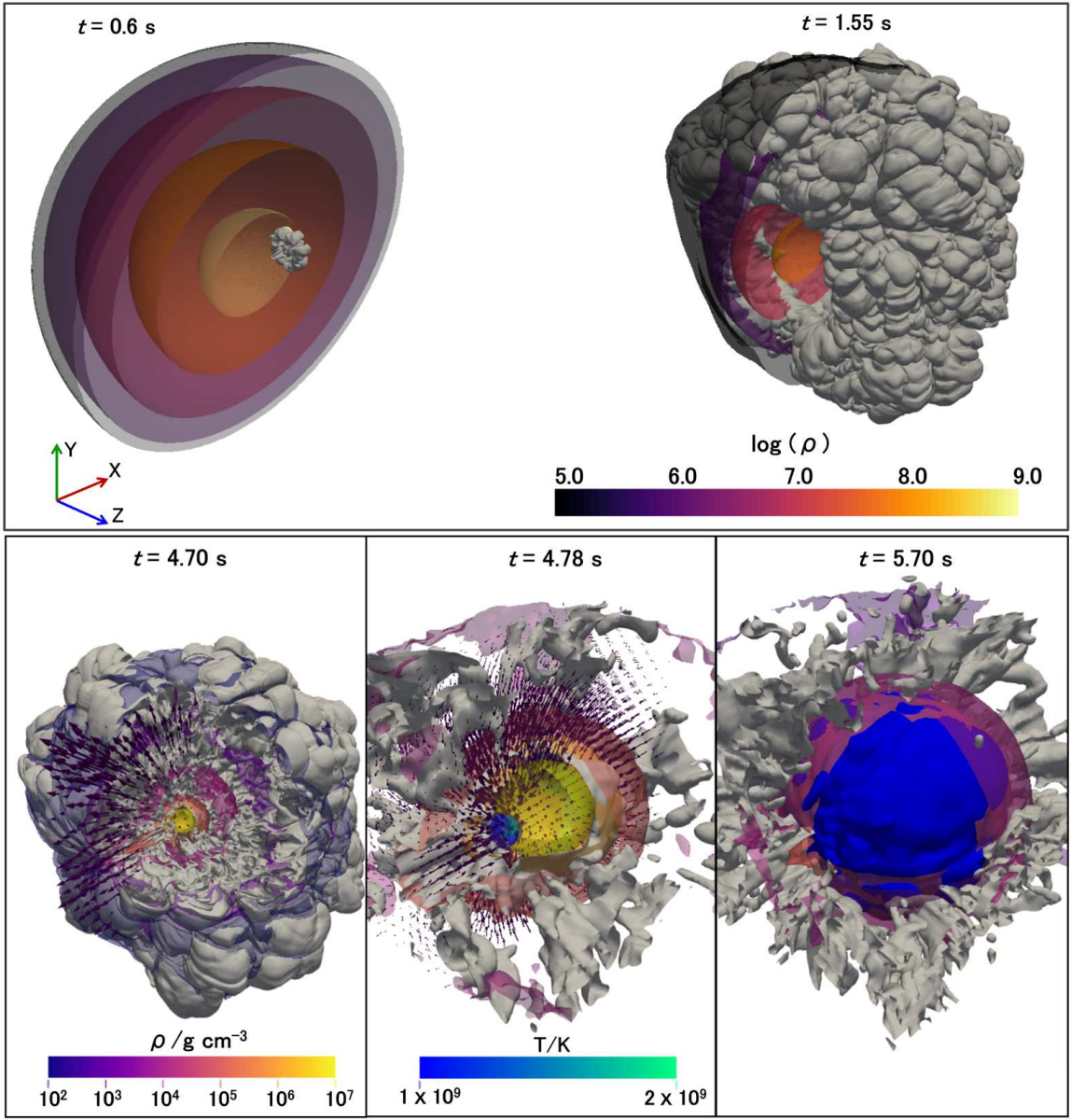}
   \caption{Examples of 3D explosion simulations for two different Chandrasekhar-mass models. \emph{Top panels}: flame surface (\emph{gray}) and  density isosurfaces (\emph{color coded}) from a 3D simulation for a Chandrasekhar-mass pure deflagration model at $t=0.6\,\mathrm{s}$ when flame is rising (\emph{left}) and $t=1.55\,\mathrm{s}$ when flame front  has almost wrapped around the WD. \emph{Bottom panels}: similar to \emph{top panels}, but for a delayed-detonation model (i.e. a gravitationally confined detonation model). Situations prior to detonation initiation, at the moment of detonation initiation, and at about $1\,\mathrm{s}$ after initiation are given from \emph{left} to \emph{right}. The hotspot with temperatures of $1\times10^{9}\,\mathrm{K}$ is marked by the blue-green contour in \emph{middle panel}. The blue surface in \emph{right panel} corresponds to detonation front. Note that the illustration is not to scale. The figure is reprinted from \citet[][see their Figure~3]{Lach2022def} and \citet[][see their Figure~1]{Lach2022gcd} with permission \copyright\ ESO.}
   \label{Fig:Ch-mass}
   \end{figure*}

\subsubsection{Chandrasekhar-mass pure deflagrations}
\label{sec:def}

Near Chandrasekhar-mass explosions in the SD scenario have long been proposed as a potential model for SNe~Ia because they could reproduce some observational features such as the light curves and spectra \citep[e.g.][]{Nomoto1984,Branch1985,Hoeflich1995, Hoeflich1996, Mazzali2007,Kasen2009, Blondin2011,Sim2013}. Moreover, \citet{Yamaguchi2015} suggested that the detection of strong K-shell emission from stable Fe-peak elements in SN~Ia remnant 3C~397 requires electron captures at high density that can only be achieved by a near-Chandrasekhar mass explosion. In such a configuration, a supersonic prompt detonation would turn essentially the entire star into iron-group elements which is inconsistent with the observed features of SNe~Ia: To produce the intermediate-mass elements (IME), such as Si and S, observed in their spectra, burning must start out as a subsonic deflagration. The WD then expands prior to being incinerated. Compared with a prompt detonation, this reduces the production of $^{56}\mathrm{Ni}$ and can in principle increase the IME yields. The outward propagation of the subsonic deflagration flame leads to  Rayleigh-Taylor instabilities that generate turbulence at the contact between hot ashes and cold fuel. This enlarges the surface area of the burning front and accelerates it.

One of commonly used near Chandrasekhar-mass explosion models is the so-called ``W7 model'' of \citet{Nomoto1984}. The W7 model is a one-dimensional (1D) pure deflagration explosion of a Chandrasekhar-mass WD, in which a parametrized description was used for the turbulent burning process. To avoid free parameters in the model, multidimensional simulations (for an example, see top panels of Fig.~\ref{Fig:Ch-mass}) have been carried out \citep[e.g.,][]{Reinecke2002,Gamezo2003,Garcia-Senz2005,Roepke2006a,Roepke2006b,Roepke2007c,Jordan2012def,Ma2013,Long2014,Fink2014,Lach2022}. The result of these simulations is that pure deflagrations are not able to reproduce the majority of normal SNe~Ia \citep{Sim2013,Kromer2013}. In the framework of the the Chandrasekhar-mass deflagration model, it is difficult to produce the canonical $0.5\,M_{\sun}$ $^{56}\mathrm{Ni}$ for normal SNe~Ia because the flame ultimately cannot catch up with the expansion of the WD and much of its material remains unburned. Enhancing the burning efficiency with multi-spot ignitions had only limited success \citep{Roepke2006a,Roepke2006b,Long2014,Fink2014,Lach2022}. Moreover, the ignition process itself is rather uncertain and multi-spot ignition does not seem very likely according to the simulations of \citet{Nonaka2012}.

 However, off-center ignited weak deflagration models have been suggested to explain the particular sub-class of SNe~Iax \citep{Kromer2013,Kromer2015,Magee2016,Kawabata2021,McCully2022,Dutta2022}. Fig.~\ref{Fig:Ch-mass} presents an example of a 3D explosion simulation for a Chandrasekhar-mass pure deflagration model from \cite{Lach2022gcd}. In the weak pure deflagration model of Chandrasekhar mass WDs (sometimes known as ``failed detonation model''), an off-center ignited pure deflagration of a Chandrasekhar-mass CO~WD (or hybrid CONe~WD) fails to completely unbind the entire WD, leaving behind a bound WD remnant \citep{Jordan2012,Kromer2013,Ma2013,Long2014,Fink2014}. It has been shown that pure deflagrations in near-Chandrasekhar-mass CO~WDs and hybrid CONe~WDs can respectively reproduce the observational light curves and spectra of brighter SNe~Iax such as SN~2005hk \citep{Kromer2013} and, less confidently, the faint Iax event SN~2008ha \citep{Kromer2015, Lach2022def}. \citet{Bulla2020} have shown that the maximum-light polarization signal observed in SN~2005hk can be explained in the context of a weak deflagration explosion of a Chandrasekhar-mass WD if asymmetries caused by both the SN explosion itself and the ejecta--companion interaction are considered. Therefore, the weak deflagration explosion of a Chandrasekhar-mass WD seems to be a potential model for SNe~Iax \citep[but see also][]{Hoeflich1995, Stritzinger2015}, at least the brighter members of this sub-class \citep{Lach2022def}.

Interestingly, the weak pure deflagration model of Chandrasekhar-mass WDs predicts the existence of a surviving bound WD remnant which is significantly heated by the explosion and highly enriched by heavy elements from SN ejecta. Searches for such surviving WD remnants would be very helpful for assessing the validity of this explosion model (see Section~\ref{sec:survivor-remnant}; \citealt{Jordan2012def, Kromer2013,Fink2014,Shen2017,Zhang2019,Vennes2017,Lach2022}).

\subsubsection{Chandrasekhar-mass delayed detonations}
\label{sec:ddt}

Besides pure deflagration models, pure detonations of near-Chandrasekhar-mass WDs have also been proposed for SNe~Ia. As already mentioned, the first numerically studied pure detonation model of a near-Chandrasekhar-mass WD in hydrostatic equilibrium \citep{Arnett1969} showed that this model produces too much $^{56}\mathrm{Ni}$ and too little IMEs to explain the observations of normal SNe~Ia. This conflict indicates that an expansion of the WD is needed prior to the detonation in order to reduce the production of $^{56}\mathrm{Ni}$ and to increase that of IMEs. To achieve this, the \emph{``delayed detonation model''} of a near Chandrasekhar-mass WD was proposed by \citet{Khokhlov1989}: The WD expands first due to an initial deflagration and causes the subsequent detonation to burn at relatively low fuel densities \citep{Khokhlov1991,Hoeflich1995,Gamezo2005,Bravo2006,Roepke2007a, Roepke2007b, Plewa2007,Jordan2008,Seitenzahl2013}, reducing the production of $^{56}\mathrm{Ni}$ and enhancing the yields of IMEs compared with the earlier pure detonation models. This therefore makes the delayed detonation model more favourable for explaining normal SNe~Ia. Figure~\ref{Fig:Ch-mass} shows an example of 3D explosion simulations for a Chandrasekhar-mass delayed detonation model (i.e. a gravitationally confined detonation model from \citealt{Lach2022gcd}).

\begin{figure*}[ht]
   \centering
   \includegraphics[width=0.98\textwidth, angle=0]{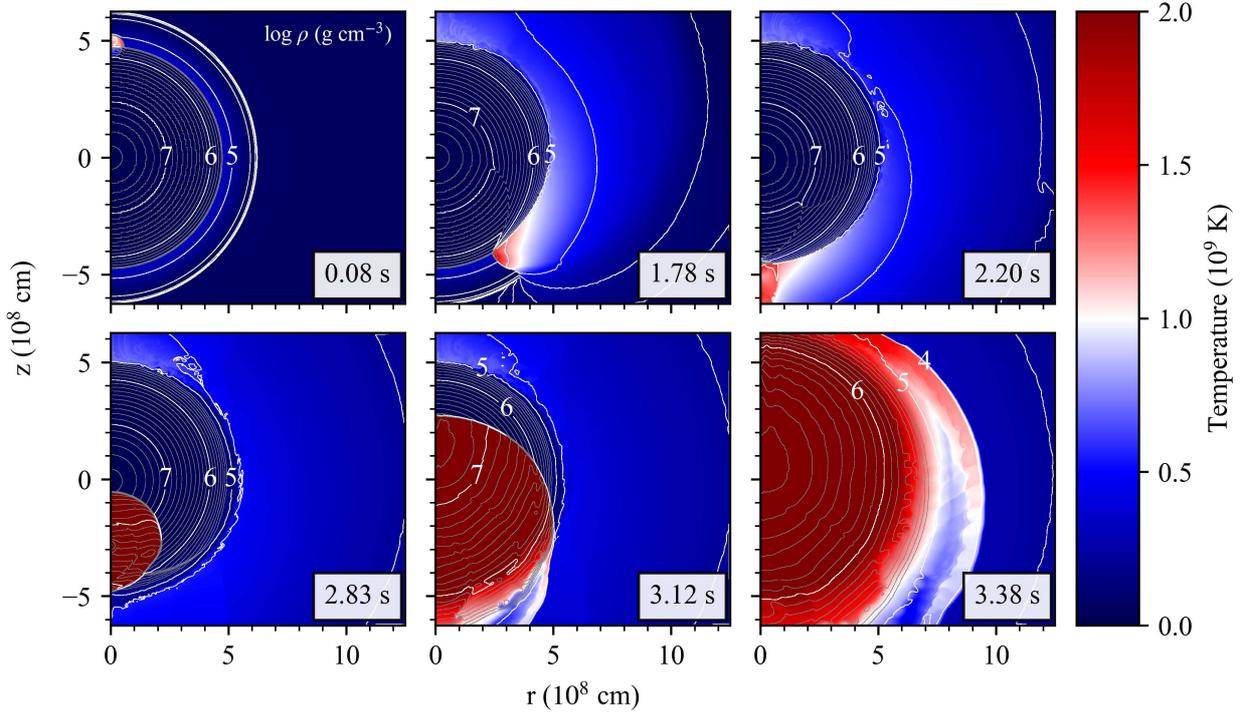}
   \caption{Time sequence of density (\emph{white and gray lines}) and temperature (\emph{color coded}) from a 2D simulation for a sub-Chandrasekhar double detonation explosion of a $1.0\,M_{\odot}$ CO~WD with a thin He shell of $0.016\,M_{\odot}$. At $0.08\,\mathrm{s}$, an initial He-ignition is observed, and the He burning starts to grow. The inward-traveling shock then surrounds the whole core and converges at the south pole at $2.20\,\mathrm{s}$. The shock wave propagates into the core to trigger the second detonation in the core at around $2.83\,\mathrm{s}$, and the core detonation proceeds to give rise to a thermonuclear runaway. The figure is reprinted from \citet[][see their Figure~2]{Boos2021} with the permission of the AAS.}
   \label{Fig:DDet}
   \end{figure*}

The key features of Chandrasekhar-mass delayed-detonation models have been summarized by \citet{Seitenzahl2013}. Several scenarios for the transition from the initial deflagration to a subsequent detonation have been proposed for SNe~Ia such as the \textit{deflagration to detonation transition} model (DDT; \citealt{Khokhlov1991,Gamezo2005, Roepke2007a, Bravo2008,Kasen2009,Seitenzahl2013,Willcox2016}), the \textit{pulsating  delayed  detonation} model (PDD; \citealt{Ivanova1974, Hoeflich1995}), \textit{gravitationally confined detonation} model (GCD; \citealt{Plewa2004,Jordan2008,Jordan2012,Townsley2007,Seitenzahl2016,Lach2022gcd}) and the \textit{pulsational reverse detonation} model (PRD; \citealt{Bravo2006,Bravo2009a,Bravo2009b}). Despite substantial effort, none of the simulations could demonstarate from first principles that the transition of the deflagration to a detonation really occurs \citep{Roepke2007d,Pan2008,Woosley2009,Woosley2011,Schmidt2010,Ciaraldi-Schoolman2013,Poludnenko2019}.

\subsubsection{sub-Chandrasekhar-mass double-detonations}
\label{sec:DDet}

Sub-Chandrasekhar mass WDs can be ignited through a \emph{double detonation mechanism} to give  rise to thermonuclear explosions in the context of either the SD or DD progenitor scenario (see Section~\ref{sec:progenitor}). The initial detonation in this model is triggered by accumulating a He shell on top of the primary WD through either \textit{stable mass transfer} (i.e. the sub-Chandrasekhar-mass double-detonation model;  \citealt{Taam1980,Nomoto1982a,Nomoto1982b,Woosley1986,Tutukov1996,Bildsten2007, Shen2009,Fink2007, Fink2010, Moll2013,Neunteufel2016,Townsley2019,Gronow2020,Boos2021}) or \textit{unstable mass-transfer} (i.e. the so-called D$^6$ model; \citealt{Guillochon2010,Dan2011,Pakmor2013,Boos2021,Shen2021,Roy2022}) from a secondary in a binary system.

 In the sub-Chandrasekhar-mass double-detonation scenario (see Fig.~\ref{Fig:explosion}), the WD accretes material from a He-burning star or a He WD companion via stable mass-transfer to accumulate a He-layer on its surface.  If the He shell reaches a critical mass of ${\sim}\,0.02$--$0.2\,M_{\sun}$ \citep[which is, however, quite uncertain;][]{Woosley1986,Bildsten2007,Woosley2011, Neunteufel2016,Polin2019}, an initial detonation of the He shell is triggered and eventually ignites a second detonation in the core. This leads to a thermonuclear explosion of the entire sub-Chandrasekhar mass WD \citep[e.g.,][]{Nomoto1982a,Woosley1986,Iben1987, Livne1990b,Woosley1994,Livne1995, Fink2007, Fink2010, Sim2010, Moll2013, Gronow2020, Gronow2021, Boos2021}. 

On the one hand, several binary systems composed of a WD and a He-rich companion star have been detected observationally (KPD~1930+2752, V445~Pup, HD~49798, CD-$30^{\circ}11223$ and PTF1~J2238+7430; \citealt{Maxted2000, Geier2007, Kato03, Kudritzki1978, Vennes2012, Geier2013,Kupfer2022}), which seems to support this scenario.  For example, CD-$30^{\circ}11223$ is a binary system containing a WD and a subdwarf-B (sdB) star, in which the WD mass is $M_{\rm{WD}}=0.76\,M_{\sun}$, the companion mass is $M_{\rm{sdB}}=0.51\,M_{\sun}$, and the orbital period is only $P_{\!\!\mathrm{orb}}\simeq1.2$ hours.  \citet{Vennes2012} and \citet{Geier2013} suggested that CD-$30^{\circ}11223$ will likely explode as a SN~Ia via the sub-Chandrasekhar double-detonation mechanism during its future evolution. Very recently, \citet{Kupfer2022} predicted that PTF1~J2238+7430 would lead to a thermonuclear explosion in the context of the sub-Chandrasekhar double-detonation scenario with a thick He shell of ${\sim}\,0.17\,M_{\sun}$.

\begin{figure*}[ht]
   \centering
   \includegraphics[width=0.86\textwidth, angle=0]{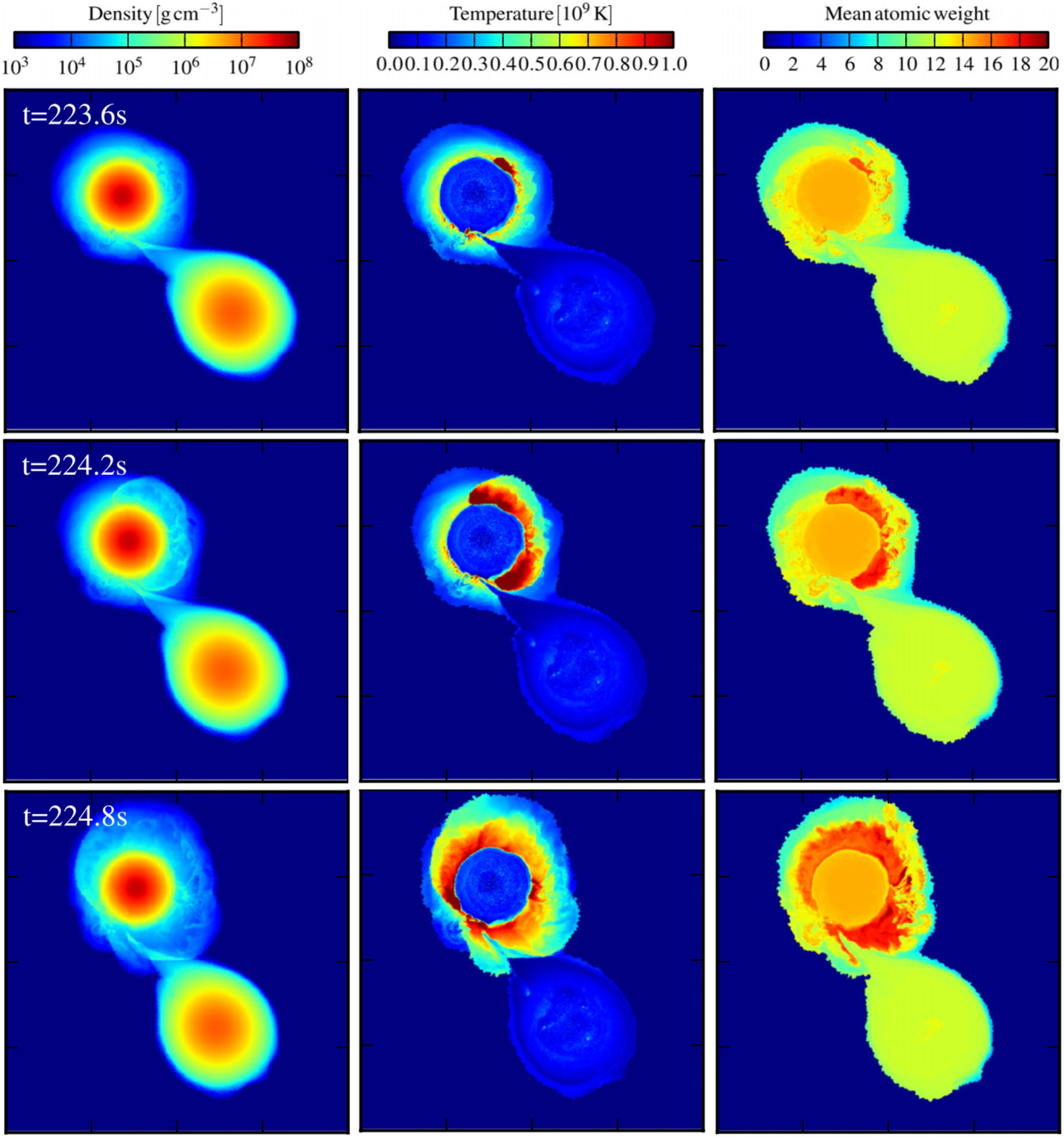}
   \caption{An example of explosion modeling for helium-ignited violent mergers of two WDs. The panels show slices of
  density, temperature, and mean atomic weight in the orbital plane (\emph{columns from left to right})  as a function of time when the He detonation forms on the surface of the primary WD in a simulation of the merger of a $1.1\,M_{\sun}$ primary CO~WD and a $0.9\,M_{\sun}$ secondary CO~WD. Note that all panels have the same length scale of $4\times10^{9}\,\mathrm{cm}$. The figure is reprinted from \citet[][see their Figure~2]{Pakmor2013} with the permission of the AAS.}
   \label{Fig:merger}
   \end{figure*}

\begin{table*}\renewcommand{\arraystretch}{2.0}
\fontsize{8}{11}\selectfont
\begin{center}
\caption{Explosion mechanisms of SNe Ia.} \label{table:2}
\centering
\begin{tabular}{|L{1.8cm}||L{2.0cm}|L{2.0cm}|L{2.2cm}|L{1.9cm}|L{2.0cm}|L{1.6cm}|} 
\hline
                 
Scenario & Explosion triggering  & Combustion mode & Imprint on nucleosynthesis  & Associated SN Ia subclass  & Reproduction of SN Ia brightness range & Reproduction of WLR \\

\hline\hline
   $\mathrm{M_{Ch}}$ deflagrations &  spontaneous in C/O material  & turbulent deflagration in C/O material & solar Mn/Fe ratio, potential overproduction of n-rich IGE & perhaps SNe Iax  & no &  no  \\
\hline
   $\mathrm{M_{Ch}}$ delayed detonations & spontaneous in C/O material  & turbulent deflagration transitioning to detonation in C/O material & solar Mn/Fe ratio & normal SNe Ia? 91T-like objects?  & yes, depending on deflagration ignition &  no  \\
\hline
   sub-$\mathrm{M_{Ch}}$ double detonations & in He shell compressional heating $\rightarrow$ leads to core detonation  & detonation in He shell, detonation in C/O core & solar Mn/Fe ratio due to He-shell detonation, low production of n-rich IGE (Z-dependent) & normal SNe Ia  & yes, depending on mass of exploding WD &  yes   \\
\hline
   He-ignited violent mergers (``$\rm{D^{6}}$'') & in He shell due to dynamical instabilities $\rightarrow$ leads to core detonation  & detonation in low-mass He shell, detonation in C/O core & possibly low Mn/Fe ratio, low production of n-rich IGE (Z-dependent) & normal SNe Ia  & yes, depending on mass of exploding WD &  yes   \\
\hline
   C-ignited violent mergers & at edge of C-core due to accretion  & detonation in C/O material & low Mn/Fe ratio, low production of n-rich IGE (Z-dependent) & normal SNe Ia? 91bg-like objects? 2010lp, 2002ex-like objects  & yes, depending on mass of exploding WD &  yes   \\
\hline
  
\end{tabular}
\end{center}
\end{table*}

On the other hand, different studies in the literature have shown that the sub-Chandrasekhar-mass double-detonation models with a thick He shell ($0.1$--$0.2\,M_{\sun}$) produce an outer layer of SN ejecta enriched with titanium (Ti), chromium (Cr), and nickel (Ni), leading to predicted spectra and light curves that are inconsistent with the observations of SNe~Ia \citep{Hoeflich1996all,Kromer2010,Woosley2011, Sim2012}. However, numerous complications remain to be solved in such a model, and both the production of IGEs in the outer layers and the predicted observables (such as spectra and color) are rather sensitive to the total mass, the thermal and the chemical conditions of the He shell \citep{Bildsten2007,Fink2010,Waldman2011,Moore2013,Shen2014,Piro2015,Townsley2019,Gronow2020,Gronow2021,Gronow2021b} and to details of the treatment of radiative transfer modeling \citep{Collins2022}. For instance, \citet{Kromer2010} showed that pollution the He shell with $^{12}\mathrm{C}$ helps to bring the predicted observables into better agreement with observations of normal SNe~Ia. More recently, some updated simulations have shown that double detonations of sub-Chandrasekhar mass WDs with a thin and C-polluted He shell holds promise for explaining SNe~Ia, including normal SNe~Ia and peculiar objects \citep{Pakmor2013,Townsley2019,Gronow2021,Boos2021,Magee2021,Shen2021a,Shen2021,Rivas2022,Collins2022}. Fig.~\ref{Fig:DDet} shows an example of the sub-Chandrasekhar-mass double-detonation simulation of  a $1.0\,M_{\odot}$ CO~WD with a thin He shell of $0.016\,M_{\odot}$ from \cite{Boos2021}.  \citet{Townsley2019} have suggested that the sub-Chandrasekhar-mass double-detonation scenario might be viable for producing spectroscopically normal SNe~Ia if the He layer is sufficiently thin (${\sim}\,0.01\,M_{\sun}$) and modestly enriched with core material. This indicates that double detonations of sub-Chandrasekhar-mass WDs may contribute the bulk of observed SNe~Ia. However, the exact critical He shell mass required for successfully initiating double detonations of the entire sub-Chandrasekhar mass WD remains uncertain. In addition, the exact He-retention efficiency of the accreting WD in the progenitor system is still poorly constrained \citep[e.g.,][]{Ruiter2014,Toonen2014}.

\subsubsection{Carbon-ignited violent mergers}
\label{sec:carbon-violent}

The ``C-ignited violent merger model'' (see Fig.~\ref{Fig:explosion}) is one of the modern versions of the DD scenario. In this model, unstable dynamical accretion of material from the secondary (less massive) WD on to the primary WD causes compressional heating sufficient to directly trigger a detonation of a CO core in primary WD, producing an SN~Ia \citep{Pakmor2010,Pakmor2011,Pakmor2012b,Sato2015,Sato2016}. While the \emph{original} DD scenario assumes an explosion of a merged object exceeding the Chandrasekhar mass limit, in the violent merger model the explosion triggers already during the merger process before the two stars are completely disrupted \citep{Guillochon2010,Pakmor2010,Pakmor2011,Pakmor2012b,Kromer201310lp,Sato2015,Sato2016}. Therefore it proceeds in sub-Chandrasekhar mass WDs. This scenario avoids the problem of a potential collapse to a neutron star in an AIC \citep{Nomoto1991,Saio1998,Timmes1994,Schwab2016}.

It has been shown that the violent mergers of two CO~WDs that involve a single carbon detonation in the primary star can generally explain the observational properties of subluminous SNe~Ia such as 1991bg-like events \citep{Pakmor2010,Pakmor2011}, SN~2010lp \citep{Kromer201310lp} and SN~2002es-like event iPTF14atg \citep{Kromer2016}.  However, the triggering of the detonation during the violent merger phase is still poorly constrained \citep{Hillebrandt2013}.

\begin{figure*}[ht]
   \centering
   \includegraphics[width=0.86\textwidth, angle=0]{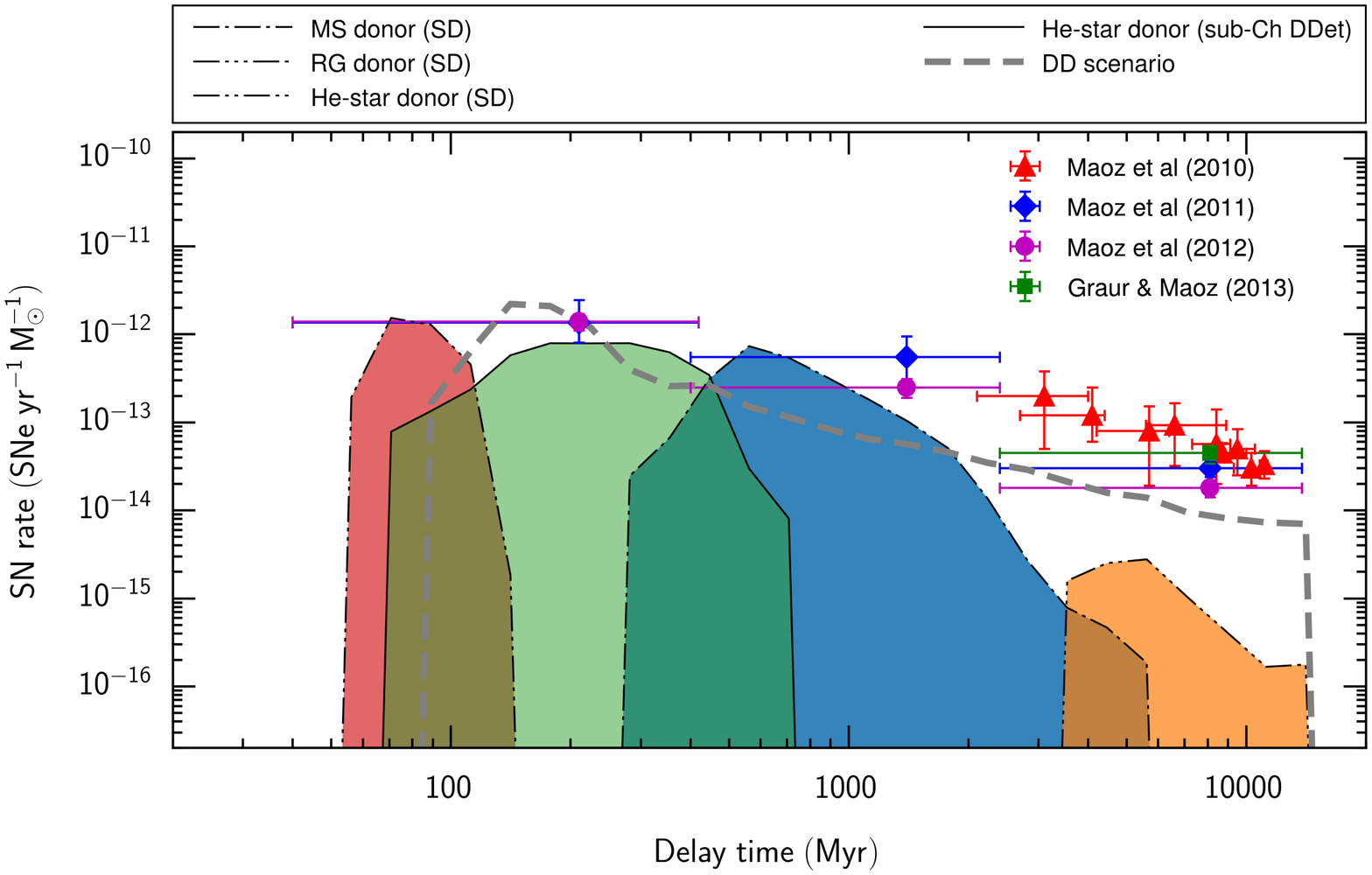}
   \caption{Distributions of delay times of different SN Ia progenitor channels predicted from binary population synthesis calculations. The observationally derived delay-time distributions are given by points with error bars \citep{Maoz2010,Maoz2011,Maoz2012,Graur2013}. Figure is reproduced based on Figure~1 of \citet[][]{Liu2015a}. The results of original DD scenario (gray dashed line) is taken from \citet[][see also \citealt{Han1998}]{Wang2010}. In the Sub-Chandrasekhar mass double-detonation (i.e. Sub-Ch DDet) model, only the results for helium-star donors are included \citep[see also][]{Wang2013}. Note that there is a large variation of the results among various BPS studies \citep[][see their Figure~8]{Maoz2014}.}
   \label{Fig:rate}
   \end{figure*}

\subsubsection{Helium-ignited violent mergers}
\label{sec:d6}

The He-ignited violent merger model -- or ``\emph{dynamically driven double-degenerate double-detonation}'' (D$^6$) model -- is another modern version of sub-Chandrasekhar explosions in the DD scenario, in which SNe~Ia are produced through the  double detonation mechanism during the merger of two WDs (see Fig.~\ref{Fig:explosion}). In the D$^6$ model, an initial He detonation triggers on the surface of a heavier CO~WD primary due to unstable dynamical He accretion from the less massive secondary which could be either another CO~WD with thin surface He layers, a He~WD, or a hybrid HeCO~WD. Via a double-detonation mechanism, the initial He detonation initiates into a detonation of CO core, producing an SN~Ia (see Fig.~\ref{Fig:merger}; e.g. \citealt{Guillochon2010,Pakmor2013,Boos2021,Shen2021,Roy2022}). Because the He detonation in this model proceeds in a dynamic stage and not in a massive He layer at hydrostatic equilibrium conditions (see Section~\ref{sec:DDet}), the impact of the He detonation products on the observables is reduced compared to the classical sub-Chandrasekhar mass double detonation scenario. For instance, \citet{Pakmor2012b} have shown that the double-detonation explosion in the violent merger of two CO WDs with masses of $0.9\,M_{\sun}$ and $1.1\,M_{\sun}$ can closely resemble normal SNe~Ia, indicating that the D$^6$ model has the potential to explain the bulk of normal SNe~Ia.

Interestingly, the secondary WD may survive from the explosion in the D$^6$ model and become a hypervelocity WD with a velocity of $\gtrsim1000\,\mathrm{km\,s^{-1}}$  (see Section~\ref{sec:survivor-remnant}; \citealt{Shen2018,Igoshev2022}). \citet{Shen2018} suggested that three hypervelocity runaway stars with a velocity of ${\gtrsim}\,1000\,\mathrm{km\,s^{-1}}$ detected in the \textit{Gaia} survey \citep[][]{Gaia2016,Gaia2018} are likely to be WD companions that survived the D$^{6}$ SNe~Ia scenario \citep[see also][]{Bauer2021}. However, the fate of the secondary WD in this model is rather unclear. \citet{Pakmor2022} have recently investigated the fate of secondary WD with self-consistent 3D hydrodynamical simulations, confirming that the primary WD can explode as an SN~Ia. But they find that there is a large uncertainty on the question of whether the secondary WD detonates or not \citep{Pakmor2022}. In contrast, \citet{Roy2022} claim that an initial He detonation does not ignite a carbon detonation in the underlying WD.

\subsubsection{Other proposed explosion models}
\label{sec:other-explosion}

In the framework of either Chandrasekhar-mass or sub-Chandrasekhar-mass explosion, some other possible explosion models have been proposed for SNe~Ia, including: \textit{\textbf{(1) The core-degenerate model}}, in which the WD merges with the core of an AGB star during the CE phase, triggering a thermonuclear explosion inside the envelope \citep{Livio2003,Kashi2011,Ilkov2012,Soker2013,Soker2014}; \textit{\textbf{(2) Tidal disruptions}}, in which the tidal interaction of a WD with a black hole triggers a thermonuclear explosion \citep{Rosswog2009a}; \textit{\textbf{(3) Head-on collisions of two WDs}}, in which two WDs collide in a binary or triple-star system, leading to a thermonuclear explosion due to the resultant shock compression \citep{Benz1989,Rosswog2009b,Raskin2009,Raskin2010,Katz2012,Kushnir2013,Papish2016}; \textit{\textbf{(4) The spiral instability model}}, in which a spiral mode instability in the accretion disk forms during the merger of two WDs and leads to a detonation on a dynamical timescale resulting a SN~Ia \citep{Kashyap2015}. 

In Table~\ref{table:2}, we present an overview of the main characteristics of different explosion mechanisms of SNe Ia. Again, the same cautionary remark as for Table~\ref{table:1} applies.

\section{Rates and delay times}
\label{sec:rates}

The observationally-inferred SN~Ia rate in our Galaxy is about $2.84\pm0.60\times10^{-3}\,\rm{yr^{-1}}$ \citep[e.g.,][]{van-den-Bergh1991, Cappellaro1997, Li2011a, Li2011b, Brown2019, Wiseman2021}. The observed delay-time distribution of SNe~Ia (DTDs, i.e.\ the distribution of durations between star formation and SN~Ia explosion) covers a wide range from ${\sim}\,10 \, \mathrm{Myr}$ to about $10\,\mathrm{Gyr}$ \citep[e.g.,][]{Maoz2010, Maoz2011, Maoz2012, Maoz2017, Graur2013}.

By comparing the expected rates and DTDs of SNe~Ia from BPS calculations for different proposed progenitor models  with those inferred from the observation, several studies attempted to place constraints on the nature of SN~Ia progenitor systems (see Fig.~\ref{Fig:rate}; e.g. \citealt[][]{Yungelson1998,Yungelson2010,Hachisu1999, Nelemans2001, Han2004, Botticella2008, Mannucci2008, Ruiter2009, Ruiter2011,Ruiter2014,Maoz2010, Meng2009, Meng2010, Wang2009, Wang2010, Mennekens2010, Toonen2012, Bours2013, Claeys2014, Graur2014a, Liu2015a, Liu2018, Liu2020,LiuDD2016,Liu2018dd,Ablimit2014,Ablimit2016,Shen2017}).  In summary,  no single proposed progenitor model is able to consistently reproduce both the observed SN~Ia rates and the DTDs (see Fig.~\ref{Fig:rate}). The DD progenitor model generally predicts a broad range of delay times that follow a $t^{-1}$ power-law, which is similar to the overall behavior of the observed DTD. But a sharp decrease of SN~Ia rates for delay times shorter than $200\,\mathrm{Myr}$ is seen in the DD model. This is inconsistent with a significant detection of prompt SNe~Ia with delay times of $35<t<200\,\mathrm{Myr}$ \citep{Maoz2010}. BPS calculations have predicted that SD models with a MS or a RG donor mainly contribute to intermediate delay times of ${\sim}\,100\,\mathrm{Myr}$ to $1\,\mathrm{Gyr}$ and long delay times of $\gtrsim3\,\mathrm{Gyr}$, respectively \citep[e.g.][]{Han2004,Ruiter2009,Meng2009,Wang2010,Claeys2014,Liu2015a}. SD models with a He star donor are expected to contribute to delay times shorter than $100\,\mathrm{Myr}$ \citep[e.g.][]{Wang2009,Claeys2014}. The SD scenario generally tends to predict much lower SN~Ia rates than those of the DD scenario (gray dashed line of Fig.~\ref{Fig:rate}). However, a large variation of the results among different BPS studies is seen \citep[][see their Fig.~8]{Maoz2014}. For recent reviews on the details of the topic, see \citet{Maoz2012PASA} and \citet{Maoz2014}.

\begin{figure*}[t]
   \centering
   \includegraphics[width=0.96\textwidth, angle=0]{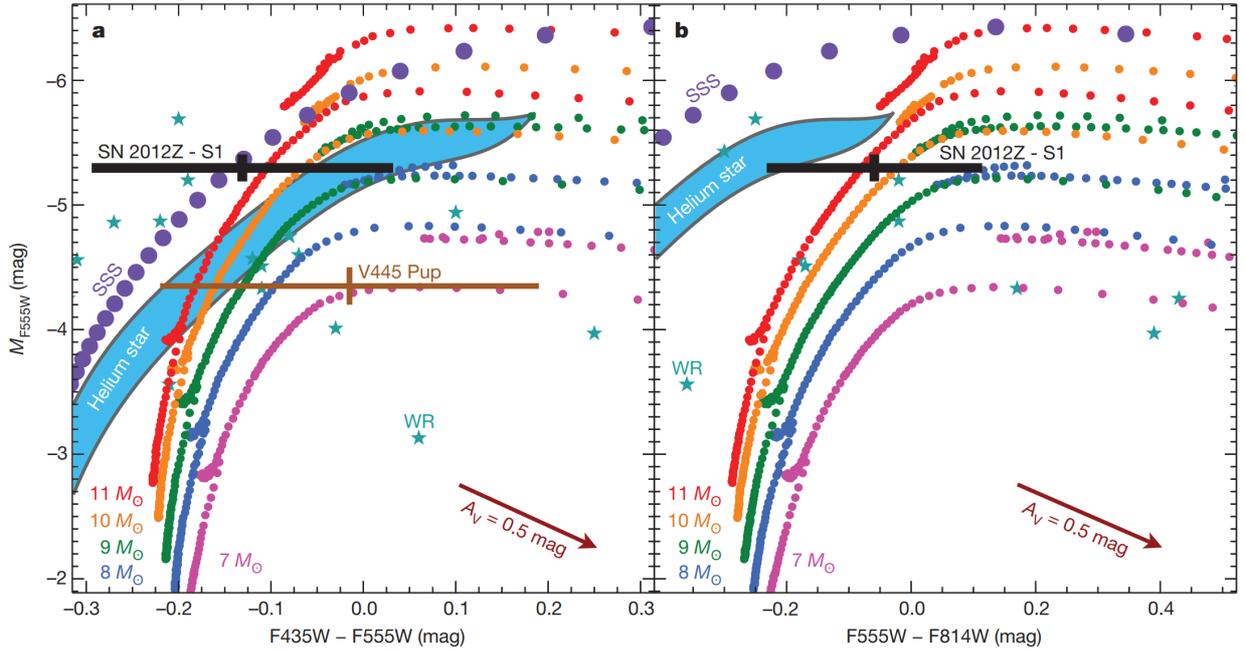}
   \caption{An example of constraints on the progenitor models of SNe Ia by searching for their pre-explosion companion stars. Here, the colour–magnitude diagrams present a comparison between detected pre-explosion luminous source of SN~2012Z (i.e. SN~2012Z-S1; \emph{black crosses}) and different theoretical models, including single stars with masses in a range of $7$---$11\,M_{\odot}$ (\emph{coloured dotted curves}), He-star donor model to a CO WD with an initial mass of $1.3\,M_{\odot}$ (\emph{shaped blue regions}), candidates of Wolf–Rayet stars (\emph{star symbols}), and thermal models for Eddington-luminosity super-soft sources (SSS; \emph{purple dots}). The observations of a nova V445 Pup (\emph{brown cross}) is also shown for comparison. The interstellar extinction of $\mathrm{A_{V}=0.05}$ (\emph{magenta arrow}) is adopted. The figure is reprinted from \citet[][see their Figure~2]{McCully2014}.}
   \label{Fig:pre-sn}
   \end{figure*}

One should always keep in mind that there are significant uncertainties in the theoretical predictions of SN~Ia rates and delayed times from BPS calculations. On the one hand, constraints on the mass-retention efficiencies in the SD scenario are still rather weak \citep{Shen2007, Wolf2013, Piersanti2014,Wang2018} yet studies show that there is a significant  impact of the mass-retention efficiencies on BPS results such as rates and DTDs \citep{Bours2013, Toonen2014, Ruiter2014, Piersanti2014}. On the other hand, the predictions of BPS calculations sensitively rely on the assumed parameters in specific BPS codes such as the CE evolution, star-formation rate and initial mass function. However, to date, strong constraints on these parameters (e.g. the CE efficiency, see \citealt{Zorotovic2010, De-Marco2011, Ivanova2013, Roepke2022})  are still lacking. This limits the predictive power of the BPS results \citep{Toonen2014, Claeys2014}. For a recent review of BPS calculations, see \citet{Han2020}.

\section{Observables of thermonuclear supernovae}
\label{sec:observables}

The approach to compare the observational features predicted by different progenitor models with observations has long been used to provide important clues to the yet poorly understood origin and explosion mechanism of SNe~Ia. Over the past decades, substantial effort in modeling SNe~Ia aimed at the prediction of optical observables (light curves and spectra; e.g.  \citealt{Kasen2006,Sim2010,Maeda2010,Kromer2010,Pakmor2010,Polin2019,Shen2021}). The main goal was to distinguish between explosions of Chandrasekhar-mass and sub-Chandrasekhar mass WD explosions as well as different mechanisms of thermonuclear combustion in these events. Despite all efforts, degeneracies make it difficult to draw firm conclusions \citep{Roepke2012}. 

Besides optical light curves and spectra predicted by radiative transfer calculations in the context of different explosion mechanisms, certain other observational signatures are also expected to be indicative for different progenitor scenarios, including the detection of pre-explosion companions, H/He lines in SN~Ia late-time spectra caused by material stripped from the companion during its interaction with the SN ejecta, early excess emission due to the ejecta--companion interaction, narrow absorption signatures of circumstellar material (CSM), radio and X-ray emission from CSM interactions, surviving companion stars (and WD remnants), polarization signals, SN remnant (SNR) morphology, etc. In this section, we will give a detailed overview to the observables predicted for different phases (from the pre-explosion phase to the SNR phase) of SNe~Ia from currently proposed progenitor scenarios and their comparisons with the observations. In particular, we focus on the question of how a binary companion star in the SD scenario shapes the observables of SNe~Ia.

\begin{figure*}[ht]
   \centering
   \includegraphics[width=0.95\textwidth, angle=0]{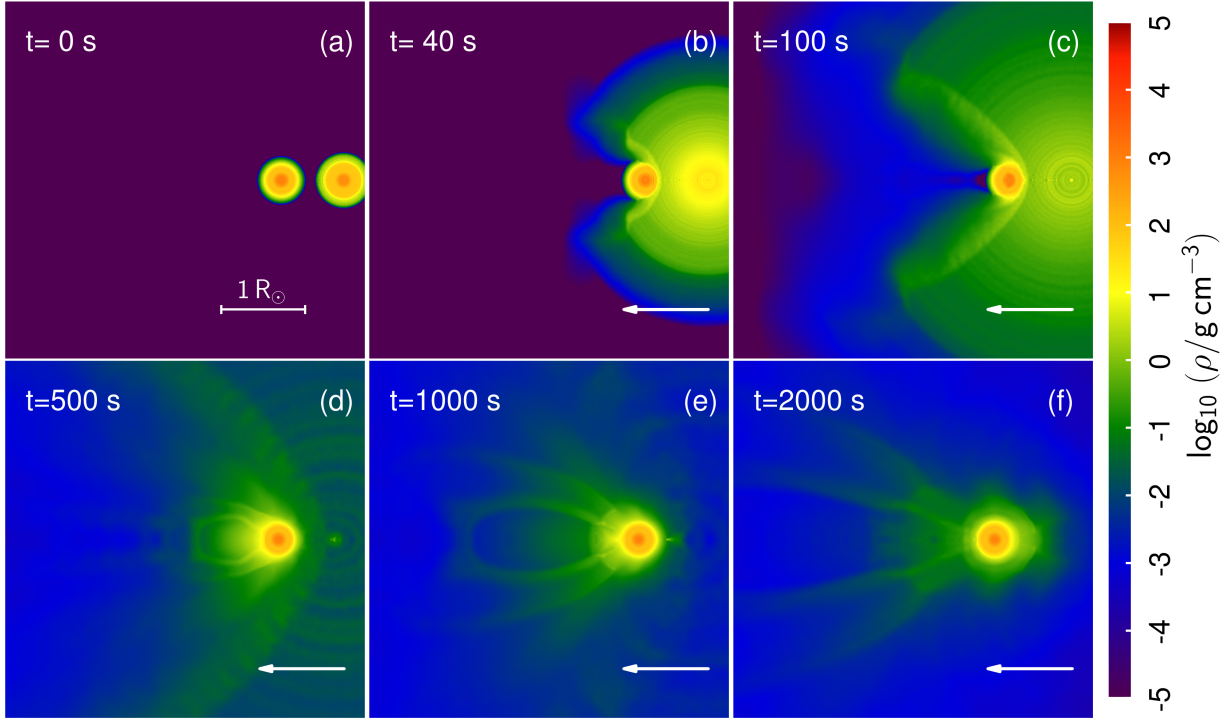}
   \caption{An example of the interaction between SN~Ia ejecta and a MS companion star in 3D hydrodynamical simulations \citep{Liu2012}. The panels show slices in the orbital plane for three different times after the SN explosion as indicated. The direction of motion of the incoming SN~Ia ejecta is from right to left (\textsl{arrow symbols}). Colours indicate density.}
   \label{Fig2}
   \end{figure*}

\subsection{Pre-explosion companion stars}
\label{sec:pre-explosion}

The companion stars in potential progenitor models of SNe~Ia fall into two categories (see Section~\ref{sec:progenitor}): (i) non-degenerate companion stars (MS, SG, RG, AGB or He-burning stars) in the SD scenario; (ii) WD companions in the DD scenario. Becuase a non-degenerate companion star is much brighter than a WD, a luminous source is expected to be detected in pre-explosion images at position of the SNe~Ia if they are generated from the SD progenitor scenario. Therefore, analyzing pre-explosion images from the SN position provides a direct way to test the SD progenitor scenario \citep[][]{Foley2010, McCully2014}.

On the theoretical side, \citet{Han2008} have comprehensively addressed the pre-explosion observable properties (luminosities, effective temperatures, masses, surface gravity, orbital and spin velocities) of MS companion stars at the moment of SN~Ia explosion by performing BPS calculations for the WD~+~MS progenitor model.  Following this work, \citet{Liu2015b} extended the calculations to present pre-explosion properties of different non-degenerate companion stars, including the MS, SG, RG companions in the SD scenario, and the He-burning companion stars from both the SD and sub-Chandrasekhar-mass double-detonation scenarios. \citet{Wong2021} also made predictions for the properties of the He-star donors at the time of explosion for a set of progenitor systems involving a CO~WD and a He star.

On the observational side, different studies have attempted to search for the expected non-degenerate companion stars by analyzing pre-explosion images at the SN position, e.g.\ those taken by the \textit{Hubble Space Telescope (HST)}. To date, however, no progenitor companion star has been firmly detected in the analysis of pre-explosion images of normal SNe~Ia \citep[e.g.][]{Maoz2008, Li2011c, Bloom2012, Kelly2014}. But there are some possible pre-explosion detections recently reported in several SNe~Iax. For instance, \citet{McCully2014} detected a blue luminous source in pre-explosion image of an SN~Iax event, SN~2012Z. As shown in Fig.~\ref{Fig:pre-sn}, the properties of this pre-explosion luminous source (i.e. SN~2012Z-S1) have been found to be consistent with those of a He-star companion to the exploding WD \citep[][]{McCully2014, Liu2015b}. Interestingly, late-time observations taken about 1400 days after the explosion by the HST have shown that SN 2012Z is brighter than the normal SN~2011fe by a factor of two at this epoch \citep{McCully2022}. Comparing with theoretical models, \citet{McCully2022} suggested this excess flux to be a composite of several sources: the shock-heated companion, a bound WD remnant that could drive a wind, and light from the SN ejecta due to radioactive decay. \citet{Foley2015a} have  analyzed pre-explosion HST images of another SN~Iax (SN~2014dt), but no source could be detected in this case.

\subsection{Ejecta--companion interaction}
\label{sec:interaction}

After the explosion in the SD scenario, the ejecta expand freely for a few minutes to hours  before hitting the non-degenerate companion star, engaging into ejecta--companion interaction (see Fig.~\ref{Fig2}). The effect of a SN explosion on a nearby companion star has been studied since the 1970s \citep[e.g.][]{Colgate1970,Wheeler1975,Fryxell1981,Marietta2000, Pakmor2008,Liu2012,Pan2012a,Boehner2017,McCutcheon2022}. There are several ways in which the SN blast wave can modify the properties of companion stars during the ejecta--companion interaction,  giving rise to observables that can be used to constrain SN~Ia progenitors. 

First, the SN ejecta significantly interact with the companion star after the explosion, stripping some H-rich and He-rich material from its surface. This effect is caused either by the direct transfer of momentum or by the conversion of the blast kinetic energy into internal heat, i.e., by evaporation/ablation. As a consequence, some H/He lines caused by the stripped material may be present in late-time spectra of SNe~Ia \citep{Wheeler1975,Mattila2005,Botyanszki2018,Dessart2020}. 

Second, the shock heating injects thermal energy into the companion star during the interaction, leading to a dramatic expansion of the surviving companion star so that it displays signatures that are different from a star without experiencing the ejecta--companion interaction. For example, it could become more luminous and have a lower surface gravity \citep{Podsiadlowski2003,Liu2022ApJ}. 

Third, radiative diffusion from shock-heated ejecta during the interaction is expected to produce an early excess in optical/UV or X-ray emission \citep{Kasen2010}. 

Fourth, the surface of a companion star may be enriched with heavy elements (e.g.\ Ni, Fe or Ca) deposited by the SN~Ia ejecta \citep{Pan2012a,Liu2013a}, which might be detectable in the spectra of a surviving companion star. 

Finally, the companion star survives from the explosion and retains its pre-explosion orbital velocity after the SN explosion, which leads a high peculiar velocity compared with other stars in the vicinity \citep{Ruiz-Lapuente2004,Schaefer2012}. The typical pre-explosion orbital velocities of the H-rich and He-rich companions in the SD Chandrasekhar-mass scenario are ${\sim}\,80$--$280\,\mathrm{km\,s^{-1}}$ \citep{Han2008} and ${\sim}\,250$--$500\,\mathrm{km\,s^{-1}}$ \citep{Wang2009}, respectively. The He-star companions in the sub-Chandrasekhar mass double-detonation scenario and the He WD (or the CO~WD which transfers its outer He layers) companions in the $\mathrm{D^{6}}$ model are respectively expected to have pre-explosion orbital velocities of ${\sim}\,400$--$1000\,\mathrm{km\,s^{-1}}$ \citep{Neunteufel2020} and ${>}1000\,\mathrm{km\,s^{-1}}$ \citep{Shen2018,Igoshev2022}.

\begin{figure}[ht]
   \centering
   \includegraphics[width=0.47\textwidth, angle=0]{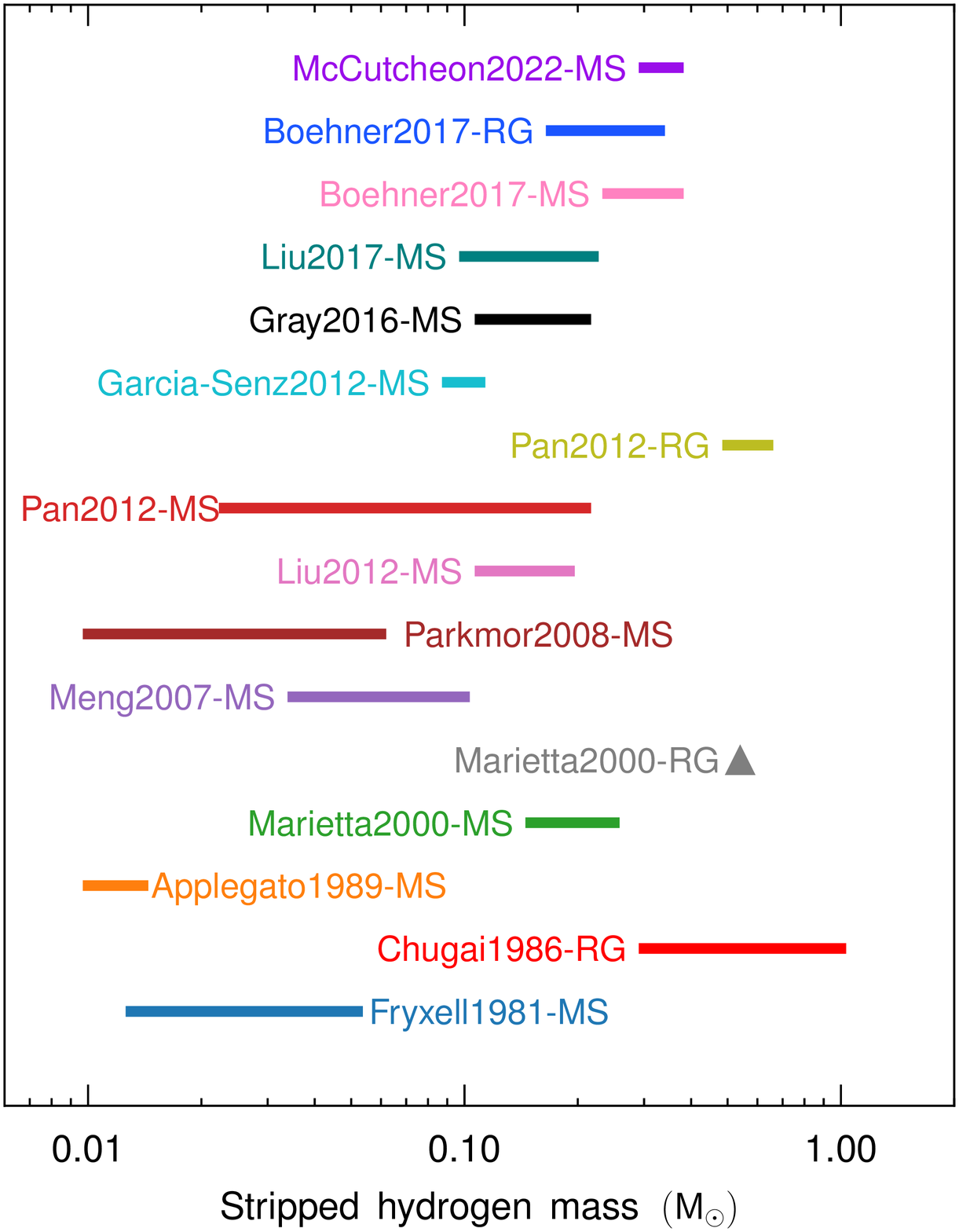}\\
   \hspace{2pt}
   \includegraphics[width=0.47\textwidth, angle=0]{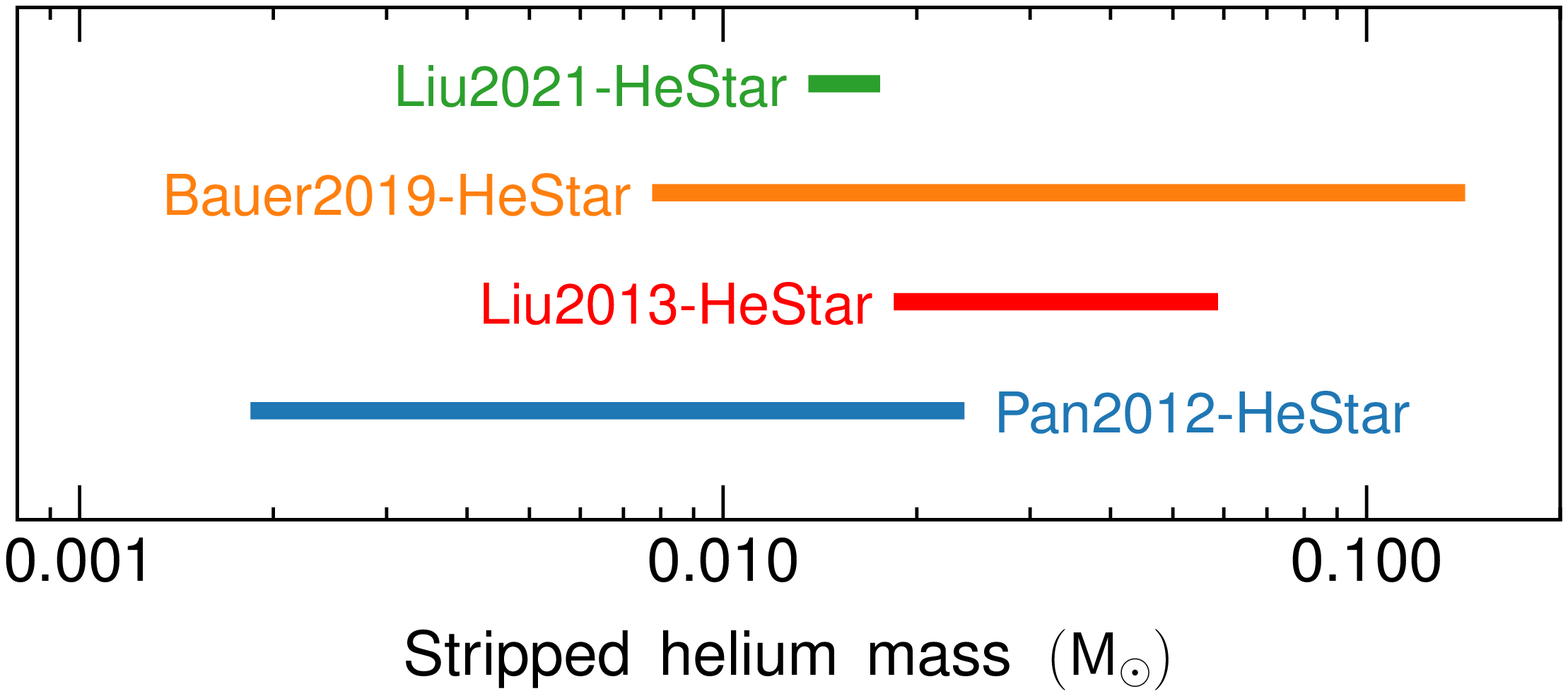}
   \caption{Theoretical predictions for the amount companion mass swept up by the SN~Ia explosion during the ejecta--companion interaction. \textsl{\textbf{Top panel}}: the swept-up H masses predicted either from analytical studies \citep{Chugai1986,Applegate1989,Meng2007} or from 2D/3D hydrodynamical simulations \citep{Fryxell1981,Marietta2000,Pakmor2008,Liu2012,Liu2017,Pan2012a,Garcia-Senz2012,Gray2016,Boehner2017,McCutcheon2022}. Suffixes ``MS'' and ``RG'' indicate that the companion star was a MS or RG star, respectively. \textsl{\textbf{Bottom panel}}: the swept-up He masses predicted by numerical simulations \citep{Pan2012a,Liu2013a,Bauer2019,Liu2021b}. }
   \label{Fig3}
   \end{figure}

\subsubsection{Searches for stripped hydrogen and helium}
\label{sec:late-spectra}

The earliest study of the effect of a SN explosion on a companion star was done by \citet{Colgate1970}. He suggested that the companion star receives a kick that is mainly caused by the evaporation from the stellar surface (i.e. the ablation), although there is also a small kick from the direct collision with the SN ejecta. \citet{Cheng1974} further investigated the impact of a SN shell onto a $2.82\,M_{\sun}$ and a $20.0\,M_{\sun}$ MS companion star for various binary separations, SN shell masses. and velocities. He concluded that the MS companion star could survive from the interaction with SN shell. Upon these two works, several analytical models were developed to estimate the amount of stripped H mass and the kick velocity received by the companion star during the ejecta--companion interaction for MS companion stars with an $n=3$ \citep{Wheeler1975} and an $n=2/3$ polytrope (which is appropriate for a low-mass MS; \citealt{Applegate1989}), and for RG companion stars \citep{Wheeler1975,Chugai1986}.

To test the analytic prescription of \citet{Wheeler1975}, several numerical simulations were performed for low-mass MS companions \citep{Fryxell1981,Taam1984} and RG stars \citep{Livne1992}. In particular, \citet{Livne1992} suggested that almost the entire envelope of a RG star could be stripped off by the SN blast, imparting a velocity to the stripped material (${\lesssim}\,10^{3}\,\mathrm{km\,s^{-1}}$) much smaller than that of SN ejecta (${\sim}\,10^{4}\,\mathrm{km\,s^{-1}}$).  Following the work of \citet{Wheeler1975}, \citet{Meng2007} semi-analytically estimated the amount of stripped H mass due to SN~Ia explosions by adopting the binary and companion properties constructed with detailed binary evolution calculations. However, they underestimated the total stripped companion masses because of neglecting the effect of the ablation on the companion surface.

\begin{figure*}[ht]
   \centering
   \includegraphics[width=0.78\textwidth, angle=0]{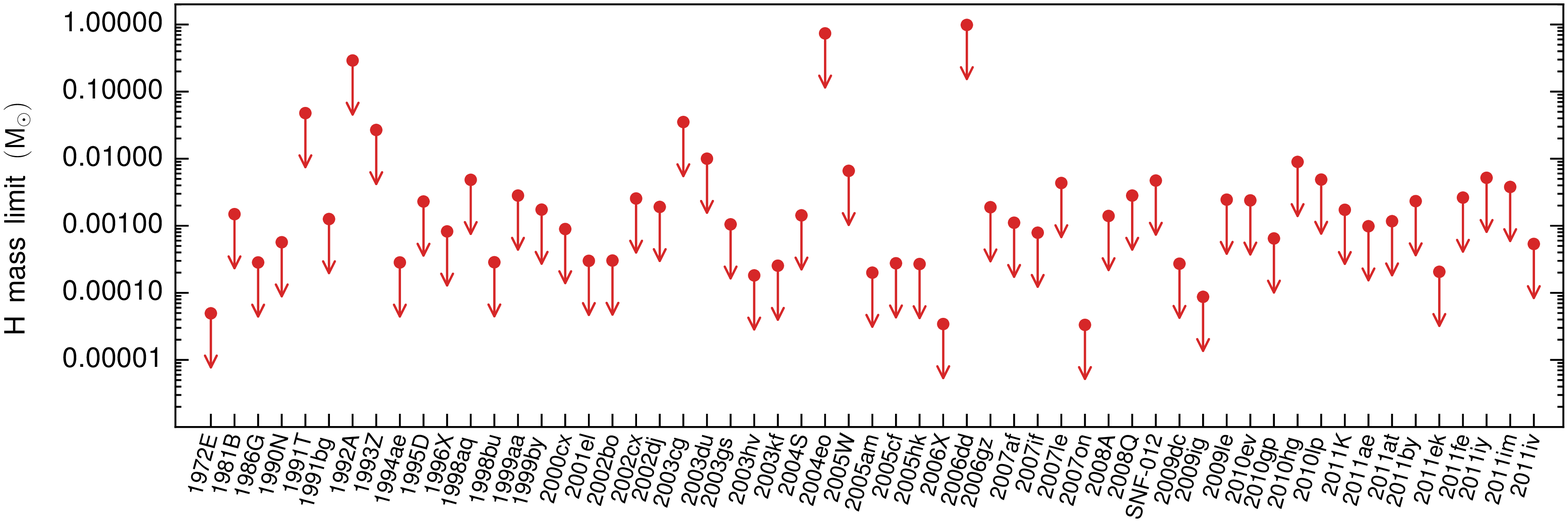}
   \includegraphics[width=0.78\textwidth, angle=0]{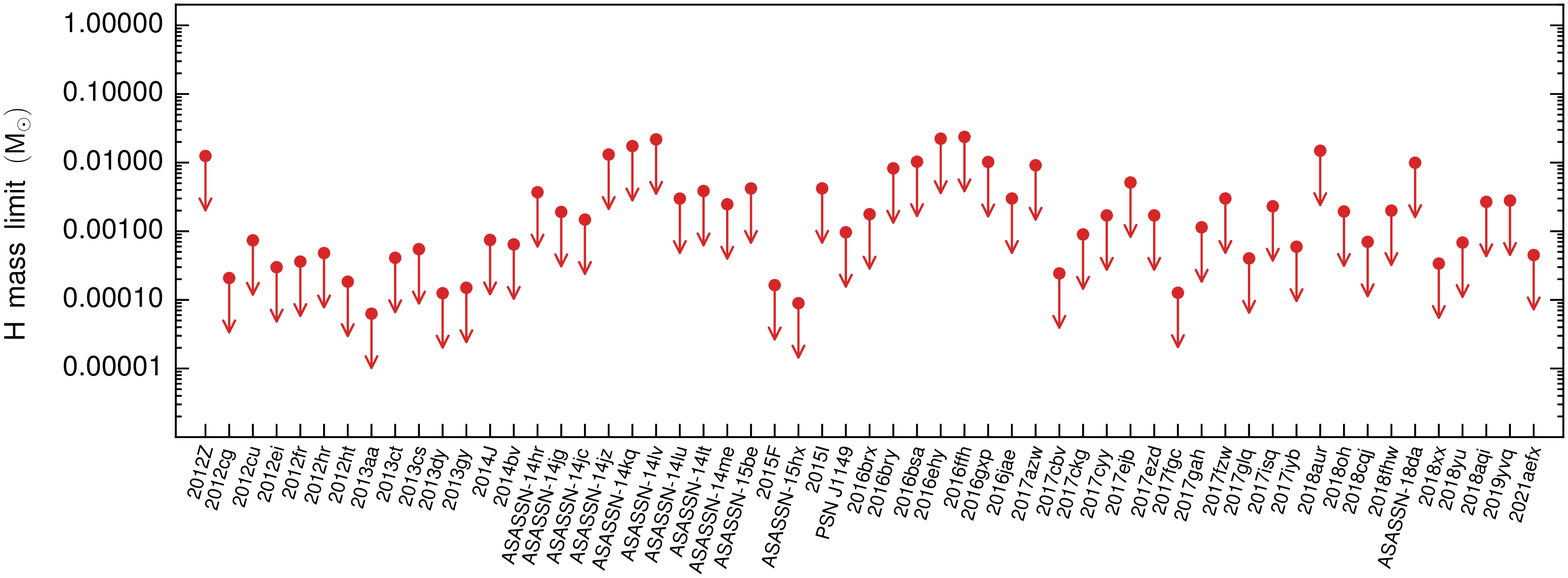}
   \includegraphics[width=0.78\textwidth, angle=0]{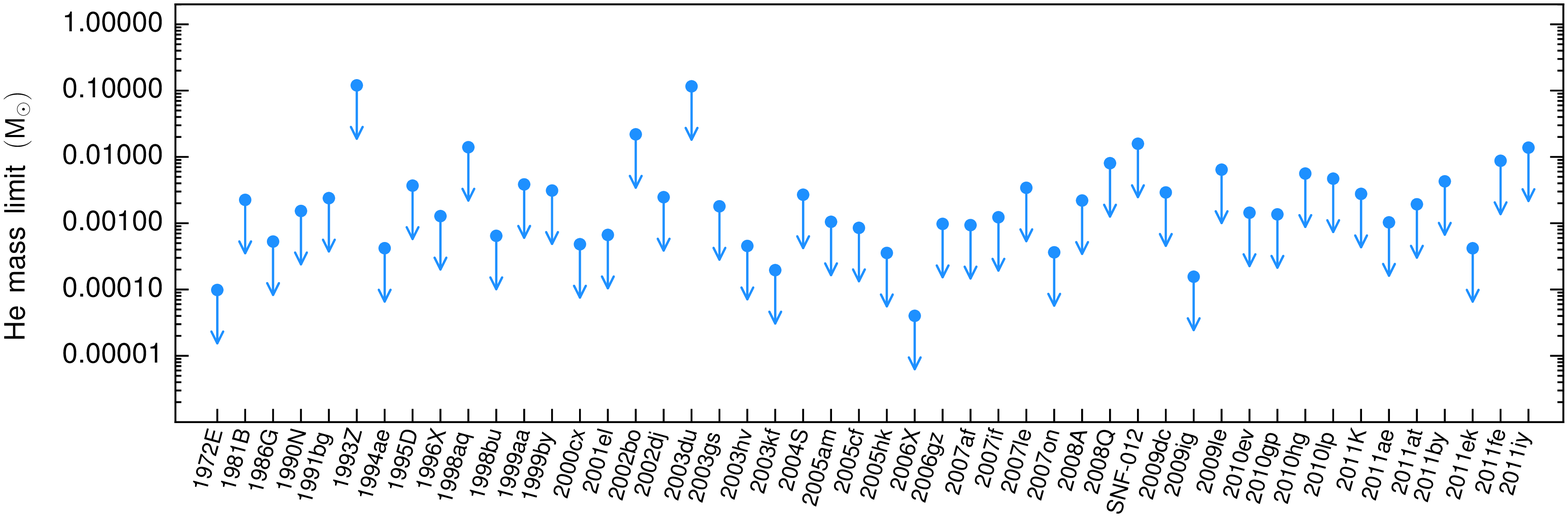}
   \includegraphics[width=0.78\textwidth, angle=0]{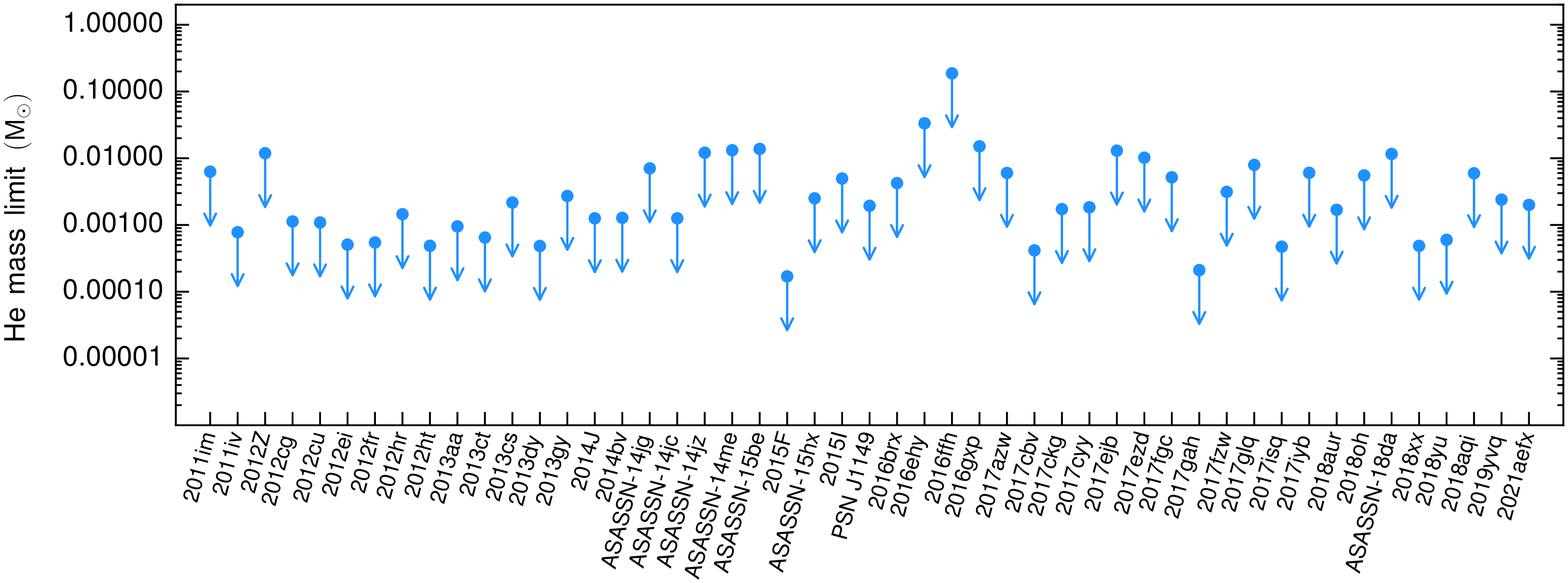}
   \caption{Distributions of observational mass limits on stripped H/He material by the explosion for a sample of SNe~Ia \citep[][and references therein]{Tucker2020}. Data points are taken from \citet{Tucker2020}.}
   \label{Fig4}
   \end{figure*}

More recently, updated two-dimensional (2D) and 3D simulations with grid-based or smoothed particle hydrodynamics (SPH) methods have been presented that investigate the details of the interaction between SN~Ia ejecta and the companion star \citep{Marietta2000,Pakmor2008,Pan2010,Pan2012a,Liu2012,Liu2013a,Liu2013b,Liu2013c, Boehner2017,Bauer2019,Zeng2020,McCutcheon2022}. For instance, \citet{Marietta2000} performed high-resolution 2D simulations to comprehensively study the interaction of SN~Ia ejecta in a variety of plausible progenitor systems with MS, SG and RG companions. However, they assumed the structure of single MS/SG/RG stars for the companion in their simulations. The study of \citet{Marietta2000} for MS companion stars was updated to 3D simulations  with the smoothed particle hydrodynamics (SPH) method by \citet{Pakmor2008}, in which they considered the effect of pre-explosion mass transfer on the structures of a companion star at the moment of SN explosion. However, they computed their companion star models by constantly removing mass while evolving a single MS star to mimick the detailed mass-transfer processes in a binary system. This makes their MS star model much more compact than one constructed from a full binary evolution calculation. Therefore, they predicted a small amount of stripped H masses of $0.01$--$0.06\,M_{\sun}$ for MS donor model. \citet{Liu2012,Liu2013a} further developed the work of \citet{Pakmor2008} by adopting more realistic companion star models constructed from detailed, state-of-the-art binary evolution calculations. They also extended simulations to cover different companion stars (MS, SG and He-star) and a range of binary separations and explosion energies \citep[see also][]{Liu2013b,Liu2013c}.  \citet{Pan2012a} employed adaptive mesh refinement (AMR)
simulations to study the ejecta--companion interaction for MS, RG and He-star companions with different binary separations and explosion energies \citep[see also][]{Pan2010}. In their simulations, however, they did not follow the full binary evolution but used initial conditions with a constant mass-loss rate when constructing their companion stars. 

The main results of ejecta--companion interaction of SNe~Ia in the literature can be summarized as follows.
\begin{enumerate}

\item[(1)] 2D or 3D hydrodynamical simulations have predicted that about 5 per cent to 30 per cent of the companion mass (i.e. $\gtrsim0.1\,M_{\sun}$) can be stripped off from the outer layers of a MS or SG companion star (see top panel of Fig.~\ref{Fig3}). For RG companions, almost the entire envelope is removed by SN~Ia blast wave. In the case of a He companion star, about one few per cent of the mass (${\sim}\,0.002$--$0.03\,M_{\sun}$) is lost in the interaction (see bottom panel of Fig.~\ref{Fig3}). 

\item[(2)] The SN impact affects not only the companion star, but also the SN ejecta themselves. The presence of a companion star strongly breaks the symmetry of the SN~Ia ejecta after the interaction (see panel~c of Fig.~\ref{Fig2}). The stripped companion material is largely confined to the downstream region behind the companion star, creating a hole in the SN debris with an opening angle of about $30^{\circ}$--$115^{\circ}$.

\item[(3)] Depending on the different stellar types, the companion stars receive kick velocities of a few ten kilometers per second to ${\sim}\,100\,\mathrm{km\,s^{-1}}$, which are lower that their pre-explosion orbital velocities. This indicates that the surviving companion star should move with a velocity which is largely determined by its pre-explosion orbital velocity.  

\item[(4)] The characteristic velocities of stripped companion material for the MS, SG, RG and He-star companions are ${\sim}\,500$--$800\,\mathrm{km\,s^{-1}}$, ${\lesssim}\,900\,\mathrm{km\,s^{-1}}$, ${\sim}\,400$--$700\,\mathrm{km\,s^{-1}}$ and ${\sim}\,800$--$1000\,\mathrm{km\,s^{-1}}$, respectively, which are slower than the maximum velocity of SN~Ia ejecta (${\sim}\,10^{4}\,\mathrm{km\,s^{-1}}$) by about one order of magnitude. This implies that H/He lines caused by stripped companion material become visible only at late-times when the photosphere recedes and moves to low velocity regions, revealing the inner SN~Ia ejecta.

\item[(5)] For a given companion model, the amount of stripped companion mass and kick velocity received by the companion star during the interaction decrease as the binary separation increases, which can be fitted by power-law relations. 

\item[(6)] The dependence of the amount of stripped mass and kick velocity on the explosion energy is in agreement with linear relations. Both quantities increase as the explosion energy increases. 

\item[(7)] The companion star is generally expected to survive the explosion and becomes a runaway or hypervelocity star. However, whether a He WD companion in the double-detonation model would survive the explosion is still unclear \citep{Pakmor2022}.

\item[(8)] The companion surface could be enriched with heavy elements (contamination) from the low-expansion-velocity tail of SN~Ia ejecta, which provides a way to observationally identify the surviving companion stars in SNRs. However, the exact level of contamination is still rather uncertain in current models because of uncertainties of mixing of the contaminants in the envelope.

\end{enumerate}

One of the key questions of the SD scenario is whether the signatures of swept-up H or He due to the interaction can be detected in late-time spectra of SNe~Ia. On the theoretical side, by performing the 1D parameterized spherically symmetric radiative transfer calculations, \citet{Mattila2005} concluded that Balmer lines should be detectable in SN~Ia nebular spectra if the stripped H masses are ${\gtrsim}\,0.03\,M_{\sun}$ \citep[see also][]{Lundqvist2013,Lundqvist2015}. In their models, they artificially added some uniform-density solar-abundance material with a low expansion velocity of $1000\,\mathrm{km\,s^{-1}}$ at the center of SN~Ia ejecta of the W7 model from \citet{Nomoto1984}. Recently, \citet{Botyanszki2018} performed, for the first time, 3D Monte Carlo simulations with a non-local thermodynamic equilibrium (NLTE) radiative transport code to determine the signatures of stripped companion material in nebular spectra of SNe~Ia as a function of viewing angle. In this study, more realistic distributions of stripped companion material and post-explosion SN~Ia ejecta structures were adopted based on 2D hydrodynamical simulations of the ejecta–companion interaction by \citet{Boehner2017}. However, the Sobolev approximation as well as the simplified treatment on line overlap and multiple scattering in \citet{Botyanszki2018} causes some uncertainties in their results. \citet{Dessart2020} further extended previous calculations with a set of 1D NLTE steady-state radiative transfer simulations  by covering a broader parameter space (a large range of masses for the ejecta, $^{56}\mathrm{Ni}$, and stripped material) and computing line overlap and line blanketing explicitly. These models adopted a 1D parameterized spherically symmetric SN ejecta structure as in \citet{Mattila2005}.

In summary, all radiative transfer calculations in the literature for SNe~Ia with stripped companion material have concluded that the ejecta--companion interaction in  the SD scenario produces significant and detectable signatures of stripped H/He in late-time spectra. They further provided the dependence of line luminosities from stripped H/He-rich material on the amount of stripped H/He mass. This indicates that searching for H/He emission due to stripped companion material in late-time spectra of SNe~Ia is promising for identifying the SD or DD nature of the progenitor system.

On the observational side, a series of observations have attempted to search for narrow, low-velocity H/He emission lines expected to be caused by swept-up H/He in late-time spectra of SNe~Ia. But to date, no strong evidence for such H/He emission has been found in late-time spectra of most SNe~Ia (even for the nearby SNe~Ia with very high quality observations, i.e., SN~2011fe and SN~2014J; \citealt[][]{Leonard2007,Lundqvist2013,Lundqvist2015,Shappee2013a,Shappee2018,Maguire2016,Graham2015,Graham2017,Sand2018,Sand2019,Dimitriadis2017,Jacobson2019,holmbo2019,Tucker2019,Tucker2020,Tucker2022a,Siebert2020,Elias-Rosa2021,Hosseinzadeh2017, Hosseinzadeh2022}). A detection was reported only for two fast-declining, sub-luminous events (SN~2018cqj and ASASSN-18tb; \citealt{Dimitriadis2019,Vallely2019,Prieto2020}). However, the $\mathrm{H\alpha}$ emission lines detected in SN~2018cqj and ASASSN-18tb have been suggested to be caused by either CSM interaction or by H material stripped from a companion star.

\begin{figure*}[t]
   \centering
   \includegraphics[width=0.95\textwidth, angle=0]{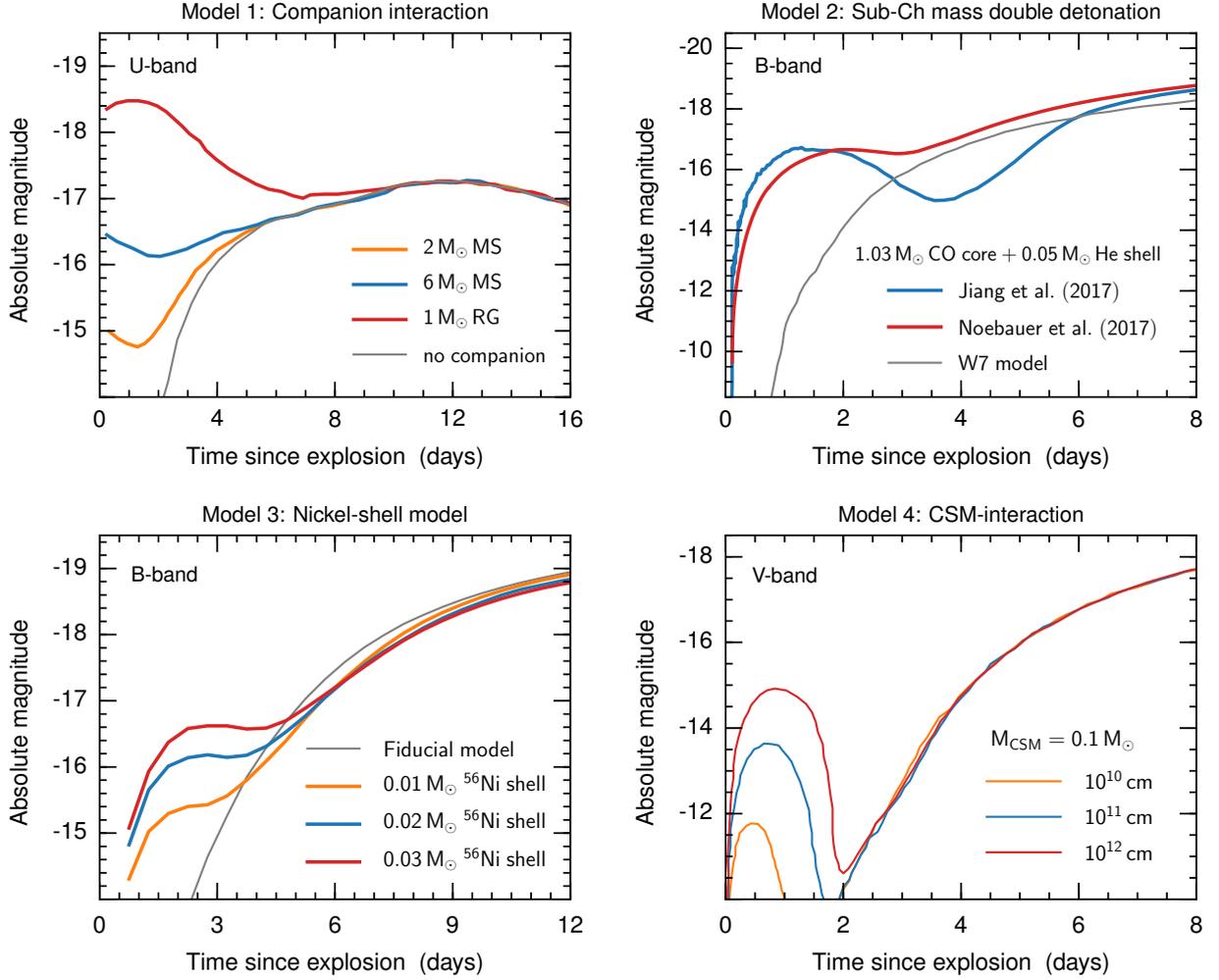}
   \caption{Examples of early excess emissions predicted from different scenarios. \emph{Upper-left panel}: $U$-band early light curves given by the interaction of SN Ia ejecta with a $1\,M_{\odot}$ RG (\emph{red line}), $2\,M_{\odot}$ MS (\emph{yellow line}) and $6\,M_{\odot}$ MS (\emph{blue line}) companion star, respectively. The figure is reproduced from \citet[][see their Figure~3]{Kasen2010} with the permission of the AAS. \emph{Lower-left panel}: $B$-band light curves predicted by the nickel-shell models with  $^{56}\,\mathrm{Ni}$ shells of $0.01\,M_{\odot}$-$0.03\,M_{\odot}$ and a given shell width of $0.06\,M_{\odot}$. The fiducial model for SN 2018oh is shown in \emph{gray line}. The figure is created by taking the data points provided by \citet[][see their Figure~3]{Magee2020a}. \emph{Upper-right panel}: $B$-band light curves of two sub-Chandrasekhar mass double detonation models presented by \citet[][\emph{red line}; see their Fig.~3]{Noebauer2017} and \citet[][\emph{blue line}; see their Fig.~3]{Jiang2017}. The result of W7 model \citep{Nomoto1984} is also given in \emph{gray line} for comparison. \emph{Lower-right panel}: $V$-band early light curves of the CSM-interaction model for a low $^{56}\,\mathrm{Ni}$ mixing level with a boxcar width of $0.05\,M_{\odot}$ given by \citet{Piro2016}, in which the mass and outer radius of CSM are set to be $M_{\mathrm{CSM}}=0.1\,M_{\odot}$ and $R_{\mathrm{CSM}}=10^{10}$-$10^{12}\,\mathrm{cm}$, respectively. The figure is reproduced from \citet[][see their Figure~10]{Piro2016} with the permission of the AAS.}
   \label{Fig:early}
   \end{figure*}

Furthermore, by analyzing late-time spectra of SNe~Ia, one can convert the line luminosity limits to limits on the mass of H/He in SN~Ia progenitors based on the current radiative transfer calculations for stripped companion material \citep{Mattila2005,Botyanszki2018,Dessart2020}. Statistical limits on stripped H/He mass by analyzing a number of SN~Ia late-time spectra have been given by \citet{Tucker2020}, and they are summarized in Fig.~\ref{Fig4}. Comparing these statistical limits from the observation with the stripped H/He masses derived from numerical simulations, we can examine the validity of SD scenario for SNe~Ia. As shown in Fig.~\ref{Fig4}, the observational constraints on the swept-up H/He masses are generally much lower than those from theoretical predictions (Fig.~\ref{Fig3}), which poses a serious challenge for the SD scenario.

Current radiative transfer simulations for SNe~Ia with stripped material are still afflicted with uncertainties, because they either simply assume parameterized spherically-symmetric SN ejecta \citep{Mattila2005,Dessart2020}, or the treat line overlap and multiple scattering in an approximative way \citep{Botyanszki2018}. For stricter predictions of the strength of H and He lines in late-epoch spectra, multi-dimensional NLTE radiative transfer calculations are needed that use the output ejecta model from 3D impact simulations and treat line overlap and multiple scatterings in detail. Moreover, there are some other possibilities that may explain to the lack of H/He emission in late-time spectra. For instance, the ``spin-up/spin-down'' model may lead to  a compact companion star whose H/He-rich envelope has been stripped before the explosion, causing the absence of stripped H/He material during the interaction (see Section~\ref{sec:sd}).

\subsubsection{Early excess emission}
\label{sec:early}

Different progenitor models and explosion mechanisms of SNe~Ia may produce distinct early light curves \citep{Kasen2010,Rabinak2012,Piro2016,Noebauer2017,Magee2020a}. Therefore, early light curves of SNe~Ia have been thought to play an important role in constraining their progenitor systems and explosion mechanism. For example, \citet{Nugent2011} used the early light curves of the nearby SN~2011fe to constrain the radius of its exploding star, confirming that it must have been a WD \citep[see also][]{Bloom2012}. In the literature, different mechanisms have been proposed to cause an excess emission (i.e. a ``bump'') in early light curves of SNe~Ia within the days following explosion, which will be described in detail below (Fig.~\ref{Fig:early}; see also \citealt{Maeda2018,Jiang2021,Magee2022}).

\textit{\textbf{Companion interaction:}}  \citet{Kasen2010} predicted that the shock caused by the ejecta--companion interaction significantly heats SN~Ia ejecta to high temperatures, which causes a strong excess emission during the first few days after the explosion that is observable in the light curves within certain viewing angles in the SD scenario  \citep[see also][]{Maeda2014,Kutsuna2015,Magee2022}. This early-time excess emission is expected to be brightest in the ultraviolet (UV) wavelengths and becomes subordinate at longer optical wavelengths. However, it can still cause a blue color evolution in the optical light curve. By applying  BPS results to the analytical model of \citet{Kasen2010}, \citet{Liu2015c} presented the distributions of expected early UV emission for different SD progenitor systems. Because the DD scenario does not predict such early UV emission, detecting early strong UV emission within the days following explosion has long been considered a smoking gun for the SD scenario of SNe~Ia \citep{Kasen2010,Hayden2010,Olling2015,Liu2015c,Magee2022}. For a given explosion model, early UV emission caused by the ejecta--companion interaction is strongly dependent on the ratio of binary separation to companion radius (assuming RLOF) at the moment of SN explosion. Therefore, the properties of this early UV emission are expected to provide a clue to the types of non-degenerate companions \citep{Cao2015,Marion2016}.

\begin{figure*}[ht]
   \centering
   \includegraphics[width=0.96\textwidth, angle=0]{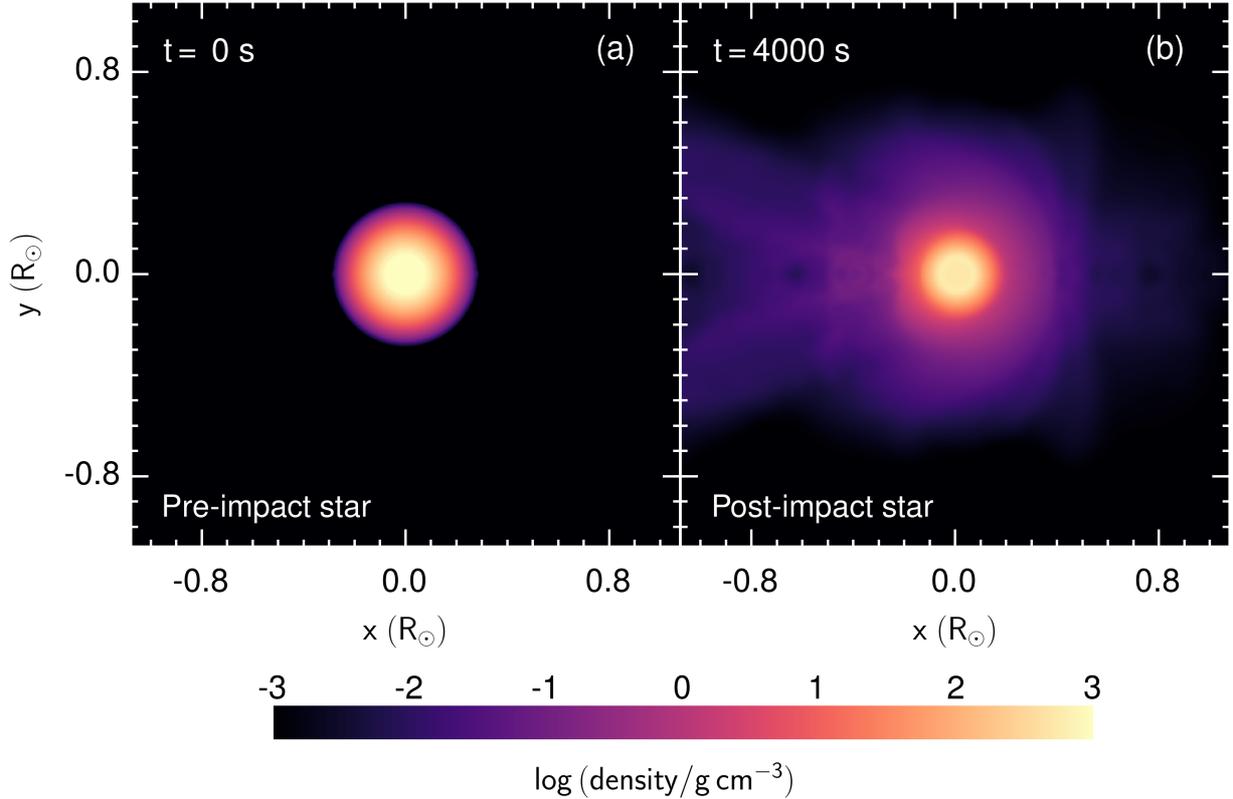}
   \caption{A He-star companion at pre-impact phase (\textsl{left panel}) and post-impact phase (\textsl{right panel}) from a 3D hydrodynamical simulation of the ejecta--companion interaction \citep{Liu2013a,Liu2022ApJ}.}
   \label{Fig5}
   \end{figure*}

\begin{figure*}[ht]
   \centering
   \includegraphics[width=0.48\textwidth, angle=0]{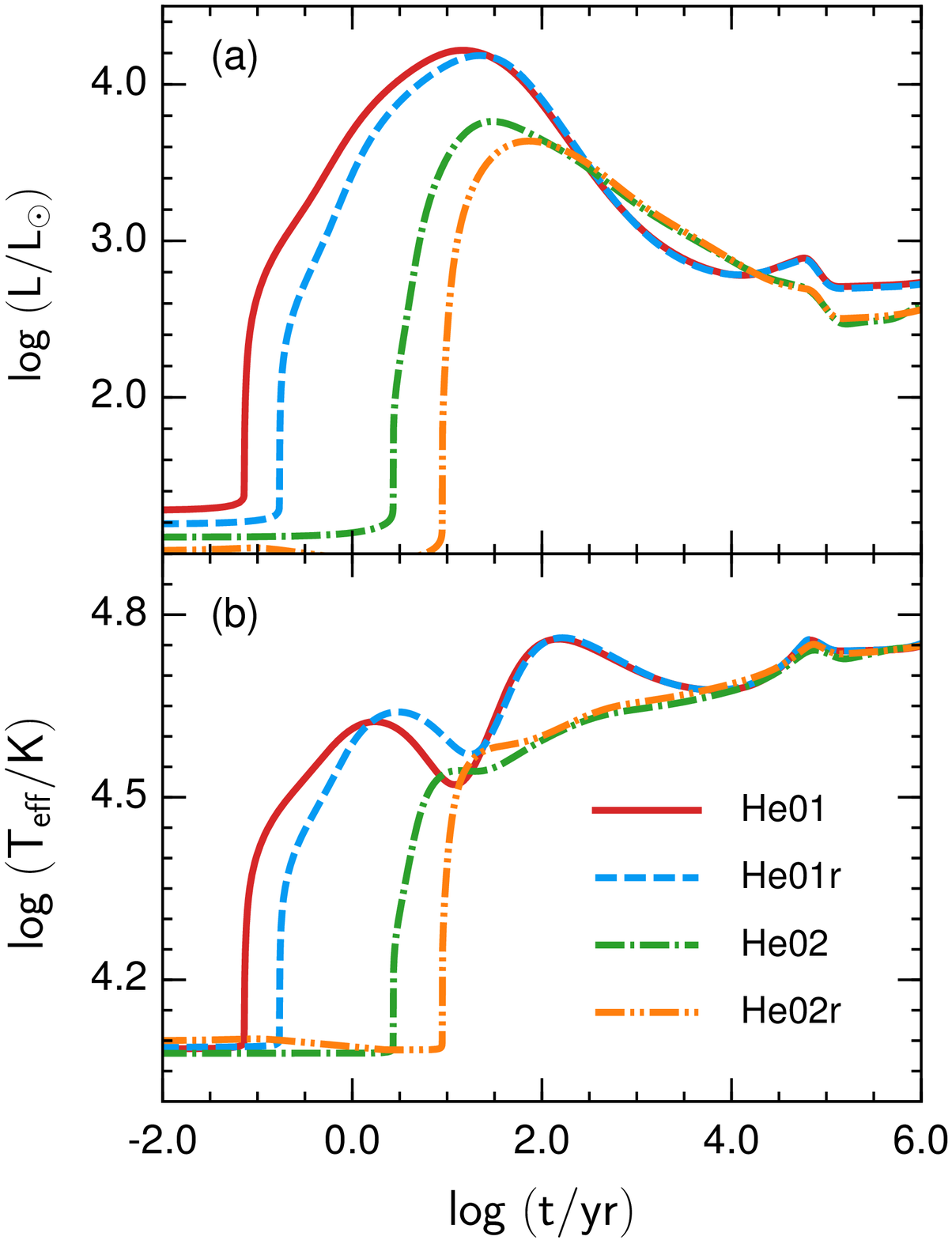}
   \includegraphics[width=0.48\textwidth, angle=0]{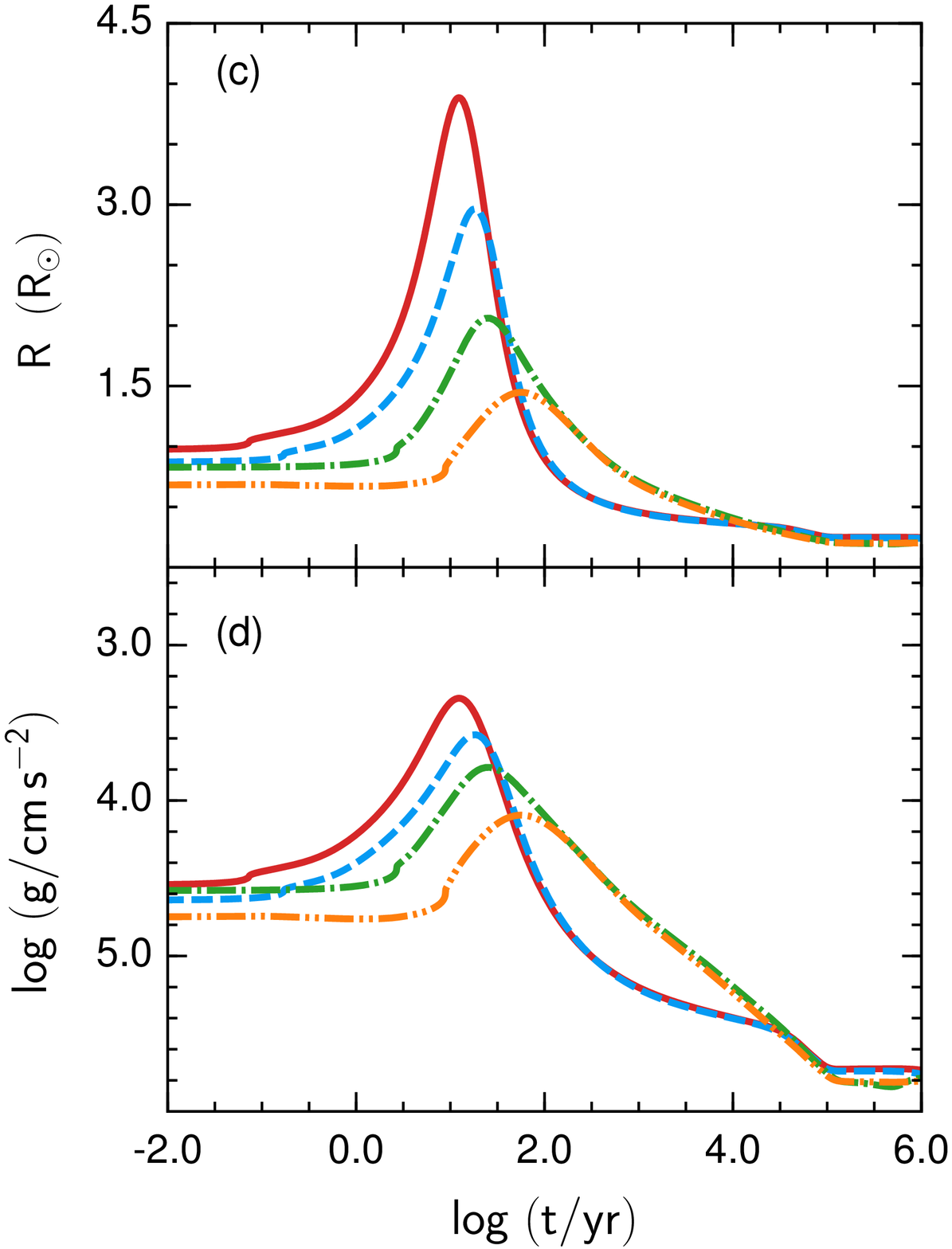}
   \caption{An example of post-evolution of the photosphere luminosity $L$, effective temperature $T_{\mathrm{eff}}$, radius $R$, and surface gravity $g$ of surviving He star companions as functions of time \citep{Liu2022ApJ}.}
   \label{Fig6}
   \end{figure*}

\textit{\textbf{Sub-Chandrasekhar-mass double-detonations:}} The burning of the initial He shell in sub-Chandrasekhar-mass double-detonation explosions can leave heavy, radioactive material in the outermost ejecta. A more massive He shell is expected to produce more radioactive material. The decay of this heavy, radioactive material could create an excess luminosity in the early light curves of SNe~Ia \citep{Sim2012,Noebauer2017,Jiang2017, Maeda2018,Polin2019,Magee2021}. This may produce the early gamma emissions detected in SN~2014J \citep{Diehl2014,Diehl2015,Isern2016}.

\textit{\textbf{Nickel-shell models:}} \citet{Piro2013} suggested that the location of $^{56}\mathrm{Ni}$ in SN~Ia ejecta could have noticeable impact on early-time light curves of SNe~Ia. \citet{Piro2016} further investigated how the distribution of $^{56}\mathrm{Ni}$ in the outer layers of the ejecta shapes early light curves of SNe~Ia. More recently, \citet{Magee2020a} comprehensively predicted early-time curves of SNe~Ia from a series of models containing $^{56}\mathrm{Ni}$ shells with different masses and widths in outer layers of SN~Ia ejecta. They have shown that a $^{56}\mathrm{Ni}$ shell in outer SN~Ia ejecta will lead to an early excess luminosity at a few days after the explosion \citep[][]{Piro2013,Piro2016,Noebauer2017,Magee2018,Magee2020a,Magee2020b}.

\textit{\textbf{CSM interaction:}} The presence of CSM is expected in different progenitor scenarios (see Section~\ref{sec:csm-interaction}), which can also significantly affect early light curves of SNe~Ia. \citet{Piro2016} have shown that the presence of CSM can lead to a significant shock cooling emission during the first few days after the explosion, which can affect the early-time rise of the light curves of SNe~Ia \citep[see also][]{Maeda2018, Moriya2023}. Depending on the degree of mixing of $^{56}\mathrm{Ni}$ in the exploding WD and the detailed configurations of the CSM, this shock cooling emission can lead to early-time signatures (such as the early colour evolution) similar to those caused by the ejecta interaction with a companion star \citep{Piro2016}. Figure~\ref{Fig:early} shows early light curves of SNe Ia predicted from different proposed models described above.

Early light curves of SNe~Ia have been studied since 2010s to search for evidence of such early excess emission \citep[][]{Hayden2010,Brown2012,Brown2012b,Olling2015,Burke2022b}. To date, nearby SNe~Ia with early excess emission have been reported and interpreted with the theoretical models discussed above: SN~2012cg \citep{Marion2016,Shappee2018}, SN~2012fr \citep{Contreras2018}, LSQ12gpw \citep{Firth2015}, SN~2013dy \citep{Zheng2013} SN~2014J \citep{Goobar2015}, SN~2015ak \citep{Jiang2018}, SN~2015F \citep{Im2015,Cartier2017}, SN~2015bq \citep{Li2022}, iPTF14atg \citep{Cao2015}, SN 2016jhr \citep{Jiang2017}), iPTF16abc \citep{Miller2018}, SN~2017cbv \citep{Hosseinzadeh2017} , SN~2017erp \citep{Jiang2018,Brown2019,Burke2022a}, SN~2018yu \citep{Burke2022a}, SN~2018oh\citep{Li201918oh,Dimitriadis2019,Shappee2019}, SN~2018aoz \citep{Ni2022NatAs,Ni2022}, SN~2019yvq \citep{Miller2020,Siebert2020}, SN~2019np \citep{Sai2022}, SN~2020hvf \citep{Jiang2021}, SN~2021aefx \citep{Ashall2022,Hosseinzadeh2022}, SN~2021zny \citep{Dimitriadis2023}, SN~2022ilv \citep{Srivastav2023} and SN~2023bee \citep{Hosseinzadeh2023,Wang2023}.

\citet{Cao2015} suggested that an early UV excess emission seen in iPTF14atg is a strong evidence of the ejecta--companion interaction in the SD scenario. Similarly, \citet{Marion2016} and \citet{Hosseinzadeh2017} respectively argued that the ejecta--companion interaction is likely to be the origin of early-excess signatures detected in SN~2012cg and SN~2017cbv. However, whether the early excess emission of these SNe~Ia can be exclusively attributed to ejecta--companion interaction is still under debate \citep{Kromer2013,Liu2016,Shappee2018}. Recently, \citet{Jiang2017} suggested that a He-detonation model provides a good explanation for early light curves of SN~2016jhr. In addition, it has been proposed that the CSM interaction and $^{56}\mathrm{Ni}$-mixing in outer layers of SN~Ia ejecta are likely to be the origin of early excess emissions of SN~2020hvf \citep{Jiang2021} and SN~iPTF16abc \citep{Miller2018}, respectively. Very recently, \citet{Dimitriadis2023} suggested that the interaction of SN ejecta with $0.04\,M_{\sun}$ H/He-poor CSM at a distance of about $10^{12}\,\mathrm{cm}$ in the context of the merger of two CO~WDs can provide an explanation for the early flux excess of SN~2021zny. In contrast to these studies, \citet{Hosseinzadeh2022} found that no available models can convincingly reproduce the early light curves of SN~2021aefx. 

In summary, it is still hard to draw a strong conclusion on the exact origin of early excess emission detected in nearby SNe~Ia. Current results seem to indicate that there may be various mechanisms at play that cause early excess emission in different SNe~Ia.

\subsection{Surviving companion stars}
\label{sec:survivor}

In the SD scenario, non-degenerate companion stars are expected to survive SN~Ia explosions \citep[see Section~\ref{sec:interaction};][]{Wheeler1975, Marietta2000, Pakmor2008, Liu2012,Liu2013c,Liu2013a,Liu2013b,Liu2021b,Pan2012a,Maeda2014,Boehner2017,Bauer2019,Zeng2020}. Searching for the surviving companion star in nearby SN remnants (SNRs) has been commonly considered to be a promising way to test the SD progenitor scenario \citep[e.g.][]{Kerzendorf2009, Ruiz-Lapuente2019}. The surviving companion stars are expected to display distinct observational signatures due to the mass-stripping, shock heating and enrichment with heavy elements from SN ejecta during the ejecta--companion interaction. In addition, as the disruption of a binary progenitor system after the explosion, surviving companion stars will be runaways stars with a high peculiar space velocity compared to other stars in the vicinity (see Section~\ref{sec:interaction}).

\subsubsection{Theoretical predictions}
\label{sec:survivor-theoretical}

Using 1D stellar evolution codes, early studies have investigated the post-impact properties of surviving companions of SNe~Ia by using simplified initial companion models instead of results frim detailed modeling of the ejecta--companion interaction. For instance, \citet{Podsiadlowski2003} and \citet{Shappee2013b} have studied the post-impact evolution of a $1.0\,M_{\sun}$ SG star and a $1.0\,M_{\sun}$ MS companion star, respectively. \citet{Podsiadlowski2003} found that the surviving SG star could become significantly overluminous by up to two orders of magnitude compared with its pre-SN luminosity for $10^{3}$--$10^{4}\,\mathrm{yr}$ after the explosion. Similar results and conclusion were derived by \citet{Shappee2013b} for their $1.0\,M_{\sun}$ MS companion star model. Both works claimed that the surviving companions become overlumious and detectable during the thermal re-equilibration phase  after the impact. However, these two works did not perform detailed hydrodynamical impact simulations, and they mimicked the ejecta--companion interaction by adopting a rapid mass loss and an extra heating during a single star evolution.

By mapping the results from 3D hydrodynamical simulations of the ejecta--companion interaction (see right-hand panel of Fig.~\ref{Fig5}) into 1D stellar evolution codes, the post-impact evolution of surviving MS star \citep{Pan2012b,Rau2022} and He-star \citep{Pan2013,Liu2022ApJ} companions of normal SNe~Ia have been followed over a long timescale of $10^{4}$--$10^{7}\,\mathrm{yr}$ after the explosion until the stars reestablish their thermal equilibrium. The same method was applied for the investigations of observational signatures of surviving companion stars from  Chandrasekhar-mass pure deflagration explosions for SNe~Iax \citep{Liu2021a,Zeng2022apj,Zeng2022raa} and from the sub-Chandrasekhar-mass double-detonation explosions \citep{Liu2021b}. The expected observational signatures of surviving companions from current studies in the literature are summarized as follows (Figs.~\ref{Fig5} and~\ref{Fig6}). 
\begin{enumerate}

\item[(1)] It is found that the ejecta--companion interaction causes an energy deposition in the companion star due to shock heating while removing some material from the companion’s surface. Because the companion stars are strongly heated and inflated during the interaction (Fig.~\ref{Fig5}), they continue to expand and become overluminous over a Kelvin--Helmholtz timescale after the impact. As the companion stars re-establish the thermal equilibrium, they continue to evolve on a track fairly close to that of an unperturbed star with the same mass. 

\item[(2)] Depending on different companion models and initial binary separations, the surviving MS and He-star companions of normal SNe~Ia expand for a time-scale of $10$--$10^{4}\,\mathrm{yr}$ and tens of years to reach a peak luminosity on the order of $10$--$1000\,L_{\sun}$ and ${\sim}\,10^{4}\,L_{\sun}$, respectively. The stars start to contract when they have radiated away the deposited energy.

\item[(3)] After the explosion, the surviving He-star companions of sub-Chandrasekhar mass double detonation SNe~Ia inflate for about $100$--$1000\,\mathrm{yr}$ and reach peak luminosities of $10^{2}$--$10^{3}\,L_{\sun}$, exceeding their pre-explosion luminosities by 2--3 orders of magnitude.

\item[(4)] The surface rotational speeds of companion stars could significantly decrease after the interaction due to the angular momentum loss and their dramatic expansion. As the deposited energy has radiated away, the stars shrink and their surface rotational speeds increase, becoming a fast-rotating object again after the thermal equilibrium is reestablished. This suggests that the surviving companion stars could rotate slowly although they are fast-rotating stars originally (assuming that the star co-rotates with its orbit due to the strong tidal interaction during the pre-SN mass-transfer phase). The expected state of rotation of surviving companions thus depends on the age of SNR.

\item[(5)] The post-impact evolution of the surviving companion star is determined strongly by the competition between the amount and depth of energy deposition into the star during the interaction.

\item[(6)] The inclusion of the orbital motion and the spin of the companion star into impact simulations does not affect the the post-impact properties of companion stars significantly.

\item[(7)] When artificially adjusting the kinetic energy of the SN ejecta by scaling the velocities based on an original explosion model (e.g. the W7 model), it is found that a higher kinetic energy leads to a higher total amount of energy deposition and a greater depth of energy deposition for a given companion star model.

\end{enumerate}

\subsubsection{Searches for surviving companion stars}
\label{sec:survivor-searches}

In the past few decades, many studies have focused on searching for surviving companion stars as predicted for the SD scenario in the Galactic SNRs, such as Tycho \citep[e.g.,][]{Ruiz-Lapuente2004,Ruiz-Lapuente2019,Fuhrmann2005,Ihara2007,Gonzalez-Hernandez2009,Kerzendorf2009,Kerzendorf2013,Kerzendorf2018a,Bedin2014}, Kepler \citep[e.g.,][]{Kerzendorf2014,Ruiz-Lapuente2018}, and SN~1006 \citep[e.g.,][]{Gonzalez2012,Kerzendorf2012,Kerzendorf2018b}. Additionally, some SNRs in the Large Magellanic Cloud (LMC): SNR~0509–67.5 \citep[e.g.,][]{Schaefer2012,Pagnotta2014,Litke2017}, SNR~0519–69.0 \citep[e.g.,][]{Edwards2012,Li2019snrs}, SNR~0505–67.9 (DEML71, \citealt{Pagnotta2015}), SNR~0509–68.7 (N103B, e.g., \citealt[][]{Pagnotta2015,Li2017}), and SNR~0548–70.4 \citep[e.g.,][]{Li2019snrs} were searched.  To date, no surviving companion could be firmly identified in these SNRs. Nevertheless, there are a few candidates show special features in line with the expectations and have therefore been proposed as surviving companions in SN~Ia events \citep[e.g.,][]{Ruiz-Lapuente2004,Ruiz-Lapuente2022,Geier2015,Shen2018}.

\textit{\textbf{Tycho~G:}} \citet{Ruiz-Lapuente2004} suggested that Tycho~G is a possible surviving companion because of its peculiar radial velocity and proper motion, and a lower surface gravity than a MS star \citep{Bedin2014}. Moreover, \citep{Gonzalez-Hernandez2009} suggested that Tycho~G has an overabundance of $\mathrm{Ni}$ relative to normal metal-rich stars, which seems to be consistent with the deposition of heavy elements from an SN~Ia explosion  \citep[but see also][]{Ihara2007}. However, some other studies cast doubts on this identification \citep{Fuhrmann2005}. For instance, \citet{Howell2011} concluded that Tycho~G is apparently not out of thermal equilibrium.  \citet{Kerzendorf2013} showed that the measured $\mathrm{[Ni/Fe]}$ ratio of Tycho~G seems to be not so unusual with respect to field stars with the same metallicity. In addition, \citet{Kerzendorf2009,Kerzendorf2012} suggested that Tycho~G is unlikely to be the surviving companion star of SN~1572 because its measured rotational velocity of ${\sim}\,6\pm1.5\,\mathrm{km\,s^{-1}}$ is much lower than rotational velocities of pre-explosion MS companion stars of $40$--$180\,\mathrm{km\,s^{-1}}$ \citep{Han2008}. However, the ejecta--companion interaction can significantly reduce the rotational velocity of a companion star \citep{Pan2012b,Liu2013a,Liu2022ApJ,Zeng2022raa}.

\textit{\textbf{US~708:}}  In the sub-Chandrasekhar double detonation scenario, the companion stars are expected to have a high orbital velocities of up to ${\simeq}\,900$--$1000\,\rm{km\,s^{-1}}$ at the moment of explosion \citep{Neunteufel2020, Neunteufel2022}. The surviving companion stars from this scenario are therefore good candidates of hypervelocity stars \citep[e.g.,][]{Geier2013,Geier2015,Shen2018, Neunteufel2019,Neunteufel2022,Igoshev2022}. The hypervelocity star US~708 has been classified as a sdO/B star. Based on a spectroscopic and kinematic analysis, \citet{Geier2015} have reported that US~708 travels with a velocity of about $1\mathord,200\,\rm{km\,s^{-1}}$, suggesting that it is one of the fastest unbound stars in our Galaxy that was ejected from the Galactic disc $14.0\pm3.1\,\mathrm{Myr}$ ago. Considering the possibilities of different acceleration mechanisms of hypervelocity stars, \citet{Geier2015}  concluded that US~708 is very unlikely to originate from the Galactic centre, but it is rather the ejected donor remnant of a sub-Chandrasekhar mass double detonation SN~Ia. Very recently, comparing the long-term evolution and properties of surviving companion stars of sub-Chandrasekhar mass double detonation SNe~Ia with the observations of US~708, \citet{Liu2021b} suggested that US~708 would require the entire pre-supernova progenitor binary to travel at a velocity of about $400\,\mathrm{km\,s^{-1}}$ if it is indeed the surviving He-star donor of a sub-Chandrasekhar mass double detonation SN~Ia. It could have been ejected from a globular cluster in the direction of the current motion of the surviving companion star \citep[see also][]{Bauer2019}.

\textit{\textbf{MV-G272:}}  \citet{Ruiz-Lapuente2022} explored the Galactic Type Ia remnant SNR~G272.2-3.2 (about $7,500\,\mathrm{yr}$ old) with the \textit{Gaia}~EDR3, finding a kinematically peculiar star (M1-M2 type dwarf) Gaia~EDR3~5323900215411075328, i.e., MV-G272. The past trajectory of MV~G272 shows that it was ejected from the centre of SNR~G272.2-3.2 about $6\,000$--$8,000\,\mathrm{yr}$ ago. The mass, radius, effective temperature, surface gravity and metallicity are measured to be $M=0.44$--$0.50\,M_{\sun}$, $R=0.446$--$0.501\,R_{\sun}$, $T_{\mathrm{eff}}=3,800\,\mathrm{K}$, $\mathrm{log}\,g=4.46$ and $[\mathrm{Fe/H}]=-0.34$, respectively. Based on kinematical characteristics of MV-G272 and its trajectory, \citet{Ruiz-Lapuente2022} proposed that MV-G272 is the surviving companion candidate of the SN~Ia that formed SNR~G272.2-3.2. They also suggested that the discovery of MV-G272 hints to the possibility of a SD SN~Ia progenitor system with an M-type dwarf companion \citep{Wheeler2012}.

\subsubsection{Searches for surviving WD remnants}
\label{sec:survivor-remnant}

In the D$^{6}$ model, the WD companion may survive the explosion (see Section~\ref{sec:d6}; \citealt{Pakmor2013,Shen2018,Tanikawa2019,Boos2021}). Such surviving WDs are expected to fly away with a velocity of ${\gtrsim}\,1000\,\mathrm{km\,s^{-1}}$. Interestingly, \citet{Shen2018} discovered three hypervelocity WDs with a velocity of $1000$-$1300\,\mathrm{km\,s^{-1}}$ in \textit{Gaia}'s second data release \citep[][]{Gaia2016,Gaia2018}. Because these three hypervelocity runaway WDs have inflated radii that do not fit into any classical spectroscopic WD subtypes, and show a much lower surface gravity and peculiar composition, \citet{Shen2018}  suggested that they are likely to be the WD companions ejected from SNe~Ia explosion via the D$^{6}$ mechanism \cite[see also][]{Chandra2022}. \citet{Shields2022} also attempted to search for fast-moving WDs predicted by the D$^{6}$ explosion in nearby SN~Ia remnant SN~1006, but none of such candidates were detected. However, the exact fate of the secondary WD in D$^{6}$ model remains uncertain because it remains unclear whether it is detonated in the SN event \citep{Pakmor2022}. By artificially igniting a carbon detonation of the secondary WD in a D$^{6}$ model, \citet{Pakmor2022} found that the explosion of secondary WD does not produce distinguishable observables (light curves and spectra) until ${\sim}\,40\,\mathrm{d}$ after the explosion compared to a simulation without an explosion of the secondary WD. Furthermore, there is still no consensus on whether the first He detonation can eventually trigger a carbon detonation of the primary WD to cause an SN~Ia \citep{Roy2022}.

It has been suggested that SNe~Iax could arise from weak deflagration explosions of Chandrasekhar-mass WDs in which the stars are not fully disrupted. Such events would leave behind a bound WD remnant that has been shock heated significantly \citep{Jordan2012def,Foley2013,Kromer2013,Fink2014,Jha2017}. If binary systems are destroyed by the SN explosion in this model, such partly burnt WD remnants would be ejected with a high velocity and a peculiar atmosphere due to enrichment with heavy elements due to the SN explosion, causing some unusual observational features \citep{Shen2017,Zhang2019}. To date, only a few candidates for such partly burnt WD remnants have been reported. \citet{Vennes2017} interpreted LP~40-365 as a partly burnt WD remnant due to its high proper motion and peculiar atmosphere that is dominated by IMEs \citep[see also][]{Raddi2018,Radddi2019,Hermes2021}. \citet{Ruffini2019} suggested that a hyper-runaway WD in Gaia DR2, LP~93-21, could be another candidate for such events. \citet{Foley2014} reported a post-explosion detection of a TPAGB-like source at the position of SN~2008ha by using \textit{HST} images obtained $4.1\,\mathrm{yr}$ after the explosion. They suggested that this source might be a candidate of either the bound remnant of the WD or its companion star if this source is indeed related to SN~2008ha.

\citet{Pakmor2021} have recently investigated the fate of a DD system containing a $0.80\,M_{\sun}$ primary CO~WD and a secondary hybrid HeCO~WD with a mass of $0.69\,M_{\sun}$. They found that the accretion from the secondary hybrid WD onto the more massive primary CO~WD gives rise to a He-detonation when the accumulated He-shell reaches a critical mass limit. This He-detonation burns through the accretion stream to also incinerates the He-shell of the secondary HeCO~WD, compressing the CO core of HeCO~WD to subsequently trigger a detonation. As a consequence, the HeCO~WD explodes as a faint transient, and the primary CO~WD survives from the explosion. This is different from the D$^6$ model, in which the surviving star is the secondary WD  \citep{Shen2018,Kawabata2021} rather than the primary CO~WD \citep{Pakmor2022}. Searches for such surviving primary WDs in future observations will help to confirm or rule out such an explosion mechanism.

\subsubsection{Single (extremely) low-mass WDs}
\label{sec:ELWDs}

In the SD scenario, depending on the actual mass of the companion star at the moment of SN~Ia explosion and the total amount of mass lost from companion surface by the mass-stripping during the interaction (see Section~\ref{sec:interaction}), the surviving companion stars could become single low-mass WDs (${<}\,0.5\,M_{\sun}$; \citealt{Justham2009,Wang2010,Liu2018}). In particular, hydrodynamical impact simulations for RG companion stars have shown that almost the whole envelope the RG companion is stripped off by the explosion, leaving a low-mass He-core remnant \citep{Marietta2000}. This suggests that the surviving companion stars could have masses less than ${\sim}\,0.3\,M_{\sun}$ \citep{Brown2010}, becoming single extremely low-mass WDs (ELWDs) if the companion stars  had been RG stars (Liu et al. in prep). However, \citet{Brown2016} have shown that all of the ELWDs ($\lesssim0.3\,M_{\sun}$) are found in binary systems rather than in isolation. If some SNe~Ia were indeed produced from the RG donor scenario (despite the low SN~Ia rates predicted by BPS calculations for this channel as shown in Fig.~\ref{Fig:rate}; see also \citealt{Maoz2012PASA,Liu2019DD,Liu2020aa}), some single ELWDs are expected to exist. Future observations might be able to test the RG donor channel for SNe~Ia by searching for such single ELWDs.

\subsection{Ejecta--CSM interaction}
\label{sec:csm-interaction}

In the SD scenario, a significant amount of CSM is generally expected to exist around the progenitor due to pre-explosion outflows \citep{Hachisu1999,Han2006,Patat2007, Sternberg2011, Dilday2012,Margutti2014}: \vspace{-\topsep}
\begin{itemize} \itemsep2pt 

\item[(a)] $\textit{Stellar winds}$. The companion star in the SD scenario could be a RG star or a AGB star. In this case, the companion star loses mass in slow stellar winds at a rate of $5\times10^{-9}\,M_{\sun}\,\mathrm{yr^{-1}}$ to $5\times10^{-6}\,M_{\sun}\,\mathrm{yr^{-1}}$ with a typical velocity of $5$--$30\,\mathrm{km\,s^{-1}}$ \citep{Seaquist1990,Vassiliadis1993,Chen2011}. Depending on models used for the description of wind accretion (e.g., Bondi–Hoyle–Lyttleton model, \citealt{Bondi1944}; slow wind accretion, \citealt{Liu2017}; wind Roche-lobe overflow,  \citealt{Mohamed2007}), mass loss rates from the binary system are expected to be ${\gtrsim}\,10^{-8}\,M_{\sun}\,\mathrm{yr^{-1}}$.  

\item[(b)] $\textit{Optically thick winds}$. At high mass-transfer rates in the SD scenario, a so-called ``optically thick wind (OTW)'' has been proposed to be driven from the WD surface if the mass transfer rate is higher than the critical accretion rate for stable H/He burning \citep[about $6\times10^{-6}\,M_{\sun}\,\mathrm{yr^{-1}}$,][]{Hachisu1999}. The WD accumulates companion material at a rate of the critical accretion rate, and the unprocessed material is lost by the OTW at a rate of ${\gtrsim}\,10^{-8}\,M_{\sun}\,\mathrm{yr^{-1}}$ with a velocity of a few $1000\,\mathrm{km\,s^{-1}}$ \citep{Hachisu1999,Han2006}.

\item[(d)] $\textit{Non-conservative mass transfer}$. At intermediate mass transfer rates, the steady mass-transfer happens through RLOF, and a small fraction of the transferred mass (${\lesssim}\,1\%$) may be lost over the outer Lagrangian points of the binary system with a velocity of several $100\,\rm{km\,s^{-1}}$ (up to ${\sim}\,600\,\rm{km\,s^{-1}}$, see \citealt{Deufel1999}) when the WD undergoes steady nuclear burning (see also \citealt{Chomiuk2012,Margutti2014}).

\item[(e)] $\textit{Nova explosions}$. At low mass transfer rates, the accreting WD is expected to experience repeated nova outbursts due to unsteady H/He burning on its surface \citep[e.g.][]{Starrfield1972,Nomoto1982a,Yaron2005,Moore2012,Wolf2013}. In these, nova shells are ejected with a typical velocity of $1000$--$5000\,\mathrm{km\,s^{-1}}$ with a typical mass of $10^{-7}$--$10^{-5}\,M_{\sun}$. Depending on the recurrence timescale and shell dynamics \citep{Hachisu2000, Moore2013}, the mass loss due to nova explosions is expected to be ${\lesssim}\,10^{-6}\,M_{\sun}\,\mathrm{yr^{-1}}$ \citep{Yaron2005}.

\end{itemize}

\begin{figure*}[t]
   \centering
   \includegraphics[width=0.96\textwidth, angle=0]{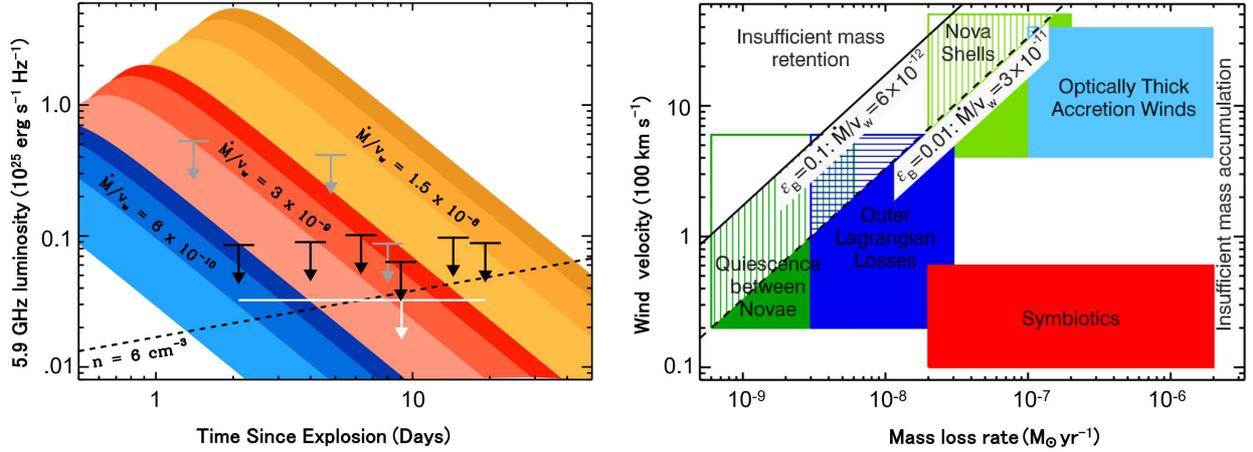}
   \caption{An example of using radio observations of SNe Ia to probe CSM properties of their progenitors \citep{Chomiuk2012}. \emph{Left panel}: A comparison between theoretical radio light curves ($colored\ swaths$) from synchrotron emissions due to the ejecta-CSM interaction and deep $3\,\sigma$ limits on the 5.9 GHz radio luminosity of SN~2011fe from Expanded Very Large Array (EVLA; $black\ arrows$). Additional data ($gray\ arrows$) from \citet[][scaled to 5.9 GHz]{Horesh2012} and a stacked limit ($white\ arrow$) are also shown. Different colored swaths give the theoretical expectations for three progenitor wind mass-loss rates highlighted in the figure in units of $M_{\odot}\,\mathrm{yr^{-1}}/\mathrm{100\,km\,s^{-1}}$ by assuming $\epsilon_{e}=0.1$. The value of $\epsilon_{B}$ ranges from 0.01 to 0.1 in each colored swath. The expected light curve from a model with uniform-density CSM of $n=6\,\mathrm{cm^{-3}}$ is presented in a \emph{dashed black line}. \emph{Righ panel}: Constraints on the parameter space (mass-loss rate vs. wind speed) for different SD progenitor models discussed above (Section~\ref{sec:csm-interaction}) for radio observations of SN~2011fe. Deep $3\,\sigma$ limits on progenitor wind mass-loss rates for the cases of $\epsilon_{B}=0.1$ and $\epsilon_{B}=0.01$ from \emph{left panel} are marked by solid and dashed line, respectively. The progenitor models located at the lower right regions of these two lines are expected to detect radio emissions by EVLA. Therefore, these progenitor models are likely to be ruled out for SN~2011fe. The figures are reprinted from \citet[][see their Figures~2 and 3]{Chomiuk2012} with the permission of the AAS.}
   \label{Fig:radio}
   \end{figure*}

While CSM interaction imprints on the observables were originally proposed as evidence for a SD progenitor evolution, it is now thought that also the DD scenario may produce significant amounts of CSM: 
\begin{enumerate}
\item[(a)] $\textit{Tidal tail ejection}$. \citet{Raskin2013} found that during the merger of two WDs (but prior to their final coalescence) ${\approx}\,10^{-4}\,$--$\,10^{-2}\,M_{\sun}$ of material can be tidally ejected (which is referred to as ``tidal tail'' mass loss) with a typical velocity of about $2\,000\,\rm{km\,s^{-1}}$ to form a CSM around the system.  Therefore, a detectable shock emission in radio, optical/UV, and X-ray wavelengths is expected to be caused by the interaction of the SN ejecta with this CSM \citep{Raskin2013}. Also, these ``tidal tail'' ejections are expected to produce relatively broad \Nai\,D absorption features at late times.

\item[(b)] $\textit{The double-detonation scenario.}$ \citet{Shen2013} suggested that multiple mass ejection episodes could happen over hundreds to thousands of years prior to the merger of a CO~WD~+~He~WD binary system in the context of sub-Chandrasekhar-mass double detonation explosion scenario \citep{Pakmor2013}. As a result, about $(3-6)\times10^{-5}\,M_{\sun}$ of material can be ejected from the binary system to the surrounding environment with a velocity of ${\sim}\,1500\,\rm{km\,s^{-1}}$.

\item[(c)]$\textit{The core-degenerate scenario.}$  \citet{Soker2013} suggested that the violent merger of a CO WD  with the core of a AGB star during the CE phase might eject some material (about $0.1\,$--$\,0.5\,M_{\sun}$) and show some signs of CSM if the SN~Ia explodes soon after the mass ejections \citep{Soker2013,Ruiter2013}.

\item[(d)]$\textit{Mass outflows during rapid accretion}$. By simulating the merger of two WDs, \citet{Dan2011} suggested that about $10^{-2}\,$ to $\,10^{-3}\,M_{\sun}$ material can be lost through the Lagrangian point with a possible velocity of about $1000\,\rm{km\,s^{-1}}$ during the rapid accretion phase of two WDs \citep[see also][]{Guillochon2010}.

\item[(e)]$\textit{Disk winds}$. \citet{Ji2013} simulated the merger of two WDs taking the the effect of  magnetic fields into account. They suggest that a rapidly rotating magnetized WD could be formed with a surrounding hot accretion disk if the WD-WD system fails to promptly detonate. They further suggested that a fraction of the disk (${\sim}\,10^{-3}\,M_{\sun}$)  could be lost with a velocity of about $2\,600\,\rm{km\,s^{-1}}$ due to a magnetically driven wind.

\end{enumerate}

\subsubsection{Radio and X-ray emission from CSM interactions}
\label{sec:radio}

After the SN explosion, the interaction of the SN~Ia ejecta with the surrounding CSM or ISM produces a shock wave, accelerating particles and amplifying the magnetic field. This leads to the emission of synchrotron radiation in radio wavelengths \citep{Chevalier1977,Chevalier1982,Chevalier1998}. Additionally, the SN shock heats the CSM or ISM to high enough temperatures ($10^{6}$--$10^{9}\,\mathrm{K}$) to produce X-ray emission by inverse Compton scattering \citep{Chevalier2006,Fransson1996}. Therefore, the detection of radio or X-ray from the CSM-interaction has long been suggested to provide a way to constrain the properties of the surrounding CSM, which can provide clues on the mass-loss history of the progenitor system prior to the SN explosion and thus place constraints on the nature of progenitor systems. However, no such radio or X-ray emission has been detectedf for most SNe~Ia \citep[e.g.][]{Immler2006, Panagia2006, Russel2012, Chomiuk2012, Chomiuk2016, Horesh2012, Margutti2012, Margutti2014, Perez-Torres2014, Kilpatrick2018, Sarbadhicary2019, Cendes2020,Lundqvist2020, Harris2021,Sand2021,Stauffer2021,Hosseinzadeh2022}, even for the nearby, well-observed events, SN~2011fe \citep[][]{Horesh2012, Margutti2012,Chomiuk2012} and SN~2014J \citep[][]{Margutti2014,Chomiuk2016,Kundu2017}.

Figure~\ref{Fig:radio} shows an example of probing the CSM properties of the progenitor of SN~2011fe with its radio observations \citep{Chomiuk12}. The lack of early-time radio detection from SN~2011fe constrains the mass-loss rate of its progenitor system to $\dot{M}=6\times10^{-10}\,M_{\sun}\,\mathrm{yr^{-1}}$ for a wind velocity of $v_{\mathrm{w}}=100\,\mathrm{km\,s^{-1}}$ (see Fig.~\ref{Fig:radio}; see also \citealt[][]{Chomiuk2012}). Moreover, an X-ray source in SN~2011fe was neitehr detected by the \emph{Swift} X-Ray Telescope nor by \emph{Chandra}, which constrains the progenitor system mass-loss rate $\dot{M}<2\times10^{-9}\,M_{\sun}\,\mathrm{yr^{-1}}$ for a wind velocity of $v_{\mathrm{w}}=100\,\mathrm{km\,s^{-1}}$ \citep{Margutti2012}. From the non-detection of early X-ray emission from SN~2014J, \citet{Margutti2014} derived an upper limit for the pre-explosion mass-loss rate of $\dot{M}<10^{-9}\,M_{\sun}\,\mathrm{yr^{-1}}$ (for wind velocity of $v_{\mathrm{w}}=100\,\mathrm{km\,s^{-1}}$) in the context of inverse Compton emission from upscattered optical photons by the SN shock \citep[see also][]{Kundu2017}. Based on a sample of 53 SNe~Ia observed with the $\textit{Swift}$ X-Ray Telescope, \citet{Russel2012} calculated a $3\sigma$ upper limit for the X-ray luminosity of $1.7\times10^{38}\,\rm{erg\,s^{-1}}$ \citep{Russel2012}, which corresponds to an upper limit on the mass-loss rate of $1.1\times10^{-6}\,M_{\sun}\mathrm{\,yr^{-1}}\times$\,$(v_{\rm{wind}})/(10\,\rm{km\,s^{-1}})$. More recently, \citet{Lundqvist2020} reported no radio detections for a sample of 21 nearby SNe~Ia with deep radio observations, which constrains their pre-explosion mass-loss rates to $\dot{M}\lesssim4\times10^{-8}\,M_{\sun}\,\mathrm{yr^{-1}}$ for a wind velocity of $v_{\mathrm{w}}=100\,\mathrm{km\,s^{-1}}$. All these results seem to disfavor the SD scenario because it usually predicts a significant amount of surrounding CSM.

In contrast, \citet{Kool2022} conducted radio searches with the \textit{electronic Multi-Element Radio Linked Interferometre Network (e-MERLIN)} with a phase-referencing mode at the $C$-band ($\mathrm{5.1\,GHz}$) for SN~2020eyj, concluding that SN~2020eyj has shown significant signatures of CSM in its spectra (narrow He emission lines). Interestingly, \citet{Kool2022} also report that radio signals were observed for SN~2020eyj in two epochs . They  argue that SN~2020eyj originated from a binary progenitor consisting of a WD and a He-star donor (i.e. WD~+~He-star model) because its spectra  show features of interaction with He-rich CSM (narrow He emission lines) and their radio detection seems to be consistent with the prediction of the WD~+~He-star model \citep{Moriya2019}. This would therefore be the first detection of radio emission from a SNe~Ia. However, whether the detected radio signals were indeed produced by SN~2020ejy still needs to be confirmed by higher-resolution radio observations.

\begin{figure*}[ht]
   \centering
   \includegraphics[width=0.88\textwidth, angle=0]{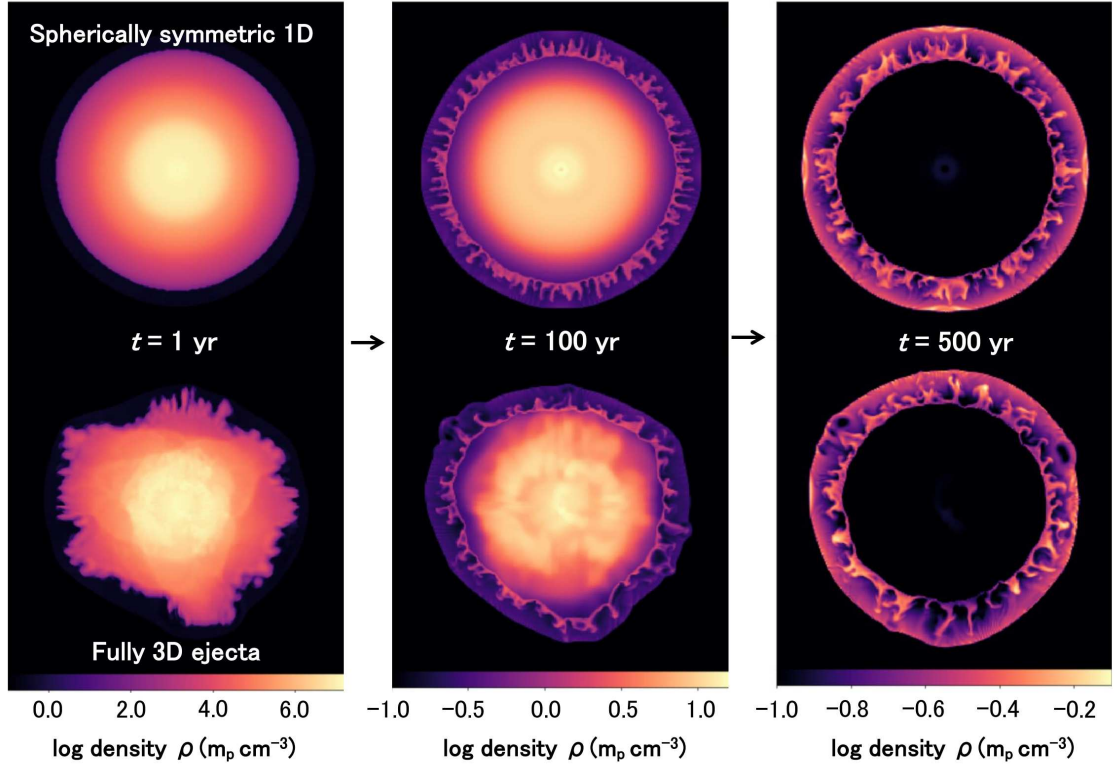}
   \caption{An example of how asymmetries in SN Ia ejecta contribute to the shaping of SNRs. The figure illustrates slices of mass density of all material at $t=1\,\mathrm{yr}$ (\emph{left}), $t=100\,\mathrm{yr}$ (\emph{middle}), and $t=500\,\mathrm{yr}$ (\emph{right}) from 3D hydrodynamical SNR simulations (i.e. the ejecta-CSM/ISM interaction) of \citet{Ferrand2019} by adopting either a 1D angle averaged spherically symmetric ejecta model (\emph{top panels}) or a fully 3D ejecta model (\emph{bottom panels}; the N100 delayed detonation model of \citealt{Seitenzahl2013}) as the initial condition. Note that different length scales are used in diagrams at different ages, which are respectively given in the order of $0.085\,\mathrm{pc}$, $5\,\mathrm{pc}$ and $13\,\mathrm{pc}$ from \emph{left to right}. The figure is reprinted from \citet[][see their Figures~3]{Ferrand2019} with the permission of the AAS.}
   \label{Fig:snrs}
   \end{figure*}

\subsubsection{Narrow spectral lines caused by CSM interaction}
\label{sec:absorption}

The CSM around SN~Ia progenitor systems is expected be ionized by the UV radiation to produce time-varying absorption features in the SN spectra, which are expected to be detected by multi-epoch high spectral resolution observations \citep{Patat2007,Sternberg2011}. The detection of such absorption features may help to place constraints on the nature of SN~Ia progenitor systems. \citet{Patat2007} reported the first detection of variable blueshifted neutral sodium (\Nai; $\lambda\lambda\,5890,5896$) absorption lines in SN~2006X. They interpreted this finding as strong evidence for a SD origin of SN~2006X. Following \citet{Patat2007}, similar time-variable absorption features associated with CSM were reported in some other SNe~Ia either from high spectral resolution observations (SN~2007le, \citealt{Simon2009}; PTF~11kx, \citealt{Dilday2012}) or from low spectral resolution observation (SN~1999cl, \citealt{Blondin2009}). \citet{Sternberg2011} studied variable \Nai absorption features in a sample of 35 nearby SNe~Ia with single-epoch high-resolution spectra. They found that 12 SNe~Ia ($\sim34\%$) displayed blueshifted absorption, and 13 SNe~Ia did not show \Nai absorption (these objects are preferentially located in early-type galaxies). They thus concluded that more than $20\%$ of SNe~Ia may result from the SD scenario. These results were confirmed by \citet{Maguire2013} who combined a new sample of 17 low-redshift SNe~Ia observed with the XShooter intermediate-resolution spectrograph on the \textit{European Southern Observatory Very Large Telescope} with the sample given by \citet{Sternberg2011}. \citet{Foley2012} suggested that SNe~Ia with blueshifted absorptions generally have higher ejecta velocities and redder colors around maximum light compared with the rest of SN~Ia population \citep[see also][]{Hachinger2017}. More recently, \citet{Graham2019} suggested that SN~2015cp was likely to have CSM properties similar to those of the well-studied event PTF11kx.

\subsubsection{Other imprints of CSM}
\label{sec:imprints}

The accreting WDs in the SD scenario are a powerful source to photoionize \Heii in the surrounding gas, producing emission in the \Heii recombination line at $\lambda~4686\,\AA$. These \Heii recombination lines can be used as a test of the nature of SN~Ia progenitors \citep{Rappaport1994,Woods2013, Johansson2014, Graur2014b,Chen2015uv}. 

If the OTW model is adopted in the SD scenario \citep{Hachisu1996,Hachisu1999}, a significant pre-explosion mass loss (or outflow) with high velocity of a few $1000\,\mathrm{km\,s^{-1}}$ is expected to modify the structure of the CSM at the time of the SN explosion. This evacuates a detectable low-density cavity around the WD \citep{Badenes2007}. However, neither a cavity around the WD nor \Heii recombination lines have been firmly detected in SNe~Ia to date \citep{Badenes2007,Johansson2014, Graur2014b}.

\subsubsection{Properties of supernova remnants}
\label{sec:SNRs}

As the SN~Ia ejecta interact with its surrounding CSM or ISM on timescales of hundreds or thousands of years, a shock wave is generated and propagates outward into the surroundings, and the SNR phase begins. The region enclosed by this shock wave, which contains both the SN ejecta and the swept-up material, is known as SNR \citep{Woltjer1972}. In general, Type Ia SNRs appear to be more spherical than those of core-collapse SNe, but some of them do show asymmetries either in their morphologies or in their emission patterns. Examples for asymmetric SNRs are Tycho \citep[SN~1574;][]{Warren2005,Vigh2011}, SN~1006 \citep{Winkler2014}, Kepler \citep[SN~1604;][]{Kasuga2018}, and the youngest known Galactic SNR~G1.9+0.3 \citep{Borkowski2013,Borkowski2017}. In particular, the observations of SNR~G1.9+0.3 demonstrated that the spatial distribution of both IMEs and Fe are significantly asymmetric \citep{Borkowski2013,Borkowski2017,Griffeth2021}.

The interaction of the SN blast wave with the CSM has been modeled in numerical simulations to make predictions on the properties of the resulting SNRs. Again, the goal is to to constrain progenitor models and explosion mechanisms  \citep[][]{Chevalier1992, Blondin2001, Badenes2007, Vigh2011, Chiotellis2012, Warren2005,Orlando2012,Borkowski2013,Burkey2013, Williams2013,Williams2017, Ferrand2019,Ferrand2022,Kasuga2021}. For instance, \citet{Badenes2007} presented 1D hydrodynamic simulations of the  ejecta--CSM interaction by assuming a pre-SN time-dependent wind. They found that the OTW from the WD surface excavates large low-density cavities around the progenitors, which is incompatible with currently observed features of known SNRs such as the Kepler, Tycho and SN~1006. \citet{Chiotellis2012} performed a 2D simulation with a spherically symmetric density profile of the SN ejecta and a wind-bubble CSM model, finding that the observed characteristics of the Kepler remnant  are in good agreement with predictions from an explosion in a dense, bow-shaped bubble formed by the wind of the AGB donor star.

Some asymmetries in SN~Ia ejecta are expected to be caused by SN explosion itself and by the presence of a companion star in the SD scenario (see Section~\ref{sec:interaction}). These asymmetries in SN~Ia ejecta might affect the subsequent ejecta-CSM interaction and therefore have an imprint on the observable properties of Type Ia SNRs such as their morphologies \citep{Ferrand2019} and the spatial distributions of different chemical elements. A long-standing question is whether asymmetries in SN~Ia ejecta caused by the explosion process contribute noticeably to the shaping of SNRs or whether the remnant properties are entirely determined by the interaction of the SN ejecta with the CSM. To test this, \citet{Ferrand2019} compared the SNR evolution in 3D simulations initialized with a spherically symmetric SN~Ia ejecta model to that an asymmetric ejecta model derived from a 3D explosion modeling (the N100 model of \citealp{Seitenzahl2013}). Fig.~\ref{Fig:snrs} shows the time evolution of the density of all material (i.e., both SN Ia ejecta and CSM) in the 3D SNR simulations of \citet{Ferrand2019}. They concluded that the impact of explosion asymmetries on the SNR morphology is still visible after hundreds of years (Fig.~\ref{Fig:snrs}). However, they neglected the effect of the ejecta--companion interaction in their SNR modeling. \citet{Vigh2011} presented 2D and 3D numerical simulations of young SNRs with an asymmetric ejecta model, in which the ejecta structure is artificially modified by considering the effect of the companion star. They concluded that the asymmetries of the SN ejecta due to the presence of a companion star can have a strong effect on the subsequent evolution of the SNR. \citet{Gray2016} found that a hole in the SN~Ia ejecta caused by the presence of a nearby companion remains visible for many centuries after the interaction between the ejecta and the ISM \citep[see also][]{Garcia-Senz2012}. \citet{Ferrand2022} performed 3D SNR simulations in the context of the D$^{6}$ model by considering the effect of the presence of a WD companion, suggesting that the WD companion produces a large conical shadow in the ejecta, visible in projection as a dark patch surrounded by a bright ring.  
 
For firm conclusions, further SNR modeling is required to comprehensively take into account the asymmetries in the ejecta caused by the ejecta--companion interaction in the context of different SD progenitor systems (e.g. the MS, SG, or He-star donor model) and to cover different explosion models. This may provide a new way to distinguish the SD and DD scenario of SNe~Ia through the SNR observations.

\begin{figure*}[ht]
   \centering
   \includegraphics[width=0.65\textwidth, angle=0]{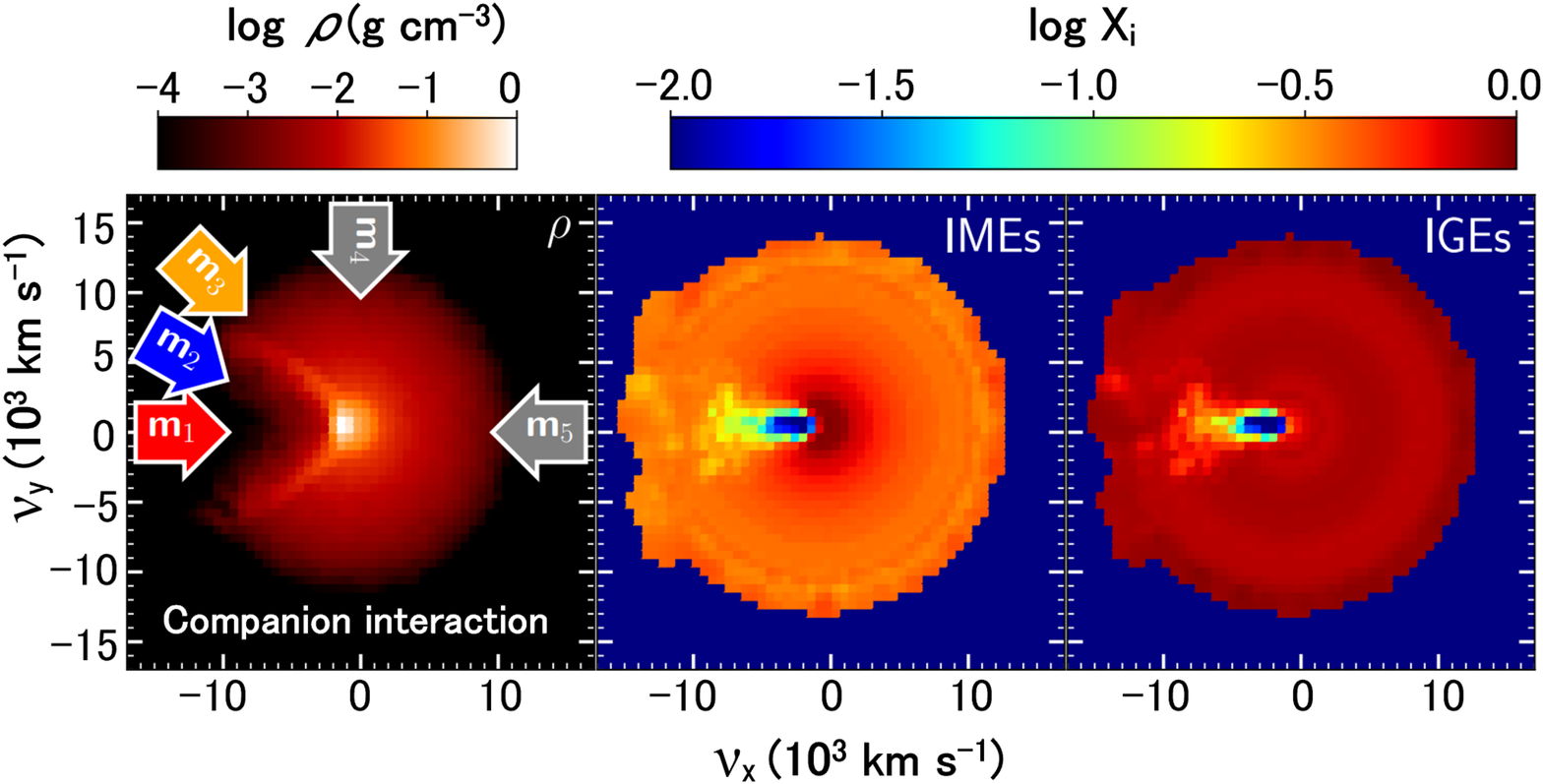}
   \includegraphics[width=0.88\textwidth, angle=0]{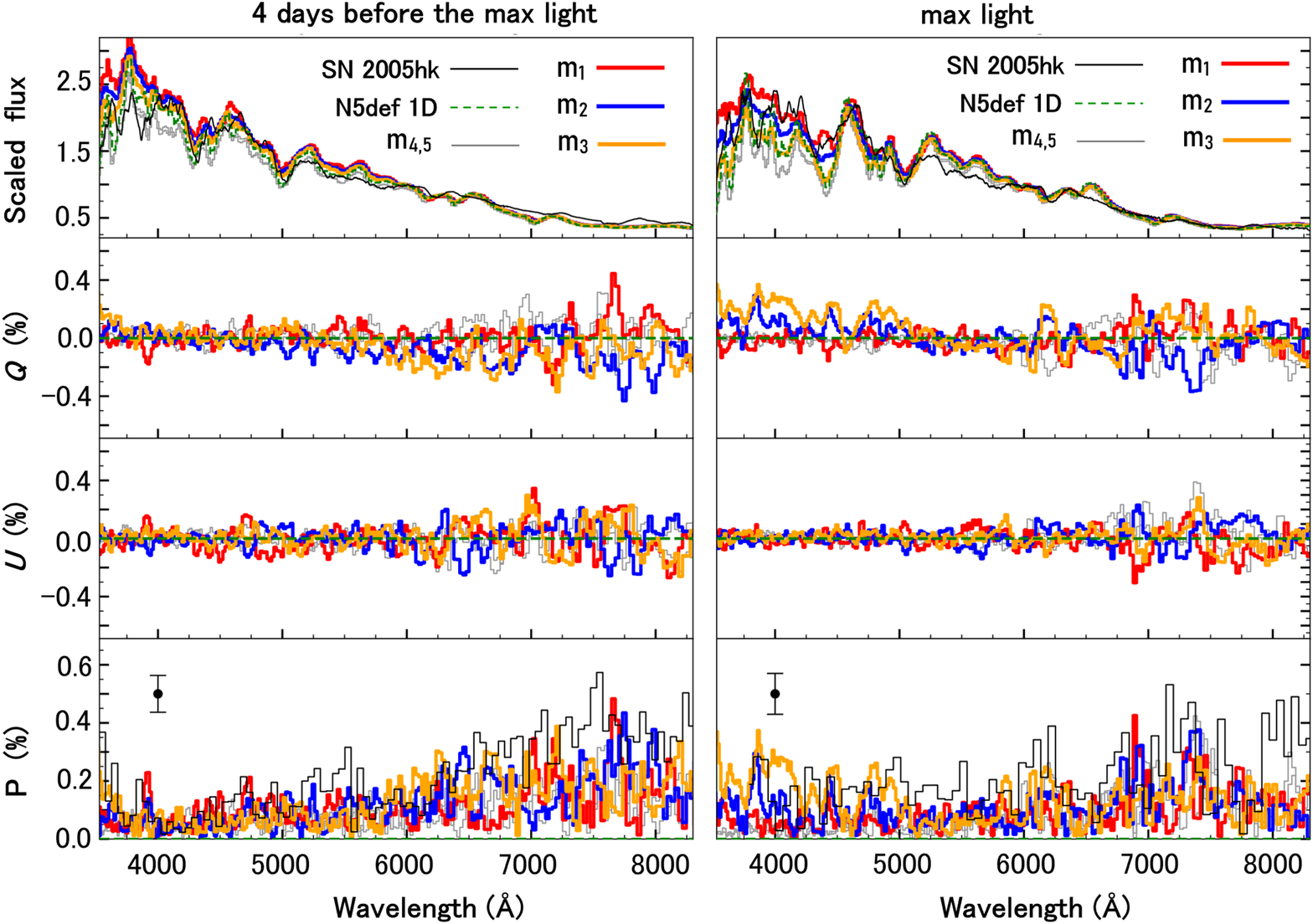}
   \caption{Theoretical flux and polarisation spectra of a weak pure deflagration explosion model (i.e. N5def model of \citealt{Fink2014}) by considering the effect of the presence of a non-degenerate companion star. \emph{Top panels}: Distribution of density (\emph{left}), IMEs (\emph{middle}) and IGEs (\emph{right}) of post-impact SN ejecta in the $x-y$ plane from a 3D hydrodynamical simulation of ejecta-companion interaction of \citet{Liu2013c}. \emph{Bottom panels}: Predicted flux, $Q$, $U$, and P spectra at pre- and around maximum light are given from \emph{top to bottom}. Different color lines give the results viewed from the five different orientations ($\mathrm{m1}$-$\mathrm{m5}$) shown in the \emph{top left panel}. Here, $Q$ and $U$ are the Stokes parameters. The results are compared with those of SN~2005hk (\emph{black solid lines}) for comparison. The figure is reproduced from \citet[][see their Fig.~3]{Bulla2020} with permission \copyright\ ESO.}
   \label{Fig:polarisation}
   \end{figure*}

\subsection{Polarization features}
\label{sec:polarization}

Scattering from electrons distributed in the ejecta of the explosion can linearly polarize light. Measuring polarization features of SNe~Ia  through spectropolarimetry is an important tool for revealing the geometry of SN~Ia ejecta and the distribution of various elements in them. This can provide constraints on the explosion geometry and thus may be helpful for disentangling progenitor models and explosion mechanisms \citep[see][for a review]{Wang2008}. Observations commonly find low levels of continuum polarization in SNe~Ia (${\lesssim}\,1\%$; \citealt{Wang2008,Cikota2019}), typically 0.2--0.3 per cent in normal SNe~Ia, and 0.3--0.8 per cent in subluminous SNe~Ia. But a few events do show higher polarization, e.g.\ SN~1999by \citep{Howell2001}, SN~2004dt \citep{Wang2006} and SN~2005ke \citep{Patat2012}. In contrast to the low-level continuum polarization, the polarization across the line profiles of spectral features associated with IMEs such as Ca, Si, sulphur (S) and Mg, has been found to be quite significant in some SNe Ia (up to $\sim\,$2 per cent; \citealt{Wang2003,Wang2006,Kasen2003,Patat2009,Patat2012,Maund2010}). This suggests significant asymmetries in  the element distribution in SN ejecta \citep{Wang2008}. Also, it has been suggested that there may exist correlations between the polarization of \Siii and either the pseudo equivalent width (pEW) or the velocity of \Siii features \citep{Meng2017,Cikota2019}. Some earlier attempts to model the spectropolarimetry of SNe~Ia assumed either an ellipsoidal ejecta shape -- which is predicted in the explosion of a rapidly rotating WD \citep{Hoflich1991,Howell2001,Wang1997,Wang2001} -- or clumped and toroidal shells in the outer ejecta layers \citep{Kasen2003} and the results have been compared to spectropolarimetry observations of SNe~Ia \citep{Hoeflich2006,Patat2012,Kasen2003}.

Asymmetric ejecta from SN~Ia explosions themselves (i.e., asymmetries in the geometry and element distribution) have been predicted by many multi-dimensional SN~Ia explosion models (\citealt{Hillebrandt2013}; see Section~\ref{sec:explosion}). Recently, synthetic spectropolarimetry has been calculated by performing Monte Carlo radiative transfer calculations for different possible explosion models, including violent mergers of two WDs \citep{Bulla2016a}, sub-Chandrasekhar mass double-detonation and Chandrasekhar-mass delayed-detonation models \citep{Bulla2016b}, and WD head-on collisions \citep{Livneh2022}. \citet{Bulla2016b} found that the sub-Chandrasekhar mass double-detonation model and the Chandrasekhar-mass delayed detonation model produce a low continuum polarization around maximum light of about $0.1$--$0.3$ per cent and it decreases after maximum light, which is in good agreement with spectropolarimetric observations of normal SNe~Ia. In contrast, \citep{Bulla2016a} showed that the violent merger model produces high  \Siii line polarization of ${\sim}\,0.5\%$--$3.2\%$, which is too strong to fit the observations of most SNe~Ia. \citet{Livneh2022} predicts a low continuum polarization of ${\sim}\,0.5\%$ for the head-on WD collision model but they find a very high \Siii line polarization of ${\sim}\,3\%$, which is inconsistent with the observations. 

All these simulations ignored the potential effect of the ejecta--companion interaction on the predicted polarization features. \citet{Kasen2004} addressed the effect of an ejecta-hole asymmetry due to the ejecta--companion interaction on the variation of the spectrum, luminosity, and polarization with viewing angle by using multi-dimensional radiative transfer calculations. They found that high continuum polarization can be seen when looking down the hole carved by the companion almost directly. However, a simplified structure and composition of the SN ejecta was used in their modeling. \citet{Bulla2020} have predicted spectropolarimetric features originating from  a Chandrasekhar-mass pure deflagration explosion simulation and the subsequent ejecta--companion interaction. They report a reasonable agreement between predicted polarization features and those observed in SN 2005hk (see Fig.~\ref{Fig:polarisation}) and suggest that the polarization seen at blue wavelengths around maximum-light was mainly caused by ejecta asymmetries due to the ejecta–companion interaction \citep{Bulla2020}. However, only a 1D averaged spherically symmetric ejecta model (which ignores the asymmetries of SN explosion itself) was used in their simulations of ejecta--companion interaction. 

For stronger conclusions on the effect of the presence of a companion star on polarization features of SNe~Ia,  realistic SN ejecta structures directly computed from simulations of ejecta--companion interaction are needed as inputs to calculate synthetic spectropolarimetry of SNe~Ia in the context of different explosion mechanisms.

\subsection{Late-Time Photometry}
\label{sec:late-photometry}

\citet{Seitenzahl2009} pointed out that the slow decay of $^{57}\rm{Co}$, $^{55}\rm{Fe}$ and $^{44}\rm{Ti}$ can contribute significantly to the light curve evolution of SNe~Ia, slowing down their decline later than ${\sim}\,900$ days after the SN explosion \citep[see also][]{Seitenzahl2017}. Different nucleosynthetic yields (e.g. $^{57}\rm{Co}$, $^{55}\rm{Fe}$ and $^{44}\rm{Ti}$) are expected from different explosion models. Therefrore,  comparing the predictions from different explosion models \citep[e.g.][]{Roepke2012} with observed late-time light curves of SNe~Ia is a potential way to constrain the explosion conditions \citep{Graur2015,Tucker2022b,Tiwari2022,Kosakowski2022}. The following important decay chains are generally considered in the study of SN~Ia late-time photometry \citep{Seitenzahl2009}:

\begin{eqnarray}\label{eq:decaychains}
& &^{56}\mathrm{Ni}  \;\stackrel{t_{1/2} = \; 6.08\,\mathrm{d}}{\hbox to 60pt{\rightarrowfill}} \; ^{56}\mathrm{Co} \; 
\stackrel{t_{1/2} = \; 77.2\, \mathrm{d}}{\hbox to 60pt{\rightarrowfill}} \; ^{56}\mathrm{Fe}, \nonumber\\*
& &^{57}\mathrm{Ni}  \;\stackrel{t_{1/2} = \; 35.60\,\mathrm{h}}{\hbox to 60pt{\rightarrowfill}}\; ^{57}\mathrm{Co} \;
\stackrel{t_{1/2} = \; 271.79\,\mathrm{d}}{\hbox to 60pt{\rightarrowfill}} \; ^{57}\mathrm{Fe},\nonumber\\*
& &^{55}\mathrm{Co}  \;\stackrel{t_{1/2} = \; 17.53\, \mathrm{h}}{\hbox to 60pt{\rightarrowfill}}\; ^{55}\mathrm{Fe} \;
\stackrel{t_{1/2} = \; 999.67\, \mathrm{d}}{\hbox to 60pt{\rightarrowfill}} \; ^{55}\mathrm{Mn},\nonumber\\
& &^{44}\mathrm{Ti}  \;\stackrel{t_{1/2} = \; 58.90\, \mathrm{yr}}{\hbox to 60pt{\rightarrowfill}}\; ^{44}\mathrm{Sc} \;
\stackrel{t_{1/2} = \; 3.97\, \mathrm{h}}{\hbox to 60pt{\rightarrowfill}} \; ^{44}\mathrm{Ca}.\nonumber
\end{eqnarray}

 To date, very late-time photometry has been observed only for a limited sample of events, including SN~2011fe \citep{Shappee2017,Dimitriadis2017,Kerzendorf2017c,Tucker2022b}, SN~2012cg \citep{Childress2015,Graur2016}, SN~2013aa \citep{Jacobson2018}, SN~2014J \citep{Yang2018,Graur2019,Li201914J}, ASASN-14lp \citep{Graur2018b} and SN~2015F \citep{Graur2018a}. The ratios  of $^{57}\mathrm{Ni}$/$^{56}\mathrm{Ni}$ and/or $^{55}\mathrm{Fe}$/$^{57}\mathrm{Co}$ of these SNe~Ia were estimated and then compared with the predictions of different explosion models. For instance, based on photometry of SN~2011fe out to about $2,400\,\mathrm{d}$ after maximum light, \citet{Tucker2022b} suggested that the late-time light curve of SN~2011fe was powered by the energy released from the decay of both $^{57}\rm{Co}$ and $^{55}\rm{Fe}$, giving rise to mass ratios of $\mathrm{log}\,(^{57}\rm{Co}/^{56}\rm{Co})\approx-1.69$ (which is reasonably consistent with the estimates for other SNe~Ia; \citealt{Graur2016,Graur2018a,Shappee2017,Dimitriadis2017,Jacobson2018,Yang2018,Li201914J}) and $\mathrm{log}\,(^{55}\rm{Fe}/^{57}\rm{Co})\approx-0.66$. They pointed out that this was the first detection of $^{55}\rm{Fe}$ in SNe~Ia. \citet{Tucker2022b} further concluded that these late-time photometry observations of SN~2011fe are consistent with the predictions either from a delayed detonation of a near-Chandrasekhar-mass WD ($1.2$--$1.3\,M_{\sun}$) with a low central density or from a sub-Chandrasekhar-mass WD ($1.0$--$1.1\,M_{\sun}$) experiencing a thin-shell double-detonation. \citet{Tiwari2022} carried out a comprehensive comparison between late-time light curves of five SNe~Ia (SN~2011fe, SN~2012cg, SN~2013aa, SN~2014J and SN~2015F) and nucleosynthetic yilds of publicly-available SN~Ia models, suggesting that SN~2015F favors the sub-Chandrasekhar progenitor model, and other SNe~Ia are consistent with both near-Chandrasekhar and sub-Chandrasekhar progenitor models.

Either a non-degenerate or WD companion star (in the SD or $\mathrm{D^{6}}$ model; see Section~\ref{sec:survivor}) or a bound WD remnant (in weak pure deflgration explosion model; see Section~\ref{sec:survivor-remnant}) are expected to survive the explosion. These surviving stars could become more luminous than their pre-explosion brightnesses after the explosion due to significant heating by the explosion and the radioactive decay of captured SN~Ia ejecta onto the star. As SNe~Ia fade at late time phases, the shocked companion star and/or WD remnant could become more luminous than SN itself  \citep{Podsiadlowski2003,Pan2012b,Pan2013,Shappee2013b,Shen2017,Liu2021a,Liu2021b,Liu2022ApJ,Zeng2022apj}. This suggests that the surviving companion star or WD remnant could begin to dominate light curves of SNe~Ia at late-time phases, causing the light curves to decline slower than those of SN explosions themselves.

\subsection{X-ray and EUV emission from accreting WDs}
\label{sec:accreting-WDs}

In the SD scenario, the accreting WDs are expected to respectively undergo an extreme UV (EUV) radiation phase and a supersoft X-ray source (SSS) phase ($0.3$--$0.7\,\mathrm{keV}$; \citep{VandenHeuvel1992, Iben1994, Kahabka1997, Yoon2003, Nomoto2007,Chen2015uv} before the SN explosions during the stable nuclear burning and the optically thick wind regimes. \citet{Chen2015uv} conducted a population synthesis study of accreting WDs in the context of SD scenario to make predictions on X-ray and EUV emissions from populations of accreting WDs.

Comparisons of the observed SSSs in galaxies of different morphological types with X-ray expectations from the SD model have been used to put constraints on SN~Ia progenitor models \citep[e.g.][]{Di-Stefano2010a, Di-Stefano2010b,Gilfanov2010, Woods2017,Woods2018,Kuuttila2019}. For instance, \citet{Gilfanov2010} reported that the observed X-ray flux from six nearby elliptical galaxies and galaxy bulges is lower than the predictions from the SD scenario by a factor of $30$--$50$. They therefore concluded that no more than five per cent of SNe~Ia in early-type galaxies can be produced by the SD scenario. However, \citet{Hachisu2010} pointed out that there are still uncertainties on the theoretical X-ray luminosity of the SSSs, such as the atmospheric models of accreting WDs and absorption of soft X-rays. Depending on the different mass-accretion rates, the accreting WDs would undergo optically thick wind evolution, SSS phases and nova outbursts. \citep{Hachisu2010} further argued that the SSS phase should be short in duration, because most of the accreting WDs in the SD scenario spend a large portion of time in the optically thick wind phase and the recurrent nova phase. However, \citet{Chen2015uv} predicted that also the optically thick wind evolution is expected to produce significant EUV emission.

\section{Summary and future perspectives}
\label{sec:summary}

Despite substantial efforts, the progenitor systems and explosion mechanisms of SNe~Ia have been established neither observationally nor theoretically, although there is consensus that SNe~Ia arise from thermonuclear explosions of WDs in binary systems. Different progenitor models and explosion mechanisms have been proposed for SNe~Ia in the literature. We have reviewed theoretical predictions for observables of SNe~Ia at different stages of the explosion and their comparison with observations in the context of currently proposed progenitor models. Despite substantial progress in the understanding of the origin and the explosion physics of SNe~Ia over the past years, there are still many open questions that need to be answered by future studies taking both theoretical and observational approaches.
\begin{itemize}
\renewcommand{\labelitemi}{\tiny$\blacksquare$}
 \setlength\itemsep{0.5em}

\item \textbf{How can different progenitor and explosion scenarios be related to the diverse zoo of SN~Ia subclasses?} SNe~Ia have been used as excellent cosmological distance indicators because of the so-called Phillips relation, an empirical tight correlation between the peak luminosity of SNe~Ia and the width of the light curve \citep{Pskovskii1977,Phillips1993,Phillips1999}. However, the physics causing the Phillips relation is still unknown. There is growing evidence for different subclasses of SNe~Ia that are in various characteristics clearly distinct from normal SNe~Ia and deviate from the Phillips relation \citep{Taubenberger2017}. Connections between potential explosion models and some subclasses of SNe~Ia have been established. For instance, SNe~Iax have been associated with off-center ignited deflagrations in Chandrasekhar-mass WDs \citep{Phillips2007, Jordan2012def, Kromer2013, Kromer2015}, a violent merger of two WDs seems to be a possible explanation of observed features of a subluminous, slowly declining SN~2002es-like event, SN~2010lp \citep{Kromer201310lp, Pakmor2013}. However, the origin of the diversity of SNe~Ia is still far from being resolved. 

\item \textbf{How do the WDs reach an explosive state? What are the progenitor systems of SNe~Ia?} Despite the substantial efforts undertaken from both the theoretical and observational side, the questions of how the WD reaches an explosive state and what progenitor systems are more likely to produce SNe~Ia remain open. No single published model is able to consistently explain the observational features and full diversity of SNe~Ia. A new progenitor paradigm might be needed if all SNe~Ia arise from the same origin. Otherwise, a contribution of several distinct progenitor channels may be an alternative. 

\item \textbf{What is the efficiency of mass accumulation  onto an accreting WD?} In the SD scenario, the accreting WD does not retain a significant fraction of the transferred material. Only a fairly narrow range of accretion rates  allows for steady mass accumulation, which makes it difficult to explain the observed SN~Ia rates in the context of that scenario. However, the exact mass-retention efficiency is still not well-constrained because of uncertainties in the novae phase at low mass transfer rates and the treatment at high mass transfer rates. BPS calculations have indicated that different retention efficiencies could have a strong influence on the predicted rates and delayed times of SNe~Ia \citep{Bours2013,Ruiter2014,Toonen2014}. The so-called optically thick wind model \citep{Hachisu1999} is generally adopted at high mass transfer rates to avoid the formation of a CE. However, the optically thick wind does not work at low metallicities, which seems to be in conflict with the detection of high-redshift ($z>2$) SNe~Ia. Also, the mixing between the accreted envelope and the
WD during the novae phase is still poorly constrained, leading to uncertainties in the mass retention during the nova outbursts \citep{Yaron2005,Denissenkov2013}. Furthermore, the donor is expected to be irradiated by nova eruptions. How the nova irradiation affects the donor star and thus the mass transfer rate is not well understood \citep{Ginzburg2021}.   

\item \textbf{What is the critical He shell mass in the sub-Chandrasekhar double-detonation model?} Early studies have suggested that double detonations of sub-Chandrasekhar mass WDs produce heavy elements in the outer ejecta layers that drive the flux too far into red colors to be consistent with the observations of normal SNe~Ia \citep{Kromer2010}. But it seems possible that a He detonation could already trigger after accreting a rather thin He shell \citep{Bildsten2007, Shen2009,Shen2018sub,Woosely2011} and then lead to a core detonation \citep{Fink2007,Fink2010,Gronow2020,Gronow2021,Boos2021}. It has been suggested that explosions caused with thin accreted He layers may be a viable scenario for normal SNe~Ia \citep{Townsley2019,Boos2021,Shen2021}. However, the critical He-shell mass that can successfully trigger double-detonations of sub-Chandrasekhar mass WDs remains uncertain, 
 as well as its dependency on the mass of the accreting WD.

\item \textbf{Can stripped H/He in the SD scenario be hidden?} Current hydrodynamic simulations predict that a significant amount of H/He can be stripped off from the companion surface by the SN blast wave. This seems to imply the presence of H/He lines in late-time spectra. However, no H/He lines have  been  detected  in  normal SNe~Ia but only in two fast-declining, sub-luminous events (see Section~\ref{sec:late-spectra}). This poses a serious challenge to the SD scenario. However, such conclusions are based on models with an analytical ejecta structure or the so-called 1D W7 model to represent a SN~Ia explosion. The properties of this simplified SN~Ia ejecta model differ from those  directly derived from 3D explosion modeling. This leads to  uncertainties not only on the amount of stripped H/He masses, but also on the prediction of late-time H/He lines and properties of surviving companion stars. To determine whether the stripped-off H/He companion material would be observable in SN~Ia late-time spectra, a new quality of impact simulations based on realistic 3D ejecta structures is required. The output ejecta structure predicted from such impact simulations should be used directly as input for subsequent 3D radiative transfer modeling to test for H/He lines in late-time spectra.

\item \textbf{Where are surviving companions in SD scenario?} Searching for the surviving companion stars in SNRs is thought to be a promising way to potentially identify SD progenitor systems of SNe~Ia. Although a few surviving WD candidates of the sub-Chandrasekhar mass double-detonation model have been reported in the literature, no promising surviving companion star predicted by the SD scenario has been identified yet beyond doubt in SNRs \citep{Ruiz-Lapuente2019}. Some results seem to support an SD origin for at least a few SNe~Ia. For instance, the detection of variable blueshifted absorption lines associated with CSM in some SNe~Ia. Also, \citet{Seitenzahl2013solar} suggested that at least a certain fraction of Chandrasekhar-mass WD explosions are needed to explain the solar manganese-to-iron ratio, which seem to favor the SD scenario. The question is whether the surviving companion stars of SD SNe~Ia do exit. The spin-up/spin-down model predicts that the companion star could either be a MS star, a RG, a subdwarf-B (sdB) star, or a WD (depending on the unknown spin-down timescale), which may explain the non-detection of surviving companion stars. Because of this, \citet{Meng2019} suggested that searches for such stars should be done preferentially in the $U/UV$ bands.  Observations are still needed to search for such surviving companions to test the SD scenario. As mentioned above, the predictions on the observational signatures should be improved by adopting realistic 3D ejecta structures in hydrodynamical impact simulations.

\item \textbf{Do early excess emissions observed in SNe~Ia share the same origin?} For a growing list of SNe~Ia early excess emission has been reported. These objects are sometimes referred to as ``early-excess SNe~Ia''. Different mechanisms have been proposed to produce the observed bumps in the early light curves of SNe~Ia (see Section~\ref{sec:early}). It remains unclear, however, whether these early-time features share the same origin. The discovery of more early excess SNe Ia in future observations will enable us to conduct statistical studies of their observational properties, and therefore provide deeper insight into the origin of this feature.

\item \textbf{What imprint leaves the companion star on SN~Ia observables from different stages of the explosion?} Over the past years, the SN~Ia explosion mechanism and the evolution of potential progenitors have been studied in quite some detail. However, the presence of a binary companion star in the explosion process and its implications for observations has received less attention. Taking this effect into account holds promise
for identifying viable progenitor systems. Conversely, its ignorance leads to a lack of realism in the observables predicted from explosion simulations. Future studies are encouraged to address the role of a companion star in shaping the observables of SNe~Ia from different phases of the explosion, such as the effect on late-time spectra, polarization signals, and SNRs. Such investigations are needed to better constrain SN~Ia progenitor systems and explosion mechanisms.

\item \textbf{How can the ``spin-up/spin-down'' model be identified observationally?} It seems that the spin-up/spin-down model is able to provide good explanations for different observational facts on SNe~Ia such as the non-detection of a pre-explosion companion star, the lack of radio/X-ray emission, and the absence of surviving companion stars and swept-up H/He in late-time spectra. On the one hand, the constraints on the exact spin-down timescale is still quite weak. On the other hand, a smoking gun is needed to observationally identify the spin-up/spin-down model. This requires numerical modeling in the context of the spin-up/spin-down model to make predictions for detectable signatures of this model and to identify the properties of the donor star.

\item \textbf{How can we improve the SN~Ia peak-brightness calibration?} With the Phillips relation, SNe~Ia have been successfully used as cosmic distance indicators to measure the accelerated expansion of the Universe. To further advance the precision of SN~Ia cosmology and to constrain the properties of dark energy, the peak-brightness standardization of SNe~Ia needs to be improved. The underlying causes of the variability of peak luminosities should be understood in order to unveil the physics behind the Phillips relation and to reduce potential systematic errors of the distance measurements. One of the keys is to understand the dependence of SN~Ia properties on the redshift, galaxy type, host environment by statistical studies with large samples of well-observed SNe~Ia.

\end{itemize}

Further understanding of the nature of SN~Ia progenitors and their explosion mechanism relies not only on the improvements to theoretical modeling (binary evolution and population synthesis calculations, impact simulations, explosion modeling and radiative transfer calculations, etc) but also on better observational constraints. This requires both individual studies of well-observed SNe~Ia and systematical investigations of different subclasses of SN~Ia, which can be realized with ongoing surveys and projects like ZTF \citep{Smith2014,Bellm2019}, \textit{Gaia} \citep{Gaia2016,Gaia2018}, JWST \citep{Gardner2006}, ATLAS \citep{Tonry2018}, PTF/iPTF \citep{Law2009, Rau2009}, ASAS-SN \citep{Shappee2014}, Pan-STARRS \citep{Kaiser2002}, YSE \citep{Jones2021}, DES \citep{Dark2016}, \textit{SkyMapper} \citep{Keller2007}, CRTS \citep{Drake2009}, OGLE-IV \citep{Wyrzykowski2014}, \textit{La Silla-QUEST} \citep{Baltay2013} and \textit{Subaru Hyper Suprime-Cam Transient Survey} \citep{Tanaka2016}, and forthcoming projects with the \textit{Vera C. Rubin Observatory LSST} \citep{Ivezic2008,Ivezic2019}. In particular, LSST is expected to discover ten million SNe during a ten-year survey, which will yield hundreds of thousands of SNe~Ia with a broad range in redshifts. The combination of statistical investigations of such large samples of SNe~Ia with individual studies of well-observed objects will allow to address many questions of SN~Ia (like the rates, the progenitor systems, the explosion mechanism, the host properties, the origin of diversity and the details of variation in peak luminosity) in more detail, and will also lead to major advances in the precision of SN~Ia cosmology \citep{LSST2009}.

\begin{acknowledgements}
We thank the anonymous referee for useful comments that helped to improve this article. We are grateful to Robert Izzard, Xuefei Chen, Ji-An Jiang, Takashi Moriya, Xiangcun Meng, and Hai-Liang Chen for very fruitful discussions. We thank Ji-An Jiang for sharing the data of their double-detonation models for partly reproducing Figure~12 of this article. Also, Figure~12 made use of the Heidelberg Supernova Model Archive (HESMA, see \url{https://hesma.h-its.org}; \citealt{Kromer2017}). This work is supported by the National Natural Science Foundation of China (NSFC, Nos.\ 12288102, 12090040/1, 11873016), the National Key R\&D Program of China (Nos.\ 2021YFA1600401 and 2021YFA1600400), the Chinese Academy of Sciences (CAS), the International Centre of Supernovae, Yunnan Key Laboratory (No. 202302AN360001), the Yunnan Fundamental Research Projects (grant Nos.\ 202201BC070003, 202001AW070007) and the ``Yunnan Revitalization Talent Support Program''—Science \& Technology Champion Project (NO.~202305AB350003). The work of FR is supported by the Klaus Tschira Foundation and by the Deutsche Forschungsgemeinschaft (DFG, German Research Foundation) -- Project-ID 138713538 -- SFB 881 (``The Milky Way System'', Subproject A10). 
\end{acknowledgements}

\bibliographystyle{raa}
\bibliography{ref}

\end{document}